\documentclass[9pt,twoside,twocolumn]{article}

\usepackage[T1]{fontenc}
\usepackage[utf8]{inputenc}
\usepackage[a4paper,
            top=20mm, bottom=20mm,
            left=14mm, right=14mm,
            columnsep=5mm]{geometry}
\usepackage{microtype}
\usepackage{amsmath,amssymb,bm}
\usepackage{graphicx}
\usepackage{booktabs}
\usepackage{array}
\usepackage{longtable}
\usepackage{multirow}
\usepackage{siunitx}
\usepackage{url}
\usepackage{xcolor}
\usepackage[numbers,compress,square]{natbib}
\usepackage[colorlinks=true,
            linkcolor=blue!60!black,
            citecolor=blue!60!black,
            urlcolor=blue!60!black]{hyperref}
\usepackage{titlesec}
\usepackage{caption}
\usepackage{float}
\usepackage{helvet}
\usepackage[symbol]{footmisc}

\DeclareSIUnit{\Molar}{M}

\newsavebox{\fitbox}
\newcommand{\fitcolumn}[1]{%
	\sbox{\fitbox}{#1}%
	\ifdim\wd\fitbox>\columnwidth
		\resizebox{\columnwidth}{!}{\usebox{\fitbox}}%
	\else
		\usebox{\fitbox}%
	\fi}
\newcommand{\fitpage}[1]{%
	\sbox{\fitbox}{#1}%
	\ifdim\wd\fitbox>\textwidth
		\resizebox{\textwidth}{!}{\usebox{\fitbox}}%
	\else
		\usebox{\fitbox}%
	\fi}

\newcommand{\fiteq}[1]{\ensuremath{\fitcolumn{$\displaystyle #1$}}}

\newlength{\figmaxh}
\titleformat{\section}{\large\bfseries}{\thesection}{0.6em}{}
\titleformat{\subsection}{\normalsize\bfseries}{\thesubsection}{0.5em}{}
\titleformat{\subsubsection}{\small\bfseries}{\thesubsubsection}{0.5em}{}
\titlespacing*{\section}{0pt}{10pt}{4pt}
\titlespacing*{\subsection}{0pt}{6pt}{2pt}
\titlespacing*{\subsubsection}{0pt}{5pt}{2pt}
\newcommand{\Phisim}{\ensuremath{\Phi_{\text{sim}}}}
\newcommand{\dpeak}{\ensuremath{d_{\text{peak}}}}
\newcommand{\Nseg}{\ensuremath{N_{\text{seg}}}}
\newcommand{\Lmax}{\ensuremath{L_{\text{max}}}}
\newcommand{\ETE}{\ensuremath{\mathrm{ETE}}}
\newcommand{\Rg}{\ensuremath{R_{\text{g}}}}
\newcommand{\Lc}{\ensuremath{L_{\text{c}}}}
\newcommand{\phimech}{\ensuremath{\varphi_{\text{mech}}}}
\newcommand{\phichem}{\ensuremath{\varphi_{\text{chem}}}}
\newcommand{\Rghat}{\ensuremath{\hat{R}_{\text{g}}}}
\newcommand{\fss}{\ensuremath{f_{\text{SS}}}}
\newcommand{\Ccov}{\ensuremath{C}}

\begin{document}

\twocolumn[{%
    \vspace*{4mm}
    {\LARGE\bfseries
        Predicting protein structures at any folding stage from sequence\par}

    \vspace{5mm}
    {\normalsize
        Lisheng Zhang\textsuperscript{1,\dag},\quad
        Qingsong Li\textsuperscript{2,\dag},\quad
        Zinan Xie\textsuperscript{3,\dag},\quad
        Yanting Yang\textsuperscript{4,5,6,\dag},\quad
        Qikui Yang\textsuperscript{7,\dag},\quad
        Peng Lu\textsuperscript{11,12},\quad
        Zhican Fu\textsuperscript{11},\quad
        Wentao Cui\textsuperscript{11},\quad
        Runtong Hou\textsuperscript{11},\quad
        Zhengwei Xie\textsuperscript{1,8,9,10,11,*}\par}

    \vspace{3mm}
    {\footnotesize
        \textsuperscript{1}Peking University International Cancer Institute, Peking University--Yunnan Baiyao International Medical Research Center, School of Basic Medical Sciences, Peking University Health Science Center, Peking University, 38 Xueyuan Road, Haidian District, Beijing 100191, China\\
        \textsuperscript{2}Yuanpei College, Peking University, 5 Yiheyuan Road, Haidian District, Beijing 100871, China\\
        \textsuperscript{3}The Hong Kong University of Science and Technology (Guangzhou), Nansha, Guangzhou, Guangdong, China\\
        \textsuperscript{4}Tsinghua University, Beijing 100084, China\\
        \textsuperscript{5}Shenzhen Medical Academy of Research and Translation, Shenzhen, Guangdong 518107, China\\
        \textsuperscript{6}Shenzhen Bay Laboratory, Shenzhen, Guangdong 518132, China\\
        \textsuperscript{7}School of Pharmaceutical Sciences, Peking University Health Science Center, 38 Xueyuan Road, Haidian District, Beijing 100191, China\\
        \textsuperscript{8}State Key Laboratory of Natural and Biomimetic Drugs, Department of Molecular and Cellular Pharmacology, School of Pharmaceutical Sciences, Peking University Health Science Center, 38 Xueyuan Road, Haidian District, Beijing 100191, China\\
        \textsuperscript{9}Department of Prosthodontics, Peking University School and Hospital of Stomatology, Beijing 100081, China\\
        \textsuperscript{10}Beijing Key Laboratory of Intelligent Drug Discovery, Beijing Gigaceuticals Tech.\ Co.\ Ltd., Beijing 102206, China\\
        \textsuperscript{11}Beijing Life Science Academy, Beijing 102209, China\\
        \textsuperscript{12}China Tobacco Gene Research Center, Zhengzhou Tobacco Research Institute of CNTC, Zhengzhou 450001, China\\[3pt]
        \textsuperscript{\dag}These authors contributed equally to this work.\\
        \textsuperscript{*}Correspondence:
        Zhengwei Xie (xiezhengwei@hsc.pku.edu.cn)\\
        ORCID:
        Lisheng Zhang (\href{https://orcid.org/0009-0005-2277-0836}{0009-0005-2277-0836});
        Zhengwei Xie (\href{https://orcid.org/0000-0001-9572-878X}{0000-0001-9572-878X})\par}

    \vspace{6mm}
    \hrule
    \vspace{3mm}

	\noindent\textbf{Abstract.}\enspace
	The thermodynamic hypothesis holds that an amino acid sequence encodes a
	unique native structure, and deep learning has made that structure routinely
	predictable. Biological function, however, arises from the whole
	conformational continuum linking the unfolded ensemble to the folded state,
	and a predictor accurate only at the native end carries no information about
	the barrier, the collapse transition or the high-entropy unfolded side.
	Mapping that continuum today requires molecular dynamics simulation of every
	protein of interest. Here we present FoldScape, a deep learning framework
	that generates C$\alpha$ conformational ensembles conditioned on a target
	folding fraction $Q$ directly from sequence. FoldScape decomposes
	three-dimensional reconstruction into a hierarchy of one-dimensional
	inter-residue distance regressions over sliding windows, assembles them into
	a chain-level representation, and samples ensembles with a rectified-flow
	decoder trained on a multi-scale corpus of coarse-grained folding
	trajectories and all-atom mechanical-unfolding trajectories, comprising
	approximately one million protein conformations from 10{,}000 protein
	sequences. FoldScape supports controlled conformational generation from
	unfolded states through folding intermediates to native-state conformations.
	At matched sample size, FoldScape achieves median conformational coverage
	of 0.989, 0.722 and 0.877 along the radius of gyration, native-contact
	fraction $Q$ and secondary-structure fraction, respectively, compared with
	0.05, 0.05 and 0.10 for seed-varied AlphaFold~3 ensembles. The substantially
	broader coverage is consistent across the evaluated proteins, order parameters
	and binning choices. Reconstruction accuracy remains approximately stable
	across the evaluated stages of a folding trajectory, and a stride-one
	crop-and-superpose inference scheme extends prediction to chains several
	times longer than a single chain-level forward pass accepts.
	We further show that the training trajectories sample the mechanical
	unfolding pathway at residue resolution, that the ensemble conditioned on
	elongation is reproducible in its continuous observables across pulling
	velocities, and that partially
	unfolded chains decompose into geometrically coherent segments---the
	physical modularity the architecture is built to exploit. Accuracy that does
	not degrade along the folding coordinate, and an ensemble that actually
	spans that coordinate, are the two properties a conformational generator
	must hold together if it is to describe the folding process rather than only
	its endpoint, and FoldScape holds them without per-protein simulation.
	\vspace{3mm}
	\hrule
	\vspace{5mm}
}]


\section{Introduction}
 

The thermodynamic hypothesis, established by Anfinsen's work on
ribonuclease refolding, holds that the amino acid sequence of a protein
encodes a unique native conformation corresponding to the global minimum
of Gibbs free energy\cite{anfinsen1973principles}. Energy landscape theory subsequently formalized
folding as a progressive, funnel-shaped search through conformational
space, in which the fraction of native contacts $Q$ ($Q \in [0,1]$)
serves as a natural reaction coordinate connecting the high-entropy
unfolded ensemble to the low-energy native state\cite{bryngelson1995funnels,onuchic1997theory,wolynes1997folding}. Deep learning
has since transformed static structure prediction---AlphaFold 3 predicts
proteins, nucleic acids, small molecules and their complexes with
near-experimental accuracy\cite{abramson2024alphafold3}, while RoseTTAFold and ESMFold offer
complementary approaches\cite{baek2021rosettafold,lin2023esmfold}, yet these methods yield only static
snapshots of the native state. Indeed, current structure predictors have
not been shown to learn the physics of protein folding\cite{dill2012folding,outeiral2022physics}. Because
folding is an inherently dynamic process, static native-state
predictions alone cannot capture the energetics and kinetics that govern
biological function. Protein function arises not from a single structure
but from the full conformational continuum: the continuous transitions,
transient intermediates and thermodynamic distributions that connect the
unfolded ensemble to the folded state\cite{henzlerwildman2007dynamics}. Here we show
that FoldScape, a deep learning
framework trained on a multi-scale corpus of coarse-grained folding and
constant-velocity mechanical-unfolding trajectories, generates C$\alpha$
conformational ensembles conditioned on a target folding fraction $Q$ directly
from amino acid sequence, and so makes the conformational continuum from the
unfolded ensemble through folding intermediates to the native folded state
accessible without running a molecular dynamics simulation for every
protein of interest.

All-atom molecular dynamics (MD) simulations can in principle map
folding landscapes, but reaching the millisecond timescales required for
even small proteins demands extraordinary computational
resources---purpose-built hardware such as Anton\cite{shaw2010atomic,lindorfflarsen2011fastfolding} or months of GPU
time. Coarse-grained models, including Gō-type potentials\cite{clementi2000topological} and
the AWSEM force field\cite{davtyan2012awsem}, reduce cost but sacrifice transferability and
still require independent per-protein simulations. Enhanced-sampling
techniques such as umbrella sampling and WHAM reconstruct free-energy
profiles along predefined reaction coordinates\cite{torrie1977umbrella,kumar1992wham}, but demand
extensive per-protein simulations and careful selection of collective
variables. Furthermore, few standardized databases of complete folding
trajectories exist, leaving the field without a common training resource
for data-driven approaches. Recent generative deep learning methods have
begun to sample equilibrium conformations with growing fidelity:
Boltzmann generators employ normalizing flows to sample equilibrium
states\cite{noe2019boltzmann}, BioEmu emulates protein equilibrium ensembles with
approximately 1 kcal mol$^{-1}$ free-energy accuracy\cite{lewis2025bioemu}, AlphaFlow applies
flow matching to generate protein conformational ensembles\cite{jing2024alphaflow}, and the
Distributional Graphormer predicts equilibrium distributions conditioned
on sequence\cite{zheng2024distributional}. However, these methods target equilibrium sampling
without explicit control over the folding reaction coordinate, and thus
cannot trace the progressive structural evolution from unfolded to
folded states. Stability predictors provide scalar $\Delta G$ without
per-stage structures\cite{lee2026stability}, and iterative AlphaFold 2 approaches
approximate folding pathways but yield no conformational ensemble at a
specified folding stage\cite{chang2026pathways}.
No existing method generates three-dimensional conformational ensembles at a
prescribed value of the folding fraction $Q$ directly from sequence.

The key insight of FoldScape is to exploit the chain topology of
polypeptides: motivated by the low-dimensional manifold hypothesis\cite{das2006landscapes},
three-dimensional coordinate reconstruction is decomposed into a
hierarchy of seven one-dimensional inter-residue distance predictions,
each resolving a single additional geometric degree of freedom, reducing
a non-convex $3N$-coordinate problem to a sequence of well-posed
one-dimensional regressions. The framework is trained on a multi-scale
trajectory corpus generated
in-house: G\=o-model coarse-grained simulations near the folding transition
temperature, and constant-velocity steered all-atom simulations that drive
mechanical unfolding, spanning diverse protein topologies across the full
unfolding--refolding transition. Public molecular dynamics repositories and
OpenAWSEM trajectories are used for cross-validation and are held out of
training. The model employs a PairFormer backbone adapted from
AlphaFold 3\cite{abramson2024alphafold3} to capture both local and long-range
residue--residue interactions. FoldScape therefore decodes end to end, from an
amino acid sequence and a specified folding stage directly to three-dimensional
conformations belonging to that stage, with no per-target simulation anywhere
in the path.


\section{Methods}

\subsection{Pipeline overview}

FoldScape generates conditional ensembles of protein C$\alpha$ traces in seven sequential stages, each handing a frozen set of features to the next. (1)~A non-redundant set of AlphaFold-predicted structures is used to seed equilibrium folding/unfolding and steered-pulling molecular dynamics (MD) trajectories that supply the conformational sampling target. (2)~Conformational latent pretraining distils each conformation into a 1{,}152-dimensional latent vector $\mathbf{v}$ that decouples sequence identity from fold state. (3)~Round~1 (R1) reads the conditioning signal established in conformational latent pretraining and predicts, for every $W{=}15$-residue sliding window, the C$\alpha$--C$\alpha$ distances at all $K{=}7$ short sequence separations together with a windowed C$\alpha$ trace. (4)~Round~2 (R2) takes R1's full output as side-channel features, refines it through feature-wise linear modulation (FiLM\cite{perez2018film}) and emits residual distance and coordinate corrections, plus four auxiliary geometric and global predictions. (5)~Stage~1 lifts every per-window R1/R2 feature to the chain index by cross-window attention pooling and runs a 6-layer chain-level PairFormer\cite{jumper2021alphafold,abramson2024alphafold3} trunk that produces the single representation $\mathbf{s}\in\mathbb{R}^{N\times384}$ that conditions flow matching. (6)~A 5-layer AlphaFold3-style coordinate decoder\cite{abramson2024alphafold3,geffner2025proteina} learns a rectified-flow conditional vector field\cite{lipman2023flow,liu2023rectified,albergo2023stochastic} on $\mathbb{R}^{N\times3}$ from $(\mathbf{X}_t,t,\mathbf{s})$ and at inference produces conformational ensembles by Euler integration from a Gaussian prior. (7)~Two orthogonal back-mapping schemes recover all-atom coordinates from each predicted C$\alpha$ trace. R1, R2, Stage~1 and the flow-matching decoder are trained sequentially with all upstream stages frozen and their outputs pre-cached on disk.

\subsection{Protein structure dataset}

We retrieved predicted protein structures from the AlphaFold Protein
Structure Database (AlphaFold~DB)\cite{jumper2021alphafold,varadi2024alphafold},
covering 20 species across mammals, birds, fish, amphibians, insects,
nematodes, bacteria, fungi and plants, with \textit{Homo sapiens}
accounting for approximately 30\% of the total.
Structures were filtered by a global predicted local distance difference
test (pLDDT) score $\geq$70, a high-confidence residue fraction
(pLDDT $\geq$70) $\geq$50\%, and a sequence length of 50--2{,}000 amino
acids, leaving 391{,}027 structures (AF-Raw391K). Sequence redundancy was
removed with MMseqs2\cite{steinegger2017mmseqs2} in two stages: an
initial pass at 65\% identity with 80\% bidirectional coverage gave
116{,}405 representative sequences (AF-Core116K), followed by 50 rounds
of iterative clustering with the identity threshold lowered stepwise from
65\% to 30\%, converging at 67{,}605 non-redundant sequences (AF-NR68K),
the canonical input for all downstream simulations. Proportional subsets
of 10{,}000 (AF-Sub10K) and 1{,}000 (AF-Sub1K) sequences were drawn for
architecture prototyping and benchmarking, respectively.

A 500-protein subset (AF-Geo500) was drawn from AF-NR68K for the
segment-geometry analysis described below, sampling 11 species ---
\textit{Arabidopsis thaliana}, \textit{Anopheles gambiae}, \textit{Apis
mellifera}, \textit{Bacillus subtilis}, \textit{Bos taurus},
\textit{Caenorhabditis elegans}, \textit{Candida albicans},
\textit{Canis lupus}, \textit{Danio rerio} and \textit{Drosophila
melanogaster} (47 proteins each), plus \textit{Escherichia coli} (30
proteins) --- chosen to maximise phylogenetic distance between sampled
proteomes (two prokaryotes, one unicellular fungus, one angiosperm, four
invertebrates, three vertebrates) so that any species-level structure in
the recovered scaling coefficient would be detectable. \textit{H.\
sapiens} was excluded from this subset so that its dominant allocation in
the full dataset would not drive the cross-species comparison.

\subsection{Coarse-grained molecular dynamics simulations}

Coarse-grained molecular dynamics (CG-MD) simulations using a G\=o-like
C$\alpha$ model were performed for 848 proteins selected from AF-Sub1K;
the remaining 152 were excluded owing to failures in upstream topology
construction or energy minimisation. Predicted structures in mmCIF
format were converted to PDB with BioPython\cite{cock2009biopython}, and
native-contact topologies were built with the GoCa force-field
builder\cite{walter2024goca}, which represents each residue as a single
C$\alpha$ bead with Lennard-Jones 12--10 interactions between native
contact pairs. Each structure was placed in a cubic box with
\SI{20}{\nano\metre} padding and energy-minimised by steepest descent
(\num{50000} steps; force tolerance
\SI{100}{\kilo\joule\per\mol\per\nano\metre}). Production runs used
GROMACS 2023.4\cite{abraham2015gromacs} at a fixed reduced temperature
$T^{*}=35$ in the GoCa unit system, with the v-rescale
thermostat\cite{bussi2007canonical} ($\tau=\SI{0.05}{\pico\second}$) and
a \SI{0.004}{\pico\second} time step. This temperature lies near the
folding transition temperature for the size range represented in
AF-Sub1K, so that folding and unfolding events are both sampled within a
single run; because $T_f$ is protein-specific, a fixed $T^{*}$
necessarily sits above $T_f$ for some chains and below it for others, and
the per-protein $T_f$ determination used in the class-A validation
simulations was not applied at production scale. Simulation length was
set per protein by piecewise linear interpolation on residue count, from
\SI{3}{\nano\second} ($\leq$100 residues) to \SI{35}{\nano\second}
($\geq$2{,}000 residues). The 848 simulations were executed in parallel
across the available compute nodes and the whole set completed within
approximately \SI{85}{\minute} of wall-clock time.

\subsection{Steered molecular dynamics simulations}

To assess mechanical stability in the reduced (disulfide-free) state ---
isolating the contribution of non-covalent interactions to mechanical
unfolding --- we performed constant-velocity steered molecular dynamics
(SMD) simulations in OpenMM\cite{eastman2017openmm}. Disulfide bonds
were reduced before simulation by separating cysteine S$\gamma$ atoms,
removing the S--S covalent bond from the topology, and capping each
cysteine with a hydrogen atom to form a free thiol. Structures were
parameterised with the AMBER ff14SB force
field\cite{maier2015ff14sb} and the OBC2 generalised Born
implicit-solvent model\cite{onufriev2004exploring}, followed by energy
minimisation. The implicit-solvent treatment omits hydrodynamic drag and
explicit solvent friction, which contribute to the unfolding barrier at
high loading rates; this trade-off, made for throughput at corpus scale,
is examined in the Discussion.

Harmonic spring potentials
($k = \SI{1000}{\kilo\joule\per\mol\per\nano\metre\squared}$) were
attached to the N- and C-terminal C$\alpha$ atoms and displaced in
opposite directions along the end-to-end vector at
\SI{1}{\nano\metre\per\nano\second} per terminus (relative pulling rate
\SI{2}{\nano\metre\per\nano\second}), exceeding typical atomic force
microscopy rates ($\sim$\SI{1}{\micro\metre\per\second}) by several
orders of magnitude. SMD at such rates has previously been shown to
reproduce mechanical unfolding pathways qualitatively and to identify
force-bearing structural elements
correctly\cite{sotomayor2007single,lu1998unfolding}; the velocity analysis
reported in Section~\ref{sec:res_velocity} further establishes, for the
observables used here, that the extension-conditioned conformational ensemble
is reproducible across two orders of magnitude of velocity. Langevin dynamics was
propagated at \SI{300}{\kelvin} with a \SI{2}{\femto\second} time step
and a friction coefficient of \SI{1}{\per\pico\second}. Simulation
duration was set from the theoretical contour length
$L_c = (N-1) \times \SI{0.38}{\nano\metre}$, and force--extension curves
were recorded every 500 steps to identify unfolding force peaks.

\subsection{Coarse-grained steered molecular dynamics simulations}

To complement the all-atom SMD data, we performed constant-velocity
pulling simulations using two additional force fields spanning different
levels of physical detail.

\paragraph{Go-kit C$\alpha$ model with hydrophobic interactions.}
Single-bead C$\alpha$ topologies were generated with
Go-kit\cite{neelamraju2019gokit} under a structure-based model
configuration with non-native hydrophobic pair interactions activated
(cutoff \SI{5.5}{\angstrom}; uniform strength
$\varepsilon_{\mathrm{hp}}=1.0$) and glycine excluded from
side-chain-dependent terms; this augments the standard Lennard-Jones
12--10 native-contact potential and distinguishes the model from the
pure-topology GoCa model, allowing the contribution of non-native
hydrophobic contacts to mechanical unfolding to be assessed. Bond, angle
and dihedral potentials were derived from the native geometry
($K_b=200$, $K_\theta=40$, $K_\phi=1$, G\=o-model reduced units), and the
same constant-velocity SMD protocol described above (harmonic
restraints, $k=\SI{1000}{\kilo\joule\per\mol\per\nano\metre\squared}$,
\SI{1}{\nano\metre\per\nano\second} per terminus) was applied.

\paragraph{OpenAWSEM coarse-grained model.}
An independent set of CG-SMD simulations used the Associative memory,
Water-mediated, Structure and Energy Model
(AWSEM)\cite{davtyan2012awsem} as implemented in
OpenAWSEM\cite{lu2021openawsem}, which represents each residue by three
sites (C$\alpha$, C$\beta$, O) and combines direct-contact,
water-mediated and burial potentials with a bioinformatically derived
fragment-memory term\cite{papoian2004role}. Unlike the G\=o-type models,
AWSEM does not require the native-contact map as input; folding
specificity instead emerges from sequence-dependent statistical
potentials and local structural biases. The same constant-velocity
pulling protocol (spring constant, rate, terminus definitions) was
applied (Supplementary Methods~S1.4).

Together, the GoCa, Go-kit, AWSEM and all-atom (OpenMM) trajectories
form a hierarchy of increasing force-field complexity --- from pure
native-contact topology, through topology plus hydrophobic interactions
and sequence-dependent statistical potentials, to full atomic detail ---
against which the consistency of mechanical unfolding pathways can be
evaluated systematically.

\subsection{External molecular dynamics data}
\label{sec:meth_externalmd}

To broaden the structural and functional space covered by downstream
analyses, we incorporated MD trajectories from five public repositories,
each curated around a distinct target class: DynaRepo\cite{mokhtari2026dynarepo}
($\sim$720 macromolecular systems --- protein--protein, protein--DNA and
single-chain proteins from PDBbind and SAbDab --- simulated in triplicate
at \SI{500}{\nano\second} under a standardised all-atom protocol,
$>$\SI{1100}{\micro\second} aggregate); GPCRmd\cite{rodriguezespigares2020gpcrmd},
covering membrane-embedded MD data for 100\% of GPCR classes, 71\% of
receptor subtypes and 80\% of GPCR families; MDRepo\cite{roy2025mdrepo},
a FAIR-compliant, petabyte-scale warehouse of community-contributed
protein MD trajectories on CyVerse; BioExcel-CV19\cite{beltran2024bioexcelcv19},
aggregating $>$\num{10000} SARS-CoV-2 protein trajectories (Spike, RBD,
ACE2, 3CLpro, RdRp) from conventional and enhanced-sampling regimes; and
DBAASP\cite{pirtskhalava2021dbaasp}, a curated antimicrobial-peptide
resource with automated MD trajectories for $>$\num{3200} peptides.
These resources extend dataset coverage across globular, membrane,
antimicrobial and virally relevant proteins, and provide independent
ensembles for cross-validating features learned from our trajectories.
Trajectory provenance and processing for these repositories are described
in Supplementary Methods~S1.5.

\subsection{Data splits}
\label{sec:meth_splits}

Trajectory frames were partitioned into training and validation sets at the
level of the non-redundant sequence clusters of AF-NR68K rather than at the
level of individual frames or windows: every frame of every trajectory of a
given chain, and every sliding window derived from it, is assigned in its
entirety to one side of the split, so that no window seen during training
reappears in validation. Because AF-NR68K is already non-redundant at 30\%
sequence identity, assigning whole chains in this way also means that no pair
of near-duplicate sequences straddles the split. The same partition is reused
unchanged by conformational latent pretraining, R1, R2, Stage~1 and the
flow-matching decoder, so the validation curves reported for the cascade refer
throughout to the same held-out chains, and the window-level errors of
Rounds~1 and~2 are held-out quantities.

Three boundaries of this arrangement should be stated. The external molecular
dynamics repositories of the preceding section are held out in their
entirety and enter neither side of the split. The twenty proteins used for
the external comparison of Section~\ref{sec:meth_coverage} are drawn from the
\textit{Homo sapiens} proteome of AlphaFold~DB, and we did not establish
whether any of them is homologous to a chain in the training corpus;
because \textit{H.\ sapiens} contributes a substantial share of AF-NR68K,
that comparison is presented as a comparison of conformational spread
between two methods and not as a held-out generalisation test. Finally, the
chains displayed in Sections~\ref{sec:res_recon} and~\ref{sec:res_longchain}
are drawn from the trajectory corpus and are not identified here as held-out
chains. What those panels are used to support is the \emph{uniformity} of
reconstruction accuracy along $Q$---a comparison internal to one chain, in
which every conformation of that chain is treated alike---rather than an
absolute accuracy figure transferable to unseen sequences; the held-out
quantities in this work are the window-level validation errors of
Section~\ref{sec:res_training}.

\subsection{Conformational latent pretraining}
\label{sec:meth_pretrain}

Conformational latent pretraining pretrains a conformational latent representation that decouples sequence identity from fold state and supplies a stable conditioning signal to every subsequent phase. Concretely, we learn a single latent vector $\mathbf{v}\in\mathbb{R}^{1152}$ that jointly encodes protein identity and three-dimensional fold state. The conformational latent pretraining module comprises (i)~a non-learnable order-parameter calculator, (ii)~a FiLM-based\cite{perez2018film} encoder (ConformExpander, 4.27~M parameters; Supplementary Table~\ref{tab:conformexpander}) and (iii)~two training-only validation probes---a coordinate-reconstruction Decoder (3.50~M) and a Q-Predictor (0.56~M)---totalling 8.33~M trainable parameters. After conformational latent pretraining, only the ConformExpander is retained and frozen; the two probes are discarded. The retained encoder supplies the conformation-conditioned global embedding $\mathbf{v}$, which takes the place of the raw ESM-C sequence mean $\bar{\mathbf{e}}$ wherever a chain-level global embedding enters the conditioning encoders of R1, R2 and Stage~1. It therefore contributes an estimated 4.3~M frozen parameters to the inference forward pass, whereas the 4.1~M parameters of the two probes exist only during pretraining (Supplementary Table~\ref{tab:pipelineparams}).

\subsubsection{Input representations}

Each protein is represented by per-residue embeddings $\mathbf{e}_i\in\mathbb{R}^{N\times 1152}$ and a mean-pooled global embedding $\bar{\mathbf{e}}\in\mathbb{R}^{1152}$ from ESM-C\cite{hayes2025esm3}, precomputed and fixed throughout training. Training samples are individual snapshots from MD trajectories, each paired with C$\alpha$ coordinates $\mathbf{X}_\mathrm{true}\in\mathbb{R}^{N\times 3}$ and the fraction of native contacts $Q$. $Q$ quantifies folding progress as the proportion of native inter-residue contacts preserved in a given conformation\cite{best2013native}. The native contact set is defined from a reference (folded) structure as
\begin{equation}
\mathcal{C}_\mathrm{native}=\bigl\{(i,j)\;\big|\;|i-j|>3,\;d_{ij}^{\mathrm{ref}}<r_c\bigr\}
\end{equation}
with cutoff $r_c=8.0$\,\AA. For a given conformation,
\begin{equation}
	\fiteq{Q=\frac{1}{|\mathcal{C}_\mathrm{native}|}\sum_{(i,j)\in\mathcal{C}_\mathrm{native}}\frac{1}{1+\exp\!\bigl[\beta(d_{ij}-d_{ij}^{\mathrm{ref}}-\lambda)\bigr]},}
	\label{eq:Q}
\end{equation}
with steepness $\beta=5.0$\,\AA$^{-1}$ and tolerance $\lambda=1.8$\,\AA. This additive-tolerance variant preserves the sigmoidal behaviour of the original multiplicative formulation of Best \emph{et al.}\cite{best2013native} while making the tolerance dimensionally explicit; it agrees with the original to within numerical precision for the short-range native contacts that dominate $\mathcal{C}_\mathrm{native}$.

\subsubsection{Order-parameter extraction}

Two conformational descriptors are computed from $\mathbf{X}_\mathrm{true}$ without learnable parameters. The radius of gyration $R_g=\sqrt{N_\mathrm{eff}^{-1}\sum_i\|\mathbf{x}_i-\bar{\mathbf{x}}\|^2}$ is normalised to $[0,1]$ via
\begin{equation}
\tilde{R}_g=\mathrm{clamp}\!\left(\frac{R_g-R_g^{\min}}{R_g^{\max}-R_g^{\min}},\,0,\,1\right),
\end{equation}
with $R_g^{\min}=2.5(N/100)^{1/3}$ and $R_g^{\max}=2.0\,N^{0.6}$, both in \aa ngstr\"om, the latter following the excluded-volume scaling of Kohn \emph{et al.}\cite{kohn2004random}. These two bounds fix a monotone, length-aware rescaling of $R_g$ onto $[0,1]$ and are not intended as physical estimates of the compact-globule and random-coil limits; the clamp absorbs conformations that fall outside them. The secondary-structure fraction $f_{ss}$ classifies consecutive C$\alpha$ triplets as helix ($\theta\in[78^\circ,100^\circ]$, $d_{1,3}\in[4.8,6.0]$\,\AA) or strand ($\theta\in[105^\circ,145^\circ]$, $d_{1,3}\in[6.0,7.5]$\,\AA) using a C$\alpha$-only geometric criterion derived from canonical helix/strand backbone virtual angles and $i,i{+}2$ distances\cite{kneller2015ss}, and reports the fraction assigned to either class. During training $(\tilde{R}_g,f_{ss})$ are computed from true coordinates (teacher forcing); at inference they are user-specified.

\subsubsection{ConformExpander encoder}

The encoder produces $\mathbf{v}$ by modulating $\bar{\mathbf{e}}$ with the conformational conditions through FiLM\cite{perez2018film}. Scalars $(\tilde{R}_g,f_{ss},Q)$ are mapped to high-dimensional features through deterministic Fourier embedding\cite{tancik2020fourier,vaswani2017attention} (Supplementary~S2.1) and processed by a two-layer SiLU\cite{elfwing2018silu} MLP to yield $\mathbf{c}\in\mathbb{R}^{512}$. Affine parameters $\bm{\gamma}=W_\gamma\mathbf{c}+\mathbf{1}$ and $\bm{\beta}=W_\beta\mathbf{c}$ transform the layer-normalised embedding,
\begin{equation}
\mathbf{e}_\mathrm{mod}=\bm{\gamma}\odot\mathrm{LN}(\bar{\mathbf{e}})+\bm{\beta},
\end{equation}
and a learned gate $\alpha=\sigma(W_\alpha\mathbf{c})\in[0,1]$ interpolates between $\mathbf{e}_\mathrm{mod}$ and a non-linear residual,
\begin{equation}
\mathbf{v}=\alpha\,W_{r2}\mathrm{SiLU}(W_{r1}\mathbf{e}_\mathrm{mod})+(1-\alpha)\,\mathbf{e}_\mathrm{mod}.
\end{equation}
Near-zero initialisation of $W_\gamma,W_\beta$ ($\sigma=0.01$) ensures $\mathbf{v}\approx\bar{\mathbf{e}}$ at the start of training\cite{dhariwal2021diffusion}.

\subsubsection{Validation probes}

Two training-only probes shape the latent space during conformational latent pretraining; both are discarded afterwards. The \emph{Decoder} (3.50~M) projects $\mathbf{v}$ and $\mathbf{e}_i$ to 256~dimensions, processes them through 4 self-attention blocks with SwiGLU transitions\cite{shazeer2020glu} (Supplementary~S2.3), and outputs reconstructed coordinates $\hat{\mathbf{X}}\in\mathbb{R}^{N\times 3}$ together with seven global-descriptor targets $\hat{\mathbf{y}}\in[0,1]^7$ ($Q$, normalised $R_g$, $f_{ss}$, normalised mean pairwise distance, contact density, normalised distance variance, normalised end-to-end distance; Supplementary~S2.4). The \emph{Q-Predictor} (0.56~M) uses a parallel FiLM architecture conditioned on $(\tilde{R}_g,f_{ss})$ to predict $\hat{Q}$ and its uncertainty $\hat{\sigma}$ from $\bar{\mathbf{e}}$ alone (Supplementary~S2.5).

\subsubsection{Loss and optimisation}

The conformational latent pretraining loss is
\begin{equation}
	\fiteq{\mathcal{L}_{\mathrm{pretrain}}=\underbrace{\mathrm{MSE}_\mathrm{Kabsch}(\hat{\mathbf{X}},\mathbf{X}_\mathrm{true})}_{\mathcal{L}_\mathrm{recon}}+0.5\,\underbrace{\mathrm{MSE}(\hat{\mathbf{y}},\mathbf{y}_\mathrm{true})}_{\mathcal{L}_\mathrm{struct}}+0.3\,\underbrace{\left(\frac{(\hat{Q}-Q)^2}{2\hat{\sigma}^2}+\ln\hat{\sigma}\right)}_{\mathcal{L}_Q}.}
\end{equation}
$\mathcal{L}_\mathrm{recon}$ is computed after optimal rigid-body superposition via the Kabsch algorithm\cite{kabsch1976solution} (SVD in float32; Supplementary~S2.6); $\mathcal{L}_Q$ is a heteroscedastic Gaussian negative log-likelihood\cite{nix1994estimating} with $\hat{\sigma}$ floored at $0.01$ and the per-sample NLL clamped at $-1.0$ to keep the $\ln\hat{\sigma}$ term from dominating the total loss. Optimisation uses AdamW\cite{loshchilov2019decoupled} ($\beta_1=0.9$, $\beta_2=0.999$) with peak learning rate $10^{-4}$, weight decay $0.01$, 2{,}000-step linear warm-up followed by cosine annealing\cite{loshchilov2017sgdr}, batch size 8, 50{,}000~steps, bfloat16 mixed precision and gradient-norm clipping at $0.5$.

\subsection{Round 1 (R1): multi-scale local geometry prediction}

R1 takes the conformational conditioning signal established in conformational latent pretraining---per-residue ESM-C embeddings\cite{hayes2025esm3} together with the frozen conformational latent $\mathbf{v}$ and the order parameters $(Q,\tilde{R}_g,f_{ss})$---and converts it into an explicit local geometry: for every sliding window of $W=15$ residues it predicts the C$\alpha$--C$\alpha$ distance at all $K=7$ short-range sequence separations $|i-j|\in\{1,\ldots,7\}$ and a windowed C$\alpha$ trace, providing both the geometric supervision and the trunk hidden states that R2 subsequently refines.

\subsubsection{Inputs and windowing}

Each protein is processed by tiling its residue sequence with a stride-1 sliding window of length $W=15$; for a chain of length $N$ this produces $N-W+1$ windows that are forwarded as an independent batch. Each window receives (i)~per-residue ESM-C embeddings of its $W$ tokens $\mathbf{e}_i\in\mathbb{R}^{W\times 1152}$, (ii)~the frozen conformational latent $\mathbf{v}\in\mathbb{R}^{1152}$ exported by conformational latent pretraining, which takes the place of the raw sequence mean $\bar{\mathbf{e}}$ as the chain-level global embedding, (iii)~the chain-level scalars $(Q,\tilde{R}_g,f_{ss})$ broadcast across the window, and (iv)~a binary mask. R1's output for the chain is reassembled from per-window predictions by parameter-free Kabsch consensus on overlapping windows (Supplementary~S3.7).

\subsubsection{Architecture}

The single-token stream is initialised by summing a linear projection of $\mathbf{e}_i$, an absolute positional embedding\cite{vaswani2017attention} and a conditioning vector $\mathbf{c}\in\mathbb{R}^{384}$ produced by a CondEncoder that fuses $\mathbf{v}$ with deterministic Fourier embeddings\cite{tancik2020fourier} of $(Q,\tilde{R}_g,f_{ss})$. This stream is contextualised by a 6-layer pre-LN Transformer trunk\cite{xiong2020layernorm} ($c_s=384$, 16~heads, expansion~4) to yield $\mathbf{h}^{(0)}\in\mathbb{R}^{W\times c_s}$.

The core of R1 is a \emph{cascade} of $K=7$ PairFormer blocks\cite{jumper2021alphafold,abramson2024alphafold3}, each dedicated to one sequence-separation scale $k\in\{1,\ldots,7\}$. A shared pair stream $\mathbf{z}\in\mathbb{R}^{W\times W\times c_z}$ ($c_z=128$), initialised from a relative-position embedding plus $\mathbf{c}$, passes through all seven blocks and is updated at each step by an outer-product mean and triangle multiplication/attention layers. Block~$k$ receives the cumulative single-stream context from all preceding blocks---aggregated by a learned cross-scale attention pooler (MultiScaleAgg; Supplementary~S3.4)---together with a zero-initialised conditioning re-injection of $\mathbf{c}$, and emits an updated $\mathbf{h}^{(k)},\mathbf{z}^{(k)}$. Immediately after each block, scale~$k$'s dedicated \emph{anchored} regression head reads $(\mathbf{h}^{(k)}_i,\mathbf{h}^{(k)}_{i+k},\mathbf{z}^{(k)}_{i,i+k})$ and predicts
\begin{equation*}
\hat{d}_{i,i+k}=a_k+s_k\tanh\!\bigl(\mathrm{MLP}(\mathbf{h}^{(k)}_i,\mathbf{h}^{(k)}_{i+k},\mathbf{z}^{(k)}_{i,i+k})\bigr),
\end{equation*}
with anchor $a_k$ and scale $s_k$ chosen to bracket the empirical C$\alpha$ distance distribution at each separation ($a_1{=}3.80$\,\AA, $s_1{=}0.50$\,\AA{} through $a_7{=}12.5$\,\AA, $s_7{=}7.00$\,\AA; Supplementary~S3.5). After all seven blocks, the seven token streams $\{\mathbf{h}^{(k)}\}_{k=1}^{7}$ are merged by a final MultiScaleAgg, the conditioning vector is re-injected, and a 2-layer Transformer decodes the result to a centred C$\alpha$ trace via a single linear head. The complete R1 module contains 43.3~M parameters (Supplementary Table~\ref{tab:r1params}).

\subsubsection{Loss and optimisation}

R1 is trained with the smooth-L1 (Huber) loss\cite{girshick2015fastrcnn,huber1964robust} on the seven anchored distance predictions and on the windowed coordinate trace after centroid removal,
\begin{equation}
	\fiteq{\mathcal{L}_{R1}=\sum_{k=1}^{K}\frac{1}{|\mathcal{P}_k|}\sum_{(i,j)\in\mathcal{P}_k}\mathrm{SL1}\bigl(\hat{d}_{ij}^{(k)},d_{ij}^{\mathrm{true}}\bigr)+\mathrm{SL1}\bigl(\hat{\mathbf{X}}-\mathrm{COM}(\hat{\mathbf{X}}),\mathbf{X}^\mathrm{true}-\mathrm{COM}(\mathbf{X}^\mathrm{true})\bigr),}
\end{equation}
summed across windows with equal weight across the seven scales (Supplementary~S3.8). Optimisation uses AdamW\cite{loshchilov2019decoupled} (peak learning rate $3\!\times\!10^{-4}$, weight decay $0.01$, $\beta_1=0.9$, $\beta_2=0.999$), 2{,}000-step linear warm-up followed by cosine annealing\cite{loshchilov2017sgdr}, gradient-norm clipping at $1.0$ and bfloat16 mixed precision (full schedule in Supplementary~S3.9).

\subsection{Round 2 (R2): residual refinement with multi-source feature-wise modulation}

R2 inherits R1's prediction targets---the seven short-range C$\alpha$--C$\alpha$ distance scales and the windowed C$\alpha$ trace---and refines them by conditioning on R1's full output. R1's per-scale distance-head pre-activations and trunk hidden states are read in as side-channel features and re-injected into every layer of an R2 cascade through FiLM\cite{perez2018film}; on top of this, R2 emits auxiliary geometric, profile, pairwise-distance and global-descriptor predictions whose hidden states themselves modulate later layers, providing the auxiliary-task supervision\cite{caruana1997multitask} that improves representation quality. Distance heads and the coordinate head produce \emph{residuals} added to R1's predictions\cite{he2016resnet}, so the sole job of R2 is to learn the systematic error of R1.

\subsubsection{Inputs}

Each window inherits R1's inputs $(\mathbf{e}_i,\mathbf{v},Q,\tilde{R}_g,f_{ss},\text{mask})$ together with the full R1 output bundle: the seven distance predictions $\{\hat{d}^{(k)}\}_{k=1}^{7}$, the windowed C$\alpha$ trace $\hat{\mathbf{X}}_\mathrm{R1}\in\mathbb{R}^{W\times 3}$, the trunk hidden states $\mathbf{h}^{(7)}_\mathrm{R1},\mathbf{h}_\mathrm{coord,R1}\in\mathbb{R}^{W\times c_s}$ and $\mathbf{z}^{(7)}_\mathrm{R1}\in\mathbb{R}^{W\times W\times c_z}$, and the per-scale distance-head pre-activations $\mathbf{g}^{(k)}_\mathrm{R1}\in\mathbb{R}^{(W-k)\times c_g}$ ($c_g=256$). All R1 features are pre-cached at training time, so R1 weights are frozen and not back-propagated through.

\subsubsection{Architecture}

The single-token stream is initialised by summing linear projections of $\mathbf{e}_i$, $\mathbf{h}^{(7)}_\mathrm{R1}$, $\mathbf{h}_\mathrm{coord,R1}$ and $\hat{\mathbf{X}}_\mathrm{R1}$ together with absolute positional embeddings and a fresh conditioning vector $\mathbf{c}\in\mathbb{R}^{384}$ from an independent CondEncoder; this fused representation is passed through a 6-layer pre-LN Transformer trunk to yield $\mathbf{h}^{(0)}\in\mathbb{R}^{W\times c_s}$. The pair stream is initialised as $\mathbf{z}^{(0)}=\mathrm{Linear}(\mathbf{z}^{(7)}_\mathrm{R1})+\mathrm{rel\_pos}(W)+\mathrm{cond\_to\_z}(\mathbf{c})$.

R2 then runs the same cascade of $K=7$ PairFormer blocks as R1, but the entry to every block is augmented by a \emph{multi-source} FiLM that simultaneously injects (i)~all seven R1 distance-head pre-activations $\{\mathbf{g}^{(k)}_\mathrm{R1}\}$, (ii)~all R2 distance-head pre-activations from preceding blocks and (iii)~the auxiliary geometry/profile features once they become available. After block~2, an auxiliary geometry branch predicts per-residue C$\alpha$ pseudo-bond ($\theta_i$) and pseudo-torsion ($\sin\tau_i,\cos\tau_i$) angles\cite{yang2020trrosetta}; after block~4, a profile branch predicts per-residue mean and standard deviation of empirical pairwise-distance distributions, a pair branch predicts the full $W\!\times\!W$ C$\alpha$--C$\alpha$ distance map\cite{senior2020alphafold}, and the pair branch's hidden state additionally modulates the pair stream of blocks~5--7 via a FiLM gate. Each block is followed by a \emph{residual} distance head $\hat{d}^{(k)}_\mathrm{R2}=\hat{d}^{(k)}_\mathrm{R1}+\delta\hat{d}^{(k)}$ in which the $\delta$-MLP is initialised so that $\delta\hat{d}\approx 0$ at the start of training. The seven block outputs are merged by a final MultiScaleAgg, refined by a 4-layer coordinate Transformer and decoded to a coordinate residual $\delta\mathbf{X}$ that yields $\hat{\mathbf{X}}_\mathrm{R2}=\hat{\mathbf{X}}_\mathrm{R1}+\delta\mathbf{X}$ after centroid removal. The complete R2 module contains 77.4~M parameters (Supplementary Table~\ref{tab:r2params}).

\subsubsection{Loss and optimisation}

R2 is trained with a multi-task smooth-L1\cite{girshick2015fastrcnn,huber1964robust} loss
\begin{equation}
	\fiteq{\mathcal{L}_{R2}=\sum_{k=1}^{7}\lambda_k^d\,\mathrm{SL1}(\hat{d}^{(k)}_\mathrm{R2},d^{(k)}_\star)+\lambda^x\,\mathrm{SL1}(\hat{\mathbf{X}}_\mathrm{R2},\mathbf{X}_\star^\mathrm{aligned})+\sum_{a\in\{\theta\tau,\mu\sigma,D,\mathrm{glob}\}}\lambda^a\,\mathcal{L}_a,}
\end{equation}
combining the seven refined distances, the windowed coordinate residual and the four auxiliary predictions (combined auxiliary weight $0.25$; per-term weights in Supplementary~S4.6). The coordinate target $\mathbf{X}_\star^\mathrm{aligned}$ is obtained by Kabsch alignment\cite{kabsch1976solution} of the ground-truth window onto the R1 prediction so that the residual term is supervised purely on the post-R1 mismatch. Optimisation uses AdamW\cite{loshchilov2019decoupled} with peak learning rate $2\!\times\!10^{-4}$, $\beta_1=0.9,\beta_2=0.999$, weight decay $0.01$, 2{,}000-step linear warm-up, cosine annealing\cite{loshchilov2017sgdr}, gradient-norm clipping at $1.0$ and bfloat16 mixed precision; R1 is frozen and its outputs are pre-cached as bfloat16 memory-mapped tensors, and only R2 parameters are updated (Supplementary~S4.7).

\subsection{Stage~1: window-to-chain assembly and full-chain regression}

Stage~1 closes the gap between R1/R2's per-window predictions and the chain-length representation consumed by flow matching. It does so in three steps: cross-window attention pooling\cite{bahdanau2015attention,vaswani2017attention} lifts every R1/R2 feature with window-local support into a chain-indexed feature; those pooled features, together with the parameter-free Kabsch C$\alpha$ consensus trace\cite{kabsch1976solution} (R1, Supplementary~S3.7), initialise the chain-level single and pair streams of a 6-layer chain-level PairFormer trunk\cite{jumper2021alphafold,abramson2024alphafold3}; the trunk produces the \emph{single representation} $\mathbf{s}\in\mathbb{R}^{N\times c_s}$ that conditions the downstream flow-matching decoder, while a parallel coordinate head emits a residual $\delta\mathbf{X}$ that refines the consensus into the Stage~1 trace $\hat{\mathbf{X}}_{S1}$.

\subsubsection{Inputs}

Stage~1 is trained on chains of at most $N_{\max}=200$ residues, and a single forward pass accordingly accepts at most $N_{\max}$ residues; longer chains are processed at inference by the crop-and-superpose scheme of Section~\ref{sec:meth_longchain}. For a chain of length $N\leq N_{\max}$, Stage~1 receives the chain-level inputs $(\mathbf{e}_i\in\mathbb{R}^{N\times 1152},\mathbf{v},Q,\tilde{R}_g,f_{ss},\text{mask})$, the full per-window R1 and R2 output bundles over all $N-W+1$ windows, the parameter-free consensus trace $\hat{\mathbf{X}}_\mathrm{cons}\in\mathbb{R}^{N\times 3}$, and a precomputed gather index that maps every global pair $(i,j)$ with $|i-j|<W$ to the windows that contain it. R1 and R2 are frozen and their outputs are pre-cached; Stage~1 is the only trainable component at this stage.

\subsubsection{Architecture}

A \emph{chain-assembly module} converts every per-window feature into a chain-indexed representation by attention pooling over windows. For each global C$\alpha$ pair $(i,j)$ with $|i-j|=k$, all $\min(W-k,N-k)$ windows containing the pair contribute a feature vector drawn from R1 and R2 (anchored distance predictions, coordinate-derived distances, distance-head hidden states $\mathbf{g}^{(k)}$, pair-stream features $\mathbf{z}_{i,j}^{(7)}$ and a window-position embedding); a learned attention pooler with one shared query (CrossWindowPool, Supplementary~S5.1) compresses these contributions to a single $c_a=256$ vector per pair. Two CrossWindowPool variants populate the chain pair stream---one for short separations ($k\leq 7$, where anchored distance heads exist) and one for long separations ($8\leq k<W$, using only coordinate-derived signals)---and a third pool aggregates per-residue trunk hidden states $\mathbf{h}^{(7)},\mathbf{h}_\mathrm{coord}$ from R1 and R2 over every window that contains the residue. The pooled pair features---together with relative-position embeddings and Gaussian-RBF features of the consensus distances---initialise a chain-level pair stream $\mathbf{z}^{(0)}\in\mathbb{R}^{N\times N\times c_z}$; the pooled residue features, an ESM-C projection, the conditioning vector $\mathbf{c}$ and a learned projection of $\hat{\mathbf{X}}_\mathrm{cons}$ are merged by a five-source gated fusion (Supplementary~S5.5) into the chain-level single stream $\mathbf{s}^{(0)}\in\mathbb{R}^{N\times c_s}$.

A 6-layer chain-level PairFormer trunk\cite{jumper2021alphafold,abramson2024alphafold3} ($c_s=384$, $c_z=128$, identical block design to R1/R2) processes $(\mathbf{s}^{(0)},\mathbf{z}^{(0)})$. Each layer interleaves two CondFiLM\cite{perez2018film} re-injections---one for the global conditioning vector $\mathbf{c}$, one for the pooled cross-window distance signal---with a zero-initialised linear re-injection of the ESM-C projection; every second layer also re-injects the conditioner's pair output into $\mathbf{z}$. The final-block single-stream output is layer-normalised and exported as $\mathbf{s}\in\mathbb{R}^{N\times c_s}$ that conditions flow matching, while a parallel linear head decodes a coordinate residual $\delta\mathbf{X}$ that yields $\hat{\mathbf{X}}_{S1}=\hat{\mathbf{X}}_\mathrm{cons}+\delta\mathbf{X}$ after centroid removal. Stage~1 contains 28.0~M parameters (Supplementary Table~\ref{tab:s1params}).

\subsubsection{Loss and optimisation}

Stage~1 is supervised by smooth-L1 loss\cite{girshick2015fastrcnn,huber1964robust} on $\hat{\mathbf{X}}_{S1}$ against ground-truth chain coordinates after Kabsch alignment\cite{kabsch1976solution}, with R1 and R2 frozen. Optimisation uses AdamW\cite{loshchilov2019decoupled} (peak learning rate $1\!\times\!10^{-4}$, $\beta_1=0.9$, $\beta_2=0.999$, weight decay $0.01$), 1{,}000-step linear warm-up with cosine annealing\cite{loshchilov2017sgdr}, gradient-norm clipping at $1.0$ and bfloat16 mixed precision; full schedule and feature-caching strategy in Supplementary~S5.8.

\subsection{Flow matching: conditional C$\alpha$ ensemble generation}

Flow matching closes the FoldScape pipeline by turning Stage~1's deterministic, point-estimate single representation $\mathbf{s}\in\mathbb{R}^{N\times c_s}$ into a conditional ensemble of C$\alpha$ traces. We use rectified-flow conditional flow matching\cite{lipman2023flow,liu2023rectified,albergo2023stochastic} on coordinate space $\mathbb{R}^{N\times 3}$: a noise prior $\mathbf{X}_0\sim\mathcal{N}(\mathbf{0},\mathbf{I}_3)^N$ is interpolated linearly with the data $\mathbf{X}_1$ as $\mathbf{X}_t=(1-t)\mathbf{X}_0+t\mathbf{X}_1$ for $t\sim\mathcal{U}(0,1)$, and a 5-layer AlphaFold3-style decoder\cite{abramson2024alphafold3,geffner2025proteina} predicts the clean target $\hat{\mathbf{X}}_1$ from $(\mathbf{X}_t,t,\mathbf{s})$. At inference, the trained velocity field $\hat{\mathbf{v}}=(\hat{\mathbf{X}}_1-\mathbf{X}_t)/(1-t)$ is integrated by Euler steps from $t=0$ to $t=1$, conditioned on the same $\mathbf{s}$ throughout, producing diverse C$\alpha$ samples that respect the conformational order parameters $(Q,\tilde{R}_g,f_{ss})$ encoded in $\mathbf{s}$.

\subsubsection{Inputs and conditioning}

For each sample, the decoder receives the current interpolated coordinates $\mathbf{X}_t\in\mathbb{R}^{N\times 3}$, a flow time $t\in[0,1]$, the chain mask and the Stage~1 single representation $\mathbf{s}\in\mathbb{R}^{N\times c_s}$ ($c_s=384$); $\mathbf{s}$ enters the network solely through the conditioning pathway (i.e.\ via AdaLN modulation\cite{peebles2023dit}) rather than the token stream, so the decoder treats it as the full structural prior delivered by upstream stages. Optional self-conditioning\cite{chen2023bitdiff} provides $\mathbf{X}_\mathrm{sc}=\hat{\mathbf{X}}_1^{\,\mathrm{prev}}$ from the previous integration step. Conformational latent pretraining, R1, R2 and Stage~1 are all frozen during flow-matching training; only the decoder is updated.

\subsubsection{Architecture}

The decoder is a 5-layer AF3-style Transformer\cite{abramson2024alphafold3,geffner2025proteina} with model width $c_s=384$, 8~attention heads, attention pair bias and adaptive layer normalisation\cite{peebles2023dit} for time and conditioning injection. The token stream is initialised by summing a linear projection of $\mathbf{X}_t$ (and a separate linear projection of $\mathbf{X}_\mathrm{sc}$ when self-conditioning) with a small set of structural input features (chain-break indicator, self-conditioning coordinates), and the conditioning vector $\mathbf{c}_\ell\in\mathbb{R}^{384}$ is built by adding a Fourier time embedding\cite{tancik2020fourier} of $t$ to a per-residue broadcast of $\mathbf{s}$. A pair representation $\mathbf{P}\in\mathbb{R}^{N\times N\times 128}$ is constructed once from relative sequence separation, Gaussian-RBF features of $\|\mathbf{X}_{t,i}-\mathbf{X}_{t,j}\|$ and (when self-conditioning) Gaussian-RBF features of $\|\mathbf{X}_{\mathrm{sc},i}-\mathbf{X}_{\mathrm{sc},j}\|$, and supplies the attention pair bias inside every layer; the pair representation is not updated across layers, in contrast to the AF3 trunk. Thirty-two learned register tokens\cite{darcet2024registers} are prepended and removed before the final readout to absorb high-norm artefact tokens. The single-stream output of the last block is decoded by a single linear $c_s\to 3$ head to the clean-coordinate prediction $\hat{\mathbf{X}}_1$. The decoder contains $\approx\!13.0$~M trainable parameters (Supplementary Table~\ref{tab:pipelineparams}).

\subsubsection{Loss and optimisation}

The decoder is trained with a composite objective combining the rectified-flow translation loss in coordinate space with auxiliary geometric and physical constraints,
\begin{equation}
	\fiteq{\mathcal{L}_\mathrm{FM}=\lambda_\mathrm{trans}w(t)\|\hat{\mathbf{X}}_1-\mathbf{X}_1\|^2+\lambda_\mathrm{global}\,w_g(t)\,\mathcal{L}_\mathrm{global}+\lambda_\mathrm{local}\,w_l(t)\,\mathcal{L}_\mathrm{local}+\lambda_\mathrm{neutral}\,\mathcal{L}_\mathrm{phys},}
\end{equation}
where $w(t)=1/(1-t)^2$ converts coordinate MSE to the rectified-flow velocity target, the cosine pair $w_g(t)=\tfrac{1}{2}(1+\cos\pi t)$, $w_l(t)=1-w_g(t)$ schedules the auxiliary losses so that global structure (RMSD, lDDT\cite{mariani2013lddt}, contact map, $R_g$, $f_{ss}$, end-to-end distance) dominates early denoising and local geometry ($d_{i,i+k}$ consistency for $k=1,\ldots,7$) dominates late denoising, and $\mathcal{L}_\mathrm{phys}$ adds always-on physical penalties on bond length, bond angle and steric clashes (Supplementary~S6.6). Outer weights are $\lambda_\mathrm{trans}=1.0$, $\lambda_\mathrm{global}=0.5$, $\lambda_\mathrm{local}=1.0$, $\lambda_\mathrm{neutral}=1.0$. Optimisation uses AdamW\cite{loshchilov2019decoupled} (peak learning rate $3\!\times\!10^{-4}$, $\beta_1=0.9$, $\beta_2=0.95$, weight decay $0.01$), 5{,}000-step linear warm-up with cosine annealing\cite{loshchilov2017sgdr}, gradient-norm clipping at $1.0$ and bfloat16 mixed precision; full training schedule and sampling settings are in Supplementary~S6.7--S6.8.

\subsection{Inference on chains longer than the chain-level training ceiling}
\label{sec:meth_longchain}

Stage~1 and the flow-matching decoder are trained on chains of at most
$N_{\max}=200$ residues, so a single chain-level forward pass accepts at most
$N_{\max}$ residues. Longer chains are handled at inference without retraining,
by the same parameter-free consensus principle already used to assemble R1 and
R2 windows into a chain (Supplementary~S3.7). A chain of length $N>N_{\max}$ is
tiled with overlapping crops of $N_{\max}$ consecutive residues at stride~1,
giving $N-N_{\max}+1$ crops; a 706-residue chain therefore yields 507 crops and
an 857-residue chain 658. Each crop is passed independently through the frozen
R1/R2 window stack, through Stage~1 and through the flow-matching decoder, so
that every crop yields a complete predicted C$\alpha$ trace of its own
$N_{\max}$ residues.

Crops are then merged into a single chain. They are visited in sequence order;
each crop is rigidly aligned by Kabsch superposition\cite{kabsch1976solution} onto the
partially assembled chain over the $N_{\max}-1$ residues it shares with its
predecessor, and per-residue coordinates are accumulated by averaging over every
crop that contains the residue. Because an interior residue is contained in up
to $N_{\max}$ crops, its final position is the average of up to $N_{\max}$
independent predictions, whereas residues within $N_{\max}$ of either terminus
are covered fewer times and are correspondingly less well determined. Two
properties of the scheme are worth stating explicitly. First, no
relative-position embedding and no entry of the codebook of size $\max_N=200$ is
ever asked to represent a separation larger than it was trained on, because all
pair features are computed inside a crop; the extrapolation in chain length is
carried entirely by the merging step and not by any learned component. Second,
long-range order across the chain is therefore assembled from overlapping local
solutions rather than predicted directly, which is the same mechanism that
operates at the window level inside R1 and R2, applied one level up. For
ensemble generation the whole procedure is repeated with independent prior
samples, each repetition producing one chain-level conformation.

\subsection{All-atom reconstruction}
\label{sec:meth_backmap}

To recover atomic detail from the predicted C$\alpha$-only ensembles, we performed back-mapping with two orthogonal schemes whose error modes are largely uncorrelated\cite{wassenaar2014backward}. Deep-learning reconstruction was carried out with \textbf{cg2all}\cite{heo2024oneparticle}, an SE(3)-equivariant graph neural network that maps a C$\alpha$ trace to a full heavy-atom model in a single deterministic forward pass, guaranteeing roto-translational consistency by construction. As a geometry-based control, the same C$\alpha$ ensembles were independently rebuilt with \textbf{PULCHRA}\cite{rotkiewicz2008pulchra}, which reconstructs backbone atoms from a curated tetrapeptide-fragment library and places side chains using a backbone-dependent rotamer library\cite{shapovalov2011rotamer}. Input C$\alpha$ coordinates were preserved in both pipelines to retain the statistics of the upstream coarse-grained predictions. Stereochemical quality of every reconstructed frame was assessed with MolProbity\cite{williams2018molprobity}, and per-residue agreement between the two reconstructions was used to flag conformations for which downstream all-atom observables should be interpreted with caution. A detailed description of the back-mapping workflow, validation protocol and cross-method consistency analysis is provided in Supplementary~S7.


\subsection{Validation simulations}

To verify that the different simulation approaches yield physically
consistent folding and unfolding pathways, we performed targeted
validation simulations on four small, experimentally well-characterised
two-state proteins: ubiquitin (PDB 1UBQ, 76 residues), the titin~I27
immunoglobulin domain (PDB 1TIT, 89 residues), protein~L (PDB 2PTL, 61
residues) and the B1 domain of protein~G (PDB 1PGB, 56 residues). Each
has published experimental $\varphi$-value
data\cite{fowler2001mapping,kim2000specific,best2002mechanical,nauli2001crystal,went2005ubiquitin}
and single-molecule force spectroscopy
measurements\cite{carrionvazquez1999mechanical,schlierf2004force,brockwell2005pulling,cao2007polyprotein,shen2012bihistidine},
providing orthogonal benchmarks for the folding transition-state
ensemble and the mechanical unfolding pathway. The complete validation
matrix is summarised in Table~\ref{tab:valmatrix}.

\begin{table*}[t]
\centering
\small
\caption{Validation simulation matrix. All simulations are performed for
the four validation proteins (ubiquitin, titin~I27, protein~L,
protein~G) unless otherwise noted. $F(Q)$ and $F(x)$ denote free-energy
profiles computed from the simulations themselves, used here only to locate
the transition-state ensemble and to compare force fields; FoldScape makes no
free-energy prediction anywhere in this work.}
\label{tab:valmatrix}
\begin{tabular}{@{}lllll@{}}
\toprule
Class & Method & Proteins & Direction & Primary output \\
\midrule
A (equil.) & GoCa at $T_f$ & 4 & Folding & $F(Q)$, $Q_\mathrm{TSE}$, $\varphi_\mathrm{sim}$ \\
A (equil.) & Go-kit at $T_f$ & 4 & Folding & $F(Q)$, $Q_\mathrm{TSE}$, $\varphi_\mathrm{sim}$ \\
B (path)   & GoCa production & 4 & Folding & Contact formation order \\
B (path)   & Go-kit production & 4 & Folding & Contact formation order \\
C (SMD)    & GoCa & 4 & Stretching & Contact rupture order \\
C (SMD)    & Go-kit & 4 & Stretching & Contact rupture order, $F(x)$ \\
C (SMD)    & AWSEM & 4 & Stretching & $F(x)$, contact rupture order \\
C (SMD)    & OpenMM & 4 & Stretching & $F(x)$, contact rupture order \\
D (rate)   & OpenMM ($4\times v$) & 4 & Stretching & Extension-binned $Q$, $R_g$, RMSD, $\Phi_i(x)$ \\
D$'$ (rate) & AWSEM ($6\times v$) & 4 & Stretching & Extension-binned $\Phi_i(x)$; rupture order (descriptive) \\
E ($\varphi$) & OpenMM ($4\times v$) & 4 & Stretching & $\Phisim$ vs.\ experimental $\varphi$ \\
F (geom.) & OpenMM production & 500 & Stretching & Binary partition; $\alpha$ in eq.~(\ref{eq:ete}) \\
\bottomrule
\end{tabular}
\end{table*}

\paragraph{Equilibrium G\=o-model simulations (class~A).}
For each protein, equilibrium folding simulations were run at the
folding transition temperature $T_f$ using both GoCa and Go-kit force
fields. $T_f$ was determined from short replica simulations at five
temperatures bracketing the expected transition followed by
heat-capacity analysis; production runs of $5\times10^7$ steps at $T_f$
yielded the free-energy profile $F(Q) = -k_BT\ln P(Q)$, from which the
TSE position $Q_{\mathrm{TSE}}$ was extracted. Simulated $\varphi$-values
were computed as the fractional native-contact formation of each residue
within the $Q_{\mathrm{TSE}} \pm 0.05$ window and compared with
experimental $\varphi$-values by Pearson correlation.

\paragraph{Production folding paths (class~B) and stretching simulations (class~C).}
Folding-path simulations at production-scale parameters (class~B) and
constant-velocity SMD stretching simulations under all four force fields
(class~C) were performed for each validation protein, yielding 8 folding
and 16 stretching trajectories, using the same pulling protocol as the
production dataset.

\paragraph{Rate-stability tests (classes~D and~D$'$).}
Class~D is the all-atom four-velocity series described under ``Choice of
pulling velocity''; class~D$'$ repeats the test in AWSEM at six
velocities ($v=0.01$, $0.05$, $0.1$, $1.0$, $2.0$,
\SI{4.0}{\nano\metre\per\nano\second}), extending above the production
setting. In both classes, the acceptance criterion is the continuous,
extension-binned observable: the RMS difference between per-residue
contact-survival curves at matched extension must fall within the band
established by class~D. Spearman rank correlation of contact-rupture
order between velocity pairs is reported as a descriptive statistic
only, since rupture order is a rank statistic obtained by thresholding a
continuous observable and falls outside the class of quantities the
velocity analysis supports.

\paragraph{$\varphi$-value discrimination (class~E).}
Described under ``Constant-velocity pulling samples the mechanical
unfolding pathway''; analysed for all four validation proteins against
published $\varphi$-values on the relevant axis: mechanical $\varphi$ for
titin~I27, protein~L and protein~G, and chemical $\varphi$ for ubiquitin
as the specificity control.

\paragraph{Segment geometry (class~F).}
Described under ``Segmental structure of the unfolding intermediate'';
performed on AF-Geo500 at the production velocity.

\paragraph{Key scientific tests.}
Protein~L and protein~G provide a particularly stringent test: despite
sharing the same $\beta$-grasp topology ($<$15\% sequence identity),
their experimental folding nuclei are mirror-symmetric, centred on the
first $\beta$-hairpin ($\beta_1$--$\beta_2$) in
protein~L\cite{kim2000specific} but the second ($\beta_3$--$\beta_4$) in
protein~G\cite{nauli2001crystal}. Correct prediction of this
polarisation by the sequence-dependent methods (Go-kit, AWSEM, OpenMM),
and its absence in the pure-topology GoCa model, directly tests whether
the training data encode information beyond native-contact
topology\cite{karanicolas2002origins}. For titin~I27, chemical\cite{fowler2001mapping}
and mechanical\cite{best2002mechanical} $\varphi$-values would in
principle allow independent validation against distinct
transition-state structures, but the two are collinear over the seven
available sites ($r=0.989$); titin~I27 therefore serves as a consistency
check rather than a discriminating test, with discrimination resting on
protein~G and ubiquitin. Further simulation settings for the classes of
Table~\ref{tab:valmatrix} are given in Supplementary Methods~S1.3, S1.4
and~S1.6.

\subsection{Constant-velocity pulling samples the mechanical unfolding pathway}

The training corpus is generated by mechanical, not chemical,
denaturation. Because the two unfolding routes proceed through different
transition states, the residue-level information they yield is not
interchangeable, so we first confirmed which pathway the pulling protocol
samples.

Mechanical and chemical unfolding proceed through structurally distinct
transition states: force applied along the end-to-end vector drives
unfolding along a shear coordinate, disrupting first the hydrogen bonds
and inter-strand interfaces that bear the
load\cite{lu1998unfolding,lu2000key,isralewitz2001steered}, whereas
chemical denaturation is driven by bulk solvation and its transition
state is set by the folding nucleus rather than the pulling
geometry\cite{fowler2001mapping,went2005ubiquitin,matouschek1989mapping}.
Direct comparison in titin~I27 confirmed that the two routes give
different unfolding barriers and structural
determinants\cite{carrionvazquez1999mechanical}. A simulation that
correctly samples the mechanical channel should therefore correlate
positively with mechanical $\varphi$-values while showing no positive
correlation with chemical folding $\varphi$-values --- a double criterion
that a simulation merely retracing the folding pathway in reverse would
fail.

For each trajectory the first force peak on the force--extension curve
was located by Savitzky--Golay smoothing (window 101 points, third
order; prominence threshold set to the larger of 15\% of the force range
and \SI{10}{\pico\newton}), giving $\dpeak$. Native contacts were defined
on the folded state (C$\alpha$--C$\alpha$ $<\SI{8.0}{\angstrom}$,
$|i-j|>3$, as in equation~(\ref{eq:Q}); see
Table~\ref{tab:contactdefs})\cite{best2013native,duarte2010optimal}, and
a contact was counted as surviving if its instantaneous
C$\alpha$--C$\alpha$ distance was below $1.5\times$ its native value. For
every residue we then computed $\Phisim$, the fraction of its native
contacts surviving, averaged over the window
$\dpeak \pm \SI{1.0}{\nano\metre}$.

\begin{table}[!tb]
\centering
\small
\caption{Contact definitions used in this work. The two definitions of $Q$
are deliberately different: equation~(\ref{eq:Q}) conditions the model and
equation~(\ref{eq:Q_cov}) evaluates it, so that the benchmark does not
inherit the contact definition used for training. In the contact-survival
analysis a native contact counts as surviving while its instantaneous
distance stays below $1.5\times$ its native value.}
\label{tab:contactdefs}
\setlength{\tabcolsep}{3pt}
\renewcommand{\arraystretch}{1.2}
\fitcolumn{\begin{tabular}{@{}llll@{}}
\toprule
Where used & Atoms & Cutoff & Separation \\
\midrule
$Q_\mathrm{cond}$, eq.~(\ref{eq:Q})    & C$\alpha$ & 8.0\,\AA           & $|i-j|>3$ \\
Contact survival (this section)        & C$\alpha$ & 8.0\,\AA           & $|i-j|>3$ \\
$Q_\mathrm{eval}$, eq.~(\ref{eq:Q_cov}) & heavy     & 4.5\,\AA           & $|i-j|\geq3$ \\
Go-kit topology (S1.4)                 & heavy     & $1.2\times4.5$\,\AA & $|i-j|\geq3$ \\
\bottomrule
\end{tabular}}
\end{table}

$\Phisim$ was compared with mechanical $\varphi$-values for titin~I27
(AFM protein engineering, 7 sites)\cite{best2002mechanical}, for
protein~L (hydrophobic-core alanine-scanning mutagenesis, 5
sites)\cite{sadler2009rheostat}, and for the B1 domain of protein~G
(engineered bi-histidine metal-chelation sites, 8 sites, reported as
$\varphi_{\mathrm{M}^{2+}\mathrm{U}}$ and converted to
$1-\varphi_{\mathrm{M}^{2+}\mathrm{U}}$ so that larger values denote
retained structure)\cite{shen2012bihistidine,cao2008chelation}, and with
chemical folding $\varphi$-values for ubiquitin (8
sites)\cite{went2005ubiquitin}. Each of the four pulling velocities was
analysed independently, so that cross-velocity reproducibility acts as an
internal robustness check. The correlations obtained, the
partial-correlation control against sequence distance to the nearer
terminus, and the ordinal scope of the agreement are reported in
Section~\ref{sec:res_phi}.

\subsection{Choice of pulling velocity}

The pulling velocity is the remaining protocol parameter that could bias
the corpus. The production velocity (\SI{1}{\nano\metre\per\nano\second}
per terminus) exceeds typical atomic force microscopy rates by several
orders of magnitude, an unavoidable consequence of the accessible
simulation timescale\cite{sotomayor2007single,lu1998unfolding,rico2013highspeed}.
Rather than ask whether the simulated velocity matches experiment, which
it cannot, we tested whether the quantities entering the model are
velocity-dependent over the computationally reachable range.

The all-atom protocol was repeated at four velocities spanning two
orders of magnitude ($v=0.01$, $0.05$, $0.1$ and
\SI{1.0}{\nano\metre\per\nano\second} per terminus) for the four
validation proteins, holding force field, implicit solvent, spring
constant, thermostat parameters and terminus definitions fixed. Because
the same conformation is reached at different times under different
velocities, time is not a comparable coordinate: all cross-velocity
comparisons were made against end-to-end distance, with each trajectory
resampled onto 40 equal-width extension bins spanning zero to the
largest extension reached by that protein at any velocity (bin width
\SIrange{0.44}{0.70}{\nano\metre}, i.e.\ approximately $0.85\,\Lc/40$)
and comparisons restricted to shared bins. This design isolates whether
velocity changes the conformation reached at a given extension,
independent of the rate at which it is reached. The quantities compared
across velocities were the fraction of native contacts $Q$, the radius of
gyration $\Rg$, C$\alpha$-RMSD from the native state, the per-residue
contact-survival curve and the per-residue domain-assignment profile;
agreement was quantified by the root-mean-square difference between
paired curves at matched extension, by Pearson and Spearman correlation
of the residue-level profiles, and by two-sample
Kolmogorov--Smirnov tests on the residue-level distributions. Results are
reported in Section~\ref{sec:res_velocity}.

Those results bound how the corpus can be used. The model is conditioned
on a reaction coordinate (extension or $Q$) and predicts conformation at
a given value of that coordinate, not the time required to reach it, so
neither collapse time nor collapse extension enters as a conditioning
variable. The velocity-sensitive quantities are thus excluded from the
model by construction, and the velocity-robust quantities are exactly the
ones it is conditioned on. Trajectories generated at different velocities
can accordingly be pooled into a single training set without introducing
systematic conformational bias, provided training targets are restricted
to continuous, reaction-coordinate-conditioned geometric and contact
observables.

\subsection{Consistency across force fields and pulling rates}

The velocity analysis above is an all-atom result on four proteins. Two
further checks bound its generality: agreement across force fields of
differing physical detail at fixed velocity, and stability of the
coarse-grained protocol over a velocity range extending beyond the
all-atom test.

Constant-velocity pulling was performed for the four validation proteins
under all four force fields (GoCa, Go-kit, AWSEM, all-atom OpenMM), and
the resulting native-contact rupture orders were compared pairwise
(Supplementary Methods~S1.4). AWSEM pulling was additionally repeated at six
velocities ($v=0.01$, $0.05$, $0.1$, $1.0$, $2.0$ and
\SI{4.0}{\nano\metre\per\nano\second}), extending the range a further
four-fold above the all-atom series and bracketing the production
setting on both sides.

Contact-rupture order is a rank statistic derived by thresholding a
continuous observable, and the velocity analysis above does not extend
to discrete labels obtained by thresholding or ranking; the stability of
such labels must therefore be established separately rather than
assumed. The primary rate-stability criterion is accordingly defined on
the continuous observable --- the root-mean-square difference between
contact-survival curves at matched extension, required to fall within
the band established by the all-atom analysis --- with the Spearman rank
correlation of rupture order reported as a secondary, descriptive
statistic rather than a pass/fail criterion. Table~\ref{tab:licence}
summarises, for each validation study, the class of quantity it
supports.

\begin{table*}[t]
\centering
\small
\caption{Scope of each validation study. Claims beyond the right-hand
column are not supported by the corresponding study.}
\label{tab:licence}
\begin{tabular}{@{}p{3.6cm}p{4.6cm}p{6.4cm}@{}}
\toprule
Study & Design & What it supports \\
\midrule
$\varphi$-value comparison &
All-atom, 4 velocities, 4 proteins with experimental $\varphi$ &
The residue-level unfolding order sampled at the force peak corresponds
to the \emph{mechanical} transition state, ordinally. Not a quantitative
$\varphi$ prediction; not informative away from the force peak. \\[3pt]
Velocity robustness &
All-atom, 4 velocities, 4 proteins, extension-binned &
$Q$ and $R_g$ at matched extension are reproducible to 2.9--7.0\% of
their dynamic range; the per-residue contact-survival and
domain-assignment profiles are reproducible in shape. Not absolute
contact probabilities; not threshold- or rank-derived discrete labels;
not collapse position. \\[3pt]
Cross-force-field &
4 force fields, fixed velocity, 4 proteins &
Rupture order is consistent across a hierarchy of physical detail.
Descriptive; no significance gate. \\[3pt]
CG rate stability &
AWSEM, 6 velocities to \SI{4.0}{\nano\metre\per\nano\second} &
The extension-conditioned contact-survival curves remain within the
all-atom band above the production velocity. \\[3pt]
Segment geometry &
500 proteins, 11 species, 557{,}550 segments &
Unfolded segments are well defined under a fixed-hyperparameter operator,
and their end-to-end extension is determined by segment length to
$\sim$\SI{3.9}{\angstrom}. Velocity-invariance of the segment
\emph{definition} is not established. \\
\bottomrule
\end{tabular}
\end{table*}

\subsection{Segmental structure of the unfolding intermediate}
\label{sec:meth_segment}

A partially unfolded chain under load is not homogeneous: folded and
unfolded regions coexist on the same chain, and whether the boundary
between them is sharp or diffuse determines whether a fragment is a
well-posed learning unit. Statistical-mechanical treatments of folding in
the Wako--Sait\^o--Mu\~noz--Eaton family represent each residue by a
binary variable and so set the folded-to-unfolded crossover width to
zero\cite{wako1978statistical1,wako1978statistical2,munoz1999simple,bruscolini2002exact,imparato2007ising},
while hydrogen-exchange experiments indicate that proteins fold and
unfold through the concerted gain and loss of cooperative sub-chain units
rather than residue by residue\cite{maity2005protein,englander2014nature}.
In both cases a sharp boundary is a model assumption rather than a
measured property, so we tested it directly on the trajectories.

Each residue was classified as folded or unfolded by Kabsch
superposition\cite{kabsch1976solution,kabsch1978discussion} of a
seven-residue C$\alpha$ window onto the corresponding window of the
native reference, with a local RMSD threshold of \SI{2.0}{\angstrom};
window width and threshold were fixed across all proteins with no
per-sample tuning. The operator was applied to every
trajectory frame of AF-Geo500 (500 proteins, 11 species) generated at the
production velocity, and maximal contiguous runs of unfolded residues of
length $\geq 3$ were retained as segments.

For each segment, the Euclidean distance between the C$\alpha$ atoms of
its first and last residues was taken as the end-to-end distance (ETE)
and compared with the fully extended contour length
$\Lmax = 3.8\,(\Nseg - 1)$~\si{\angstrom}, where \SI{3.8}{\angstrom} is
the C$\alpha$--C$\alpha$ virtual bond length under the \textit{trans}
peptide condition, a covalent constraint holding in folded and unfolded
regions alike\cite{levitt1976simplified,oldfield1994analysis}:
\begin{equation}
	\ETE = \alpha \times 3.8\,(\Nseg - 1).
	\label{eq:ete}
\end{equation}
The scaling coefficient $\alpha$ in equation~(\ref{eq:ete}) was estimated
at three levels, which are not interchangeable. The \emph{global}
estimate is a single origin-constrained least-squares fit over all
segments; because the residual is taken on ETE, this fit weights each
segment by $\propto\Lmax^{2}$ and is therefore dominated by long
segments. A two-piece variant of the same fit, with the breakpoint fixed
at $\Nseg = 30$, tests for length dependence, and the residuals of the
two-piece model give the absolute error relevant to geometric
reconstruction. \emph{Per-segment} estimates
$\alpha_k = \ETE_k/L_{\text{max},k}$ were computed for segments with
$\Nseg\geq5$ and $\Lmax>\SI{1}{\angstrom}$, and binned on a logarithmic
length grid to resolve any systematic dependence on segment length.
\emph{Per-protein} estimates were taken as the mean of $\alpha_k$ over
the segments of each protein; comparing the dispersion of the per-protein
and per-segment distributions separates variation intrinsic to individual
segments from variation between proteins and species. Fitted values are
reported in Section~\ref{sec:res_segment}.

Two limitations apply to this analysis. First, all trajectories were
pulled from the N-terminus to the C-terminus; this direction is
externally imposed rather than an intrinsic protein property, and
mechanical response is known to be anisotropic, so quantities that depend
on pulling direction should not be read as intrinsic properties. Segment
length and end-to-end distance in equation~(\ref{eq:ete}) are intrinsic
geometric quantities and are not subject to this limitation. Second, the
binary partition is by construction a threshold-derived discrete label,
and the velocity analysis described above does not extend to such labels;
the velocity-invariance of equation~(\ref{eq:ete}) and of the segment
decomposition is therefore not established by that analysis and remains
to be tested directly.

\subsection{Conformational coverage benchmark and statistical comparison}
\label{sec:meth_coverage}

The preceding sections define the training corpus, the architecture of the
cascade and the simulations used to validate the protocol that generates the
corpus. This section introduces an external reference and answers a different
question: relative to a publicly available and widely used structure
predictor, how much of the folding coordinate does a FoldScape ensemble
actually span. We therefore construct AlphaFold~3 baseline ensembles,
define three complementary order parameters, and compare the two methods
through a conformational coverage statistic evaluated at matched sample
size.

\subsubsection{AlphaFold 3 baseline ensembles}

As an external reference for conformational spread we assembled
AlphaFold~3 (AF3)\cite{abramson2024alphafold3} ensembles for twenty human
proteins, whose sequences were taken from the \textit{Homo sapiens}
proteome of AlphaFold~DB\cite{varadi2024alphafold} and span 36 to 857
residues. Each protein was submitted as a single-chain job in the native
AF3 JSON format with the multiple-sequence-alignment and
structural-template fields left unset, so that AF3 built both through its
default data pipeline, and with twenty random model seeds supplied per
protein. Each seed returns five samples, giving 100 predicted structures
per protein and 2{,}000 conformations in total. All seeds for a given
protein share the same alignment and template stack, and the seed is the
only source of variation between predictions, so that the resulting
spread reflects the structural uncertainty AF3 itself expresses for a
fixed evolutionary input. Predictions from all seeds were pooled per
protein and superposed on C$\alpha$ atoms\cite{kabsch1976solution} to give
one AF3 ensemble per target.

The status of these ensembles should be stated plainly. They serve as an
empirical, reproducible baseline rather than as a description of solution
dynamics: the authors of AF3 note that varying the random seed does not
yield an approximation to the solution
ensemble\cite{abramson2024alphafold3}, and comparable strategies for
AlphaFold~2 instead perturb the
alignment\cite{waymentsteele2024conformations,delalamo2022sampling}.

\subsubsection{FoldScape ensembles}

For the same twenty sequences, FoldScape generated conformational ensembles
with the complete trained pipeline, sampling the flow-matching decoder along
the folding coordinate to give 300 conformations per protein. Each conformation
was obtained by integrating the decoder from an independent Gaussian prior at
one setting of the conditioning triple $(Q,\tilde{R}_g,f_{ss})$, the settings
being swept along the folding coordinate; as at any other inference time,
$(\tilde{R}_g,f_{ss})$ are user-specified rather than read off the target
conformation (Section~\ref{sec:meth_pretrain}). The fourteen
chains of at most 200 residues were generated in a single chain-level pass; the
six longer chains (216--857 residues) were generated through the stride-one
crop-and-superpose scheme of Section~\ref{sec:meth_longchain}. For B3KPR5, 79
conformations were discarded at the parsing stage because their residue count
did not match the reference structure, leaving 221 valid conformations; the
remaining nineteen proteins each contributed 300, for 5{,}921 conformations in
total.

\subsubsection{Per-conformation order parameters}

FoldScape predicts C$\alpha$ traces only, so every predicted conformation
was first back-mapped to heavy atoms with cg2all\cite{heo2024oneparticle}
(Section~\ref{sec:meth_backmap}) before any order parameter was computed;
AF3 models were used as returned. Structures from both methods were then
parsed through an identical pipeline. All hydrogen atoms were discarded and
only heavy atoms retained, so that back-mapped frames carrying explicit
hydrogens and hydrogen-free AF3 models present exactly the same atom set and
can be compared directly. Only the first MODEL of each PDB file and only the
first altloc were used. Residues left unresolved by the back-mapping step are
omitted rather than extrapolated (Supplementary~S7.4), so a reconstructed
frame can carry fewer residues than its reference; frames whose residue count
differed from that of the protein's reference structure were discarded rather
than imputed. Because both back-mapping schemes carry a secondary-structure
prior learned from folded structures, \fss{} is the axis of the three most
exposed to reconstruction bias. Three complementary order
parameters were computed for every conformation. These are evaluation
quantities computed on all-atom coordinates and are defined
independently of the C$\alpha$-level conditioning descriptors of
Section~\ref{sec:meth_pretrain}; the differences are stated explicitly
below.

\paragraph{Radius of gyration.}
Mass-weighted over heavy atoms and reported in nanometres,
\begin{equation}
	R_\mathrm{g}=\sqrt{\frac{\sum_i m_i\lVert \mathbf{x}_i-\mathbf{x}_\mathrm{com}\rVert^2}{\sum_i m_i}},
	\qquad
	\mathbf{x}_\mathrm{com}=\frac{\sum_i m_i\mathbf{x}_i}{\sum_i m_i}.
	\label{eq:rg_cov}
\end{equation}
The radius of gyration of equation~(\ref{eq:rg_cov}) distinguishes compact
from extended conformations and is the most sensitive of the three to the
folding collapse. Because
$R_\mathrm{g}$ carries physical units whereas $Q$ and \fss{} are already
dimensionless and bounded, we write \Rghat{} for the radius of gyration
when it appears on the normalised coverage axis defined below; \Rghat{}
is not the fixed-bound normalisation $\tilde{R}_g$ used for model
conditioning in Section~\ref{sec:meth_pretrain}.

\paragraph{Fraction of native contacts.}
We use the soft definition of Best, Hummer and
Eaton\cite{best2013native}. The reference state is the first model of
that protein's AF3 ensemble, and the native contact set comprises
heavy-atom pairs separated by at least three residues in sequence whose
reference distance is below \SI{4.5}{\angstrom}, tabulated with a KD
tree\cite{bentley1975kdtree} so that large chains remain tractable. For
any conformation,
\begin{equation}
	\fiteq{Q=\frac{1}{|C_\mathrm{native}|}\sum_{(i,j)\in C_\mathrm{native}}\frac{1}{1+\exp\!\big[\beta\,(r_{ij}-\lambda r^{\mathrm{ref}}_{ij})\big]},}
	\label{eq:Q_cov}
\end{equation}
with $\beta=5.0$\,\AA$^{-1}$ and $\lambda=1.8$ (dimensionless), following the
original parameterisation of Best \emph{et al.}\cite{best2013native}. Across the twenty proteins the native
contact count ranges from 258 to 10{,}765. $Q$ is the folding reaction
coordinate and measures how far a conformation lies from the folded
state. Equation~(\ref{eq:Q_cov}) is the original multiplicative
formulation of Best \emph{et al.}\ on heavy atoms, and is used here in
preference to the additive-tolerance C$\alpha$ variant of
equation~(\ref{eq:Q}) so that the benchmark does not inherit the contact
definition used to train the model (Table~\ref{tab:contactdefs}). The two are
numerically different functions of the same conformation, and a value computed
under one is not directly comparable with a value computed under the other;
where the distinction matters we write $Q_\mathrm{eval}$ for
equation~(\ref{eq:Q_cov}) and $Q_\mathrm{cond}$ for equation~(\ref{eq:Q}).

\paragraph{Stable secondary-structure fraction.}
Residues were assigned using the DSSP electrostatic hydrogen-bond
criterion\cite{kabsch1983dssp}. The amide hydrogen is rebuilt from the
preceding peptide plane following the original DSSP procedure, so the
presence or absence of explicit hydrogens in the input file does not
affect the result. The hydrogen-bond energy is
\begin{equation}
	\fiteq{E=0.42\times0.20\times332\left(\frac{1}{r_\mathrm{ON}}+\frac{1}{r_\mathrm{CH}}-\frac{1}{r_\mathrm{OH}}-\frac{1}{r_\mathrm{CN}}\right)\ \mathrm{kcal\,mol^{-1}},}
	\label{eq:dssp}
\end{equation}
and $E<-0.5$\,kcal\,mol$^{-1}$ counts as a bond. An $\alpha$ helix
requires two consecutive overlapping 4-turns, and only contiguous helical
runs of at least four residues are retained. $\beta$ structure is
identified through the parallel and antiparallel bridge criteria of
Kabsch and Sander\cite{kabsch1983dssp}, and only ladders built from at
least two consecutive bridges are retained, so isolated bridges and
single turns do not count. \fss{} is the fraction of residues assigned to
either class. It measures secondary-structure content and is
complementary to the other two quantities; it is the all-atom,
hydrogen-bond-based counterpart of the C$\alpha$-only geometric
$f_{ss}$\cite{kneller2015ss} used for conditioning in
Section~\ref{sec:meth_pretrain}.

\subsubsection{Conformational coverage}

Coverage answers one question: along a given order-parameter axis, how
much of the accessible range does a method's conformations actually span.
It measures breadth of exploration, not accuracy, and not Boltzmann
weight.

For each protein and each order parameter, the conformations of both
methods are first pooled, and the pooled minimum and maximum are used for
min--max normalisation onto $[0,1]$. The two methods therefore share a
single ruler whose endpoints are set by their joint extremes. The
interval $[0,1]$ is divided into $K$ equal bins, with $K=20$ for the main
result. Coverage is the fraction of bins that are occupied,
\begin{equation}
	\Ccov=\frac{1}{K}\sum_{k=1}^{K}\mathbb{1}\big[\,\text{bin }k\text{ contains at least one conformation}\,\big].
	\label{eq:coverage}
\end{equation}
Because the two methods contribute different numbers of conformations,
comparing occupied bin counts directly would favour the larger sample.
All coverage values are therefore evaluated at a matched sample size
$n=\min(n_A,n_B,100)$, which equals 100 for all twenty proteins here. The
point estimate draws $n$ conformations without replacement from that
method's full set, evaluates equation~(\ref{eq:coverage}) on that
subsample, and averages over 2{,}000 repetitions, in the same spirit as rarefaction curves used to
compare species richness in
ecology\cite{hurlbert1971rarefaction,gotelli2001richness}. The 95\%
confidence interval comes from a non-parametric
bootstrap\cite{efron1979bootstrap} that draws $n$ conformations with
replacement over 2{,}000 repetitions and takes the 2.5\% and 97.5\%
percentiles.

\subsubsection{Hypothesis testing}

The within-protein test is a two-sided label permutation
test\cite{ernst2004permutation}. Under the null hypothesis the method
label is independent of conformation, so the pooled conformations are
randomly repartitioned into two groups of the original sizes and the same
statistic $\Delta \Ccov=\Ccov_A-\Ccov_B$ is recomputed. With 100{,}000
permutations,
$p=(1+\#\{|\Delta \Ccov^{*}|\ge|\Delta \Ccov|\})/(n_\mathrm{perm}+1)$.
The permutation statistic uses fewer resamples than the observed
statistic, so the null distribution is slightly broader and the test is
conservative. As a complementary view of the raw distributions we also
report a Mann--Whitney $U$ test\cite{mann1947utest} on the order
parameters themselves.

The across-protein test treats the protein as the pairing unit. Each
protein contributes one pair of point estimates $(\Ccov_A,\Ccov_B)$ per
order parameter, and the twenty differences are tested with a two-sided
Wilcoxon signed-rank test\cite{wilcoxon1945signedrank} using the exact
method.

Both levels treat the three order parameters as one testing family and
apply a Holm--Bonferroni correction\cite{holm1979bonferroni}; adjusted
$P$ values are reported throughout. Significance is marked as **** for
$P<10^{-4}$, *** for $P<10^{-3}$, ** for $P<10^{-2}$ and * for $P<0.05$.

\subsubsection{Robustness check}

The bin count is the only free parameter in the coverage definition, so
the entire analysis was repeated for $K\in\{10,15,20,25,30,40\}$ to
establish whether the conclusion depends on binning granularity.

\subsubsection{Implementation}

All analyses were implemented in Python~3.12 with
NumPy\cite{harris2020numpy}, SciPy\cite{virtanen2020scipy} and
matplotlib\cite{hunter2007matplotlib}, parallelised across proteins, with
the random seed fixed at 20240521. The analysis script, the
per-conformation descriptor table, the per-protein coverage table and the
complete statistics in JSON are released as described under Data
availability.


\section{Results}

The Results are organised in four blocks.
The first places FoldScape against an external reference and fixes the scale of
the result at the outset:
Section~\ref{sec:res_coverage} measures, on twenty human proteins and against
AlphaFold~3 baseline ensembles at matched sample size, how much of the folding
coordinate a FoldScape ensemble actually spans, and finds a gap of about an
order of magnitude that holds for every protein, every order parameter and
every binning setting tested.
The second reports \emph{what} FoldScape predicts at each point of that
coordinate:
Section~\ref{sec:res_recon} evaluates reconstructed C$\alpha$ geometry against
conformations drawn from explicit folding trajectories, and
Section~\ref{sec:res_longchain} follows a single 706-residue chain across the
complete folding coordinate.
The third reports \emph{how} that prediction is obtained:
Section~\ref{sec:res_training} shows that the four stages of the cascade
converge as mutually independent optimisation problems and quantifies what each
one contributes.
The fourth establishes \emph{why} the design takes the form it does:
Sections~\ref{sec:res_phi}, \ref{sec:res_velocity} and~\ref{sec:res_segment}
show, respectively, that the training trajectories sample the mechanical
unfolding pathway at residue resolution, that the conformational ensemble
conditioned on elongation is reproducible in its continuous observables across
two decades of pulling speed, and that unfolding intermediates decompose into
geometrically coherent segments --- the property that motivates fragment-level
learning.
Coverage and accuracy are logically independent quantities, and they are
established here on different material; the two are read together in
Section~\ref{sec:disc_coverage}.

\subsection{FoldScape ensembles cover an order of magnitude more of the
	folding coordinate than AlphaFold 3 baselines}
\label{sec:res_coverage}

\begin{figure*}[tp]
	\centering
	\includegraphics[width=\textwidth,height=\figmaxh,keepaspectratio]{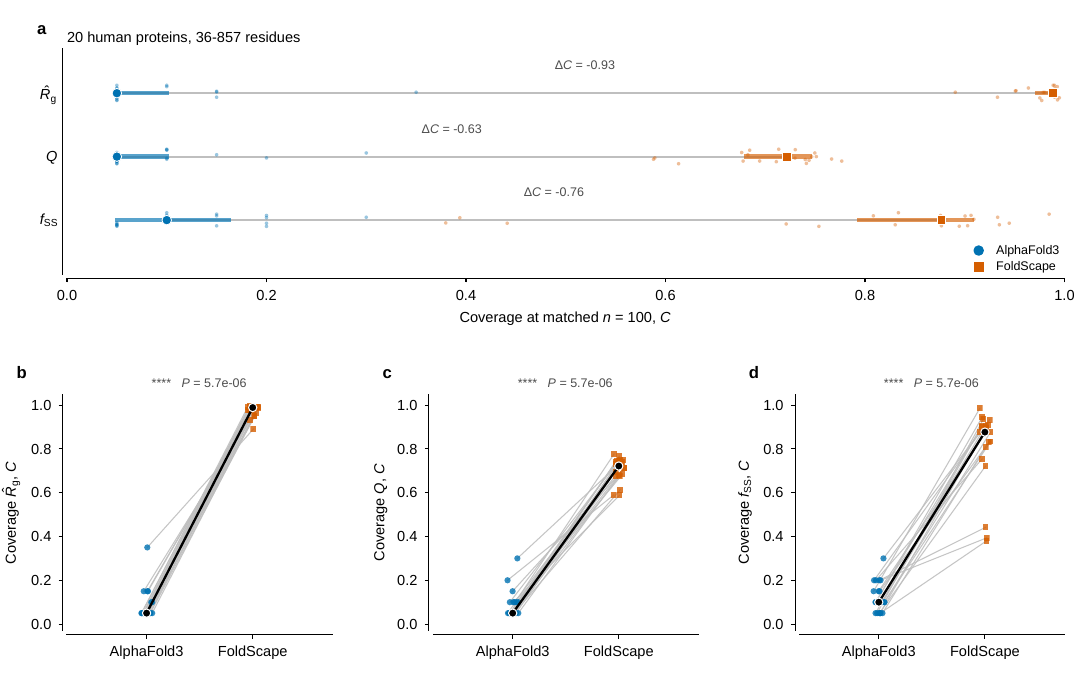}
	\caption{\textbf{At matched sample size, FoldScape ensembles show
			consistently higher conformational coverage than AlphaFold~3
			baselines across three order parameters.}
		\textbf{a,} Coverage summary for twenty human proteins
		(36--857 residues) on three order parameters. Each row is one order
		parameter; small dots are per-protein point estimates (vertically
		jittered to avoid overplotting), thick horizontal bars are the
		across-protein interquartile ranges, the large circle and square are
		the across-protein medians, and the grey line joins the two medians.
		The annotated $\Delta \Ccov$ is the median of the per-protein
		difference $\Ccov_\mathrm{AF3}-\Ccov_\mathrm{FoldScape}$, and the
		asterisks come from a two-sided Wilcoxon signed-rank test with the
		protein as the pairing unit, Holm-corrected across the three order
		parameters.
		\textbf{b--d,} Paired slope plots for each order parameter: radius of
		gyration \Rghat{} (\textbf{b}), fraction of native contacts $Q$
		(\textbf{c}) and stable secondary-structure fraction \fss{}
		(\textbf{d}). Each grey line is one protein, joining its coverage
		point estimates under the two methods, and the heavy black line joins
		the two medians. The split is 20 to 0 on all three axes, with every
		protein pointing the same way.
		All coverage values are evaluated with $K=20$ bins at matched sample
		size $n=100$, and point estimates are means over 2{,}000 subsamples
		drawn without replacement. $P$ values are Holm-corrected two-sided
		Wilcoxon signed-rank tests; **** denotes $P<10^{-4}$.}
	\label{fig:coverage}
\end{figure*}

What separates a generator of the folding process from a predictor of its
endpoint must ultimately reduce to a measurable quantity, and we begin with
that measurement. Two requirements have to be met together: the conformations
a method emits must be accurate at every point of the folding coordinate, and
they must actually be distributed across that coordinate rather than
concentrated at one end of it. This section settles the second requirement
against an external reference, and Sections~\ref{sec:res_recon}
and~\ref{sec:res_longchain} then establish the first, showing that FoldScape's
reconstruction accuracy is flat along the coordinate rather than degrading away
from the native state. Across twenty human proteins we place
FoldScape ensembles and AlphaFold~3 baseline ensembles on the same ruler
and measure conformational coverage through three complementary order
parameters (Section~\ref{sec:meth_coverage}). The panel spans 36 to 857
residues and therefore extends both below the 50-residue lower bound of the
training filter and above the 200-residue ceiling of a single chain-level
forward pass. All values of $Q$ in this section are the evaluation quantity
$Q_\mathrm{eval}$ of equation~(\ref{eq:Q_cov}), not the conditioning quantity
$Q_\mathrm{cond}$ of equation~(\ref{eq:Q}). The conclusion holds for
every protein, every order parameter and every binning setting tested
(Fig.~\ref{fig:coverage}).

\begin{table*}[!tb]
	\centering
	\caption{\textbf{Per-protein conformational coverage at $K=20$ bins and
			matched sample size $n=100$.}
		In each order-parameter column the value before the slash is the
		AlphaFold~3 coverage and the value after the slash is the FoldScape
		coverage. All 60 per-protein permutation tests are significant
		(Section~\ref{sec:res_coverage}).}
	\label{tab:coverage}
	\small
	\setlength{\tabcolsep}{6pt}
	\renewcommand{\arraystretch}{1.1}
	\begin{tabular}{@{}lrrrccc@{}}
		\toprule
		Protein & Residues & $n$ (AF3) & $n$ (FoldScape)
		& \Ccov, \Rghat{} & \Ccov, $Q$ & \Ccov, \fss{} \\
		\midrule
		E9PIP6      &  36 & 100 & 300 & 0.35 / 0.89 & 0.10 / 0.59 & 0.15 / 0.81 \\
		A0A2R8YF12  &  61 & 100 & 300 & 0.05 / 0.98 & 0.05 / 0.77 & 0.20 / 0.44 \\
		H3BM92      &  66 & 100 & 300 & 0.05 / 0.96 & 0.10 / 0.68 & 0.05 / 0.83 \\
		E5RK64      &  71 & 100 & 300 & 0.05 / 0.95 & 0.05 / 0.75 & 0.05 / 0.38 \\
		C9JYH9      &  96 & 100 & 300 & 0.15 / 0.99 & 0.05 / 0.61 & 0.05 / 0.89 \\
		A0A804HJL3  & 101 & 100 & 300 & 0.05 / 0.95 & 0.10 / 0.73 & 0.20 / 0.39 \\
		H0YMC9      & 121 & 100 & 300 & 0.15 / 0.93 & 0.10 / 0.73 & 0.20 / 0.75 \\
		F5H2L2      & 131 & 100 & 300 & 0.15 / 0.99 & 0.05 / 0.68 & 0.10 / 0.91 \\
		C9JCT1      & 141 & 100 & 300 & 0.10 / 0.99 & 0.30 / 0.75 & 0.30 / 0.93 \\
		D6R997      & 152 & 100 & 300 & 0.05 / 0.98 & 0.05 / 0.74 & 0.05 / 0.72 \\
		A0A3B3ISB1  & 157 & 100 & 300 & 0.10 / 0.99 & 0.10 / 0.71 & 0.10 / 0.83 \\
		K7EPK2      & 163 & 100 & 300 & 0.05 / 0.99 & 0.05 / 0.74 & 0.10 / 0.94 \\
		M0R0D4      & 169 & 100 & 300 & 0.05 / 0.98 & 0.05 / 0.68 & 0.10 / 0.90 \\
		F6QUN3      & 176 & 100 & 300 & 0.05 / 0.98 & 0.20 / 0.59 & 0.15 / 0.98 \\
		G3V2U3      & 216 & 100 & 300 & 0.05 / 0.99 & 0.15 / 0.68 & 0.15 / 0.88 \\
		A0A096LNW8  & 280 & 100 & 300 & 0.05 / 0.99 & 0.05 / 0.74 & 0.10 / 0.91 \\
		H0YFY2      & 323 & 100 & 300 & 0.05 / 0.99 & 0.05 / 0.78 & 0.20 / 0.88 \\
		B2R923      & 345 & 100 & 300 & 0.05 / 0.99 & 0.05 / 0.71 & 0.10 / 0.88 \\
		B3KVM9      & 675 & 100 & 300 & 0.05 / 0.99 & 0.05 / 0.75 & 0.05 / 0.93 \\
		B3KPR5      & 857 & 100 & 221 & 0.05 / 0.99 & 0.05 / 0.69 & 0.05 / 0.90 \\
		\bottomrule
	\end{tabular}
\end{table*}

\subsubsection{AlphaFold 3 ensembles occupy a narrow band near the native
	state}

At matched sample size $n=100$, the median AF3 coverage is 0.05 for
\Rghat{}, 0.05 for $Q$ and 0.10 for \fss{}. Of twenty bins, one or two are
typically occupied. Per protein, AF3 coverage ranges from 0.05 to 0.35 for
\Rghat{}, from 0.05 to 0.30 for $Q$ and from 0.05 to 0.30 for \fss{}, with
no protein exceeding 0.35 (Table~\ref{tab:coverage}). Counting occupancy
directly, all 100 AF3 predictions fall into a single \Rghat{} bin for 14 of
the 20 proteins, into a single $Q$ bin for 12, and into a single \fss{} bin
for 6; the median single-bin occupancy fraction is 1.000 for \Rghat{} and
$Q$ and 0.945 for \fss{}.

The raw scales tell the same story. The radius of gyration of an AF3
ensemble varies over a narrow interval; on the $Q$ axis the median
per-protein lower bound is 0.958, 93.1\% of all predictions have $Q>0.95$
and the 5th percentile is 0.940. The variation produced by twenty random
seeds is confined to a
narrow band around the native state. This is consistent with the design
intent of AF3: varying the seed expresses the model's own structural
uncertainty, not the conformational continuum traversed during
folding\cite{abramson2024alphafold3}.

\subsubsection{FoldScape ensembles span most of the folding coordinate}

On the same ruler, the median FoldScape coverage is 0.989 for \Rghat{},
0.722 for $Q$ and 0.877 for \fss{}, with interquartile ranges of
0.972--0.990, 0.681--0.745 and 0.795--0.907 respectively. Along the
radius-of-gyration axis the twenty bins are very nearly filled.

Raw spans give the same picture: for every protein and on every order
parameter the FoldScape ensemble occupies a wider interval than the AF3
ensemble of the same protein, without exception. At the level of individual
conformations, the median FoldScape $Q$
is 0.242 and 29.3\% of conformations have $Q<0.1$, placing them on the
high-entropy unfolded side; 25.6\% have $\fss=0$, corresponding to
complete loss of secondary structure.

The extended end admits an independent physical consistency check. A fully
straightened chain should approach the rigid-rod limit
$R_\mathrm{g}^{\max}=(N-1)\times3.8\,\text{\AA}/\sqrt{12}$. The ratio of the
most extended FoldScape conformation's $R_\mathrm{g}$ to that limit lies
between 0.95 and 1.00 for nineteen proteins, with a median of 0.98. The
single exception is B3KPR5 at 0.68, the one protein whose ensemble lost
conformations at the parsing stage (Section~\ref{sec:meth_coverage}); with
part of its ensemble unavailable this ratio is a lower bound, and we report it
unchanged rather than imputing the missing frames. The extended end of the
ensemble therefore respects the geometric ceiling set by chain connectivity.
The scope of that statement should be read exactly: the ratio is bounded above
by unity by construction, so it excludes conformations that are more extended
than a straight chain permits and nothing else. Chain-internal stereochemistry
--- clashes, bond lengths and bond angles --- is not tested by it, and is
instead assessed frame by frame with MolProbity\cite{williams2018molprobity} in
the back-mapping workflow (Supplementary~S7.4), whose per-frame score
distributions accompany the deposited analysis outputs.

\subsubsection{The coverage gap is consistent across all proteins and all
	order parameters}

The within-protein label permutation tests number 60 in total
(20 proteins $\times$ 3 order parameters) and every one is significant. In
58 of them the Holm-adjusted $P$ value is $3.0\times10^{-5}$, the
resolution floor of a test with 100{,}000 permutations, and the \fss{} test
for A0A2R8YF12 gives $4.0\times10^{-5}$. The weakest case is \fss{} for
A0A804HJL3 at $P=8.8\times10^{-3}$, which also carries the smallest
coverage gap of all 60 combinations ($\Delta \Ccov=-0.197$).

The across-protein test, with the protein as the pairing unit, agrees. The
median $\Delta \Ccov=\Ccov_\mathrm{AF3}-\Ccov_\mathrm{FoldScape}$ is
$-0.926$ for \Rghat{}, $-0.630$ for $Q$ and $-0.755$ for \fss{}, and the
split is 20 to 0 on all three axes: every protein points the same way, with
no reversal and no tie. The two-sided Wilcoxon signed-rank test gives
$P=1.9\times10^{-6}$, or $P=5.7\times10^{-6}$ after Holm correction, which
is the smallest value attainable for twenty paired samples under a
two-sided exact test. The inferential weight therefore rests on the
consistency of the direction rather than on the magnitude of any single
$P$ value. In Fig.~\ref{fig:coverage}b--d each line is one protein; every line
rises, with no exception in either direction.

\subsubsection{The conclusion does not depend on binning granularity}

Raising the bin count from 10 to 40 lowers the mean coverage of both
methods, an inevitable consequence of finer binning, but leaves the
difference stable. On all three order parameters the mean $\Delta \Ccov$ keeps
the same sign and the same order of magnitude across the whole range of $K$,
and the 20-to-0 split is unchanged at every bin count tested, indicating that
the result follows from the difference in spread between the two ensembles
rather than from any particular binning choice.

\subsubsection{FoldScape expands conformational coverage along the folding
	coordinate}

Taking the union of the two ensembles completes the picture. These occupancy
counts use the same $K=20$ binning as above but each method's complete
ensemble rather than a matched subsample, so they are bin counts of the pooled
data and are not on the same footing as the matched-$n$ coverage values.
For \Rghat{} the union fills all twenty bins, of which a median of 19 are
occupied by FoldScape alone and 0 by AF3 alone. For \fss{} the union covers a
median of 20 bins. For $Q$ the union covers a median of 17 bins, with 1 bin
occupied by AF3 alone and 15 by FoldScape alone.

The results establish broad conformational coverage across three independent
order parameters. At matched sample size, the median FoldScape coverage is
0.989 for \Rghat{}, 0.722 for $Q$ and 0.877 for \fss{}, compared with
0.05, 0.05 and 0.10 for the seed-varied AF3 baseline, respectively.
Across every evaluated protein and all three order parameters, FoldScape
occupies a wider range of the folding-coordinate space. Together, these
measurements demonstrate the ability of sequence-conditioned conformational
generation to explore diverse folding stages beyond a narrow distribution
of native-centered structural predictions
\cite{bryngelson1995funnels,onuchic1997theory,dill2012folding}.

\subsubsection{What this comparison does and does not support}

Coverage measures breadth of exploration. It does not measure the accuracy
of individual conformations, nor the thermodynamic weight of the ensemble.
A method could well show low coverage while placing every frame closer to
the truth, and this statistic cannot adjudicate that.

A second limitation concerns the conditioning variables: at inference
$\tilde{R}_g$ and $f_{ss}$ are supplied by the user
(Section~\ref{sec:meth_pretrain}), and because the radius of gyration is also
one of the three evaluation order parameters, part of the \Rghat{} coverage is
attributable to the conditioning schedule rather than to what the model has
learned, whereas the AlphaFold~3 baseline has no comparable control. The
ablation that would settle this---supplying only the sequence and $Q$, with
$(\tilde{R}_g,f_{ss})$ predicted rather than given---has not been performed, so
the \Rghat{} axis should be read as the weakest of the three, while $Q$ and
\fss{} are less exposed because both are evaluated by all-atom criteria
(equations~(\ref{eq:Q_cov}) and~(\ref{eq:dssp})) that differ from the
C$\alpha$ quantities used for conditioning.

The reference state for $Q$ is the first model of the AF3 ensemble, so AF3
enjoys a structural advantage at the native end of the $Q$ axis. Its low
coverage there reflects a concentrated distribution rather than a departure
from the reference.

The matched sample size is capped at 100, so the comparison uses 100
conformations from each method and leaves the remainder of each FoldScape
ensemble unused. This is the price paid for a fair comparison.

Finally, the construction of the AF3 baseline bounds what it can support.
It is a reproducible external reference and not an estimate of the solution
ensemble\cite{abramson2024alphafold3}, and this comparison should not be
read as a failure of AF3 on the task it was designed for.

\subsection{C$\alpha$ reconstruction along the folding coordinate}
\label{sec:res_recon}

Protein structure prediction is now considered largely solved at the native
end of the folding reaction for the majority of monomeric
domains\cite{jumper2021alphafold,baek2021rosettafold,lin2023esmfold,abramson2024alphafold3,kryshtafovych2023casp15,lane2023frontier}.
The folding process is a different target: describing it requires the
conformational ensemble at \emph{every} value of the folding coordinate $Q$,
the fraction of native contacts
formed\cite{bryngelson1995funnels,onuchic1997theory,best2013native}, so a predictor accurate only
at $Q = 1$ carries no information about the barrier region, the collapse
transition, or the expanded unfolded
side\cite{dill2012folding,noe2019boltzmann,zheng2024distributional,lewis2025bioemu}. We therefore evaluated
FoldScape against conformations drawn from explicit folding trajectories rather
than native structures alone, using $Q$ rather than simulation time as the
evaluation axis (Fig.~\ref{fig:recon}).

Figure~\ref{fig:recon}a shows C$\alpha$ traces for six proteins at
conditioning values $Q_\mathrm{cond} = 0.6$--$1.0$ (equation~(\ref{eq:Q})) (green, molecular dynamics reference; cyan, the chain-level trace
exported by Stage~1 of the trained cascade;
Kabsch-superposed\cite{kabsch1976solution}). In all six cases the prediction
reproduces the reference's global topology --- chain path, order of
secondary-structure elements, and relative domain orientation --- including
features unobtainable from a single local window: the long helical arms
projecting from the core in a(iv) and a(vi) are recovered in both direction and
length, which requires the model to place a helix correctly relative to a core
separated from it by a substantial stretch of sequence.

Residual deviations fall into two classes. The first is helix-packing
register: in the bundle at the left of a(ii), corresponding helices are
displaced laterally by roughly one helix diameter, so the bundle's envelope is
correct but its internal register is not fully resolved. The second is
solvent-exposed long loops: the long loop at the right of a(v) runs inside the
structure in the prediction but bulges outward in the reference, the largest
single deviation among the six structures. Both classes occupy positions where
long-range contact restraints are sparse --- consistent with a fold that is
already committed at this $Q$ but whose local assembly retains residual
freedom.

\subsection{A complete folding trajectory of a 706-residue chain}
\label{sec:res_longchain}

\begin{figure*}[tp]
	\centering
	\includegraphics[width=\textwidth,height=\figmaxh,keepaspectratio]{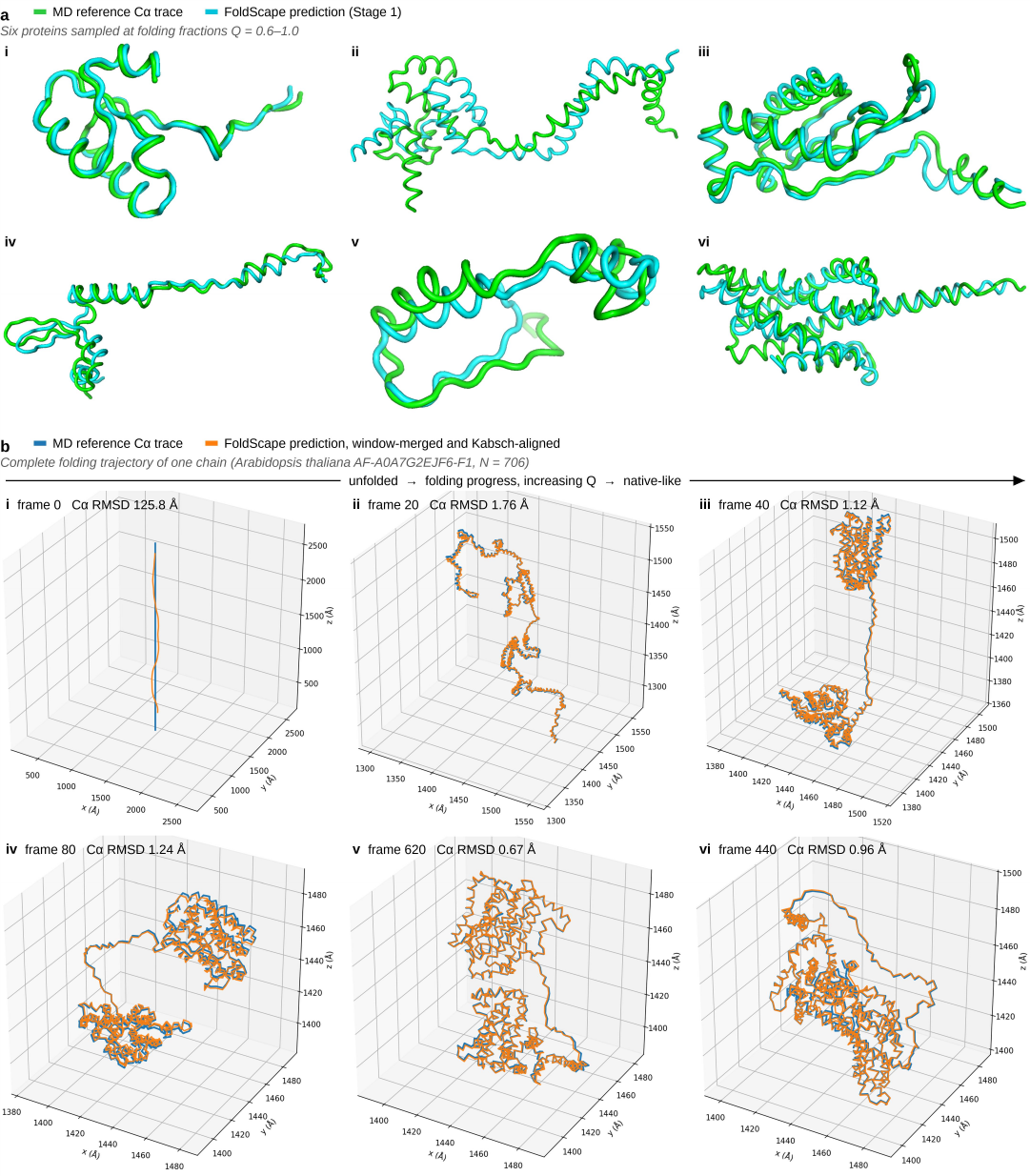}
	\caption{\textbf{FoldScape reconstructs C$\alpha$ structures across the entire
			folding coordinate.}
		\textbf{a}, Superposition of C$\alpha$ traces for six proteins sampled at
		conditioning values $Q_\mathrm{cond} = 0.6$--$1.0$ (equation~(\ref{eq:Q})). Green, molecular dynamics reference; cyan,
		FoldScape prediction, taken as the chain-level trace exported by Stage~1 of
		the trained cascade and shown here as an intermediate of the cascade rather
		than as the output of the pipeline; the two are optimally superposed by the
		Kabsch algorithm. The overall topology, the order of secondary-structure elements
		and the relative orientation of domains are reproduced in all six cases;
		residual deviations are confined to helix-packing register within helical
		bundles (for example a(ii)) and to solvent-exposed long loops (for example
		a(v)).
		\textbf{b}, Six representative conformations along the complete folding
		trajectory of a single chain (\textit{Arabidopsis thaliana}
		AF-A0A7G2EJF6-F1, $N = 706$). Blue, molecular dynamics reference; orange,
		FoldScape prediction from the complete trained pipeline including the
		flow-matching decoder, assembled from 507 overlapping 200-residue crops by
		the stride-one crop-and-superpose scheme of
		Section~\ref{sec:meth_longchain} and superposed onto the reference; the
		``window-merged and Kabsch-aligned'' label inside the panel refers to that
		crop merge, not to the sliding-window consensus internal to R1 and R2.
		Panels are
		ordered by increasing $Q$; the indices annotated in the figure identify
		frames within the stored trajectory and do not follow the $Q$ order.
		Annotations give
		the global C$\alpha$ RMSD of each conformation. From the fully extended state in
		b(i) to the compact state in b(vi) the axis range contracts from
		approximately $2\times10^{3}$\,\AA{} to approximately $10^{2}$\,\AA{};
		b(iii) and b(iv) reveal a multi-domain route in which two domains collapse
		independently before the connecting segment contracts and brings them
		together. The 125.8\,\AA{} value in b(i) arises from the ill-conditioning of
		global superposition on a nearly one-dimensional extended chain rather than
		from a loss of local geometric accuracy (see Discussion). Axis ranges differ
		between panels; coordinate values are simulation-box coordinates and carry no
		absolute meaning.}
	\label{fig:recon}
\end{figure*}

To test whether this accuracy extends across the full folding coordinate
rather than its native segment alone, we followed a single chain ---
\textit{Arabidopsis thaliana} AF-A0A7G2EJF6-F1\cite{varadi2024alphafold}, $N = 706$ ---
through a complete folding trajectory (Fig.~\ref{fig:recon}b). Six
conformations, ordered by increasing $Q$, are shown with the molecular
dynamics reference in blue and the FoldScape prediction in orange.

b(i), the fully extended conformation, has an axis range of order
$2\times10^{3}$\,\AA{} --- the same order as the \SI{2679}{\angstrom} contour
length expected for 706 residues at the canonical 3.8\,\AA{}
C$\alpha$--C$\alpha$ spacing --- and a global C$\alpha$
root-mean-square deviation (RMSD) of 125.8\,\AA{}. By b(ii) the chain has
collapsed into a compact but largely unstructured state (coordinate range
$2.5\times10^{2}$\,\AA{}, RMSD 1.76\,\AA{}); this RMSD is measured against that
one reference frame, drawn from a high-entropy region of the ensemble, and
should not be read as agreement with the unfolded ensemble as a whole. b(iii) and
b(iv) show a two-globular architecture with two
independently formed compact cores joined by an extended linker: the cores are
furthest apart in b(iii) and have begun to approach as the linker contracts in
b(iv) (RMSD 1.12\,\AA{} and 1.24\,\AA{}). In b(v) the cores are in contact with
only one long loop free, and RMSD reaches the series minimum, 0.67\,\AA{}.
b(vi), the most compact conformation, has fused cores with one protruding loop
(RMSD 0.96\,\AA{}).

The frame indices annotated in Fig.~\ref{fig:recon}b do not increase along the
panel sequence, confirming that the series is ordered by $Q$ rather than by
position in the trajectory. Read along $Q$, the RMSD remains approximately stable over the displayed
conformations: excluding b(i), it spans 0.67--1.76\,\AA{} across the
sampled partially folded and compact states, with neither a monotonic
decrease nor convergence confined to the most compact frame. b(i), at
125.8\,\AA{}, is the series' single outlier; we show below that this reflects
the metric rather than the prediction.

Two further points bear on the method's scope. First, at 706 residues this
chain exceeds by more than threefold the 200-residue ceiling that a single
chain-level forward pass accepts. It is nevertheless processed by the complete
trained pipeline --- R1, R2, Stage~1 and the flow-matching decoder --- through
the stride-one crop-and-superpose scheme of
Section~\ref{sec:meth_longchain}: 507 overlapping 200-residue crops are each
predicted in full and merged by parameter-free Kabsch superposition, so that
every interior residue is determined as the average of up to 200 independent
predictions. The result is therefore a test of the assembly scheme at a chain
length far outside the training range, not of a single module extrapolating.
Second, b(iii) and b(iv) resolve a multi-domain folding route in which the two
domains collapse independently before the linker contracts and docks
them\cite{han2007multidomain,borgia2011misfolding}, recovered correctly only because the
reconstruction is accurate simultaneously at two separated cores and along the
extended linker joining them.

\subsection{Stage-wise convergence of the FoldScape cascade}
\label{sec:res_training}

\begin{figure*}[tp]
	\centering
	\includegraphics[width=\textwidth,height=\figmaxh,keepaspectratio]{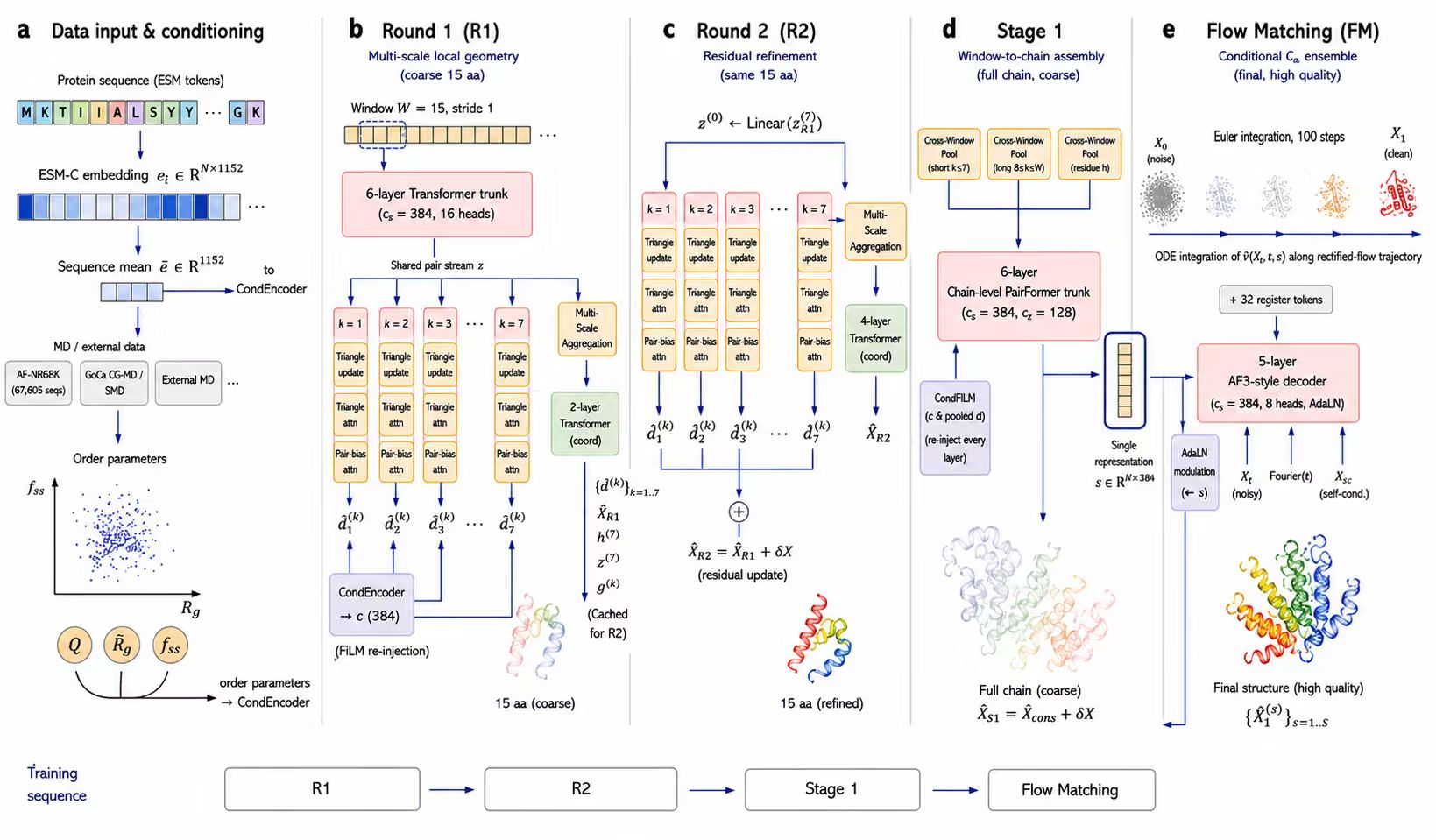}
	\caption{\textbf{Architecture of the FoldScape cascade.}
		\textbf{a}, Data input and conditioning. Residue-level ESM-C
		embeddings\cite{hayes2025esm3,lin2023esmfold} $e_i \in \mathbb{R}^{N\times1152}$ and their
		sequence mean $\bar{e} \in \mathbb{R}^{1152}$---the latter entering the cascade
		through the frozen conformational latent $\mathbf{v}$ of
		Section~\ref{sec:meth_pretrain}---are combined with order parameters
		($Q$, $\tilde{R}_g$, $f_{ss}$) computed from the trajectory frames of the
		training corpus, namely G\={o}-C$\alpha$ coarse-grained MD on the 848
		proteins drawn from AF-Sub1K and constant-velocity SMD, and are encoded by
		CondEncoder. AF-NR68K (67{,}605 sequences) is the non-redundant pool from
		which those subsets are drawn; the external MD repositories shown in the
		panel are held out for cross-validation and do not enter training.
		\textbf{b}, Round~1: sliding windows of $W = 15$ residues with stride 1 pass
		through a 6-layer Transformer trunk ($c_s = 384$, 16 heads) and a shared pair
		stream $z$; seven scale-specific PairFormer blocks (triangle update, triangle
		attention and pair-bias attention) predict the C$\alpha$--C$\alpha$
		distances $\hat{d}^{(k)}$ at sequence separations $k = |i-j| = 1$--7, which a
		multi-scale aggregation module and a 2-layer coordinate Transformer decode into
		the windowed trace $\hat{X}_{R1}$. Conditioning $c$ (384) is re-injected by
		feature-wise linear modulation\cite{perez2018film}; $h^{(7)}$, $z^{(7)}$ and
		$g^{(k)}$ are cached for Round~2.
		\textbf{c}, Round~2: the cached pair stream initialises
		$z^{(0)} \leftarrow \mathrm{Linear}(z^{(7)}_{R1})$, and a second seven-scale
		cascade with a 4-layer coordinate Transformer predicts only the residual update
		$\hat{X}_{R2} = \hat{X}_{R1} + \delta X$ on the same 15-residue windows.
		\textbf{d}, Stage~1: cross-window attention pooling (short $k \le 7$, long
		$8 \le k < W$, and residue-level $h$) lifts window-local features to the chain
		index; a 6-layer chain-level PairFormer trunk ($c_s = 384$, $c_z = 128$) with
		CondFiLM re-injection at every layer produces the single representation
		$s \in \mathbb{R}^{N \times 384}$ and, through an auxiliary head, the full-chain
		coarse trace $\hat{X}_{S1} = \hat{X}_{cons} + \delta X$.
		\textbf{e}, Flow matching: a 5-layer AlphaFold3-style decoder\cite{abramson2024alphafold3}
		($c_s = 384$, 8 heads, AdaLN) with 32 register tokens, AdaLN modulation from
		$s$, Fourier time features and self-conditioning learns the conditional velocity
		field $\hat{v}(X_t,t,s)$ along a rectified-flow
		trajectory\cite{liu2023rectified,lipman2023flow}; 100-step Euler integration from noise
		$X_0$ yields the C$\alpha$ ensemble $\{\hat{X}_1^{(s)}\}_{s=1..S}$.
		Bottom, training order: R1 $\rightarrow$ R2 $\rightarrow$ Stage~1 $\rightarrow$
		flow matching, each stage trained with its upstream frozen.}
	\label{fig:arch}
\end{figure*}

FoldScape decomposes C$\alpha$-trace reconstruction into a cascade of four
separately supervised stages (Fig.~\ref{fig:arch}). Round~1 (R1) resolves
multi-scale local geometry inside 15-residue sliding windows; Round~2 (R2)
refines the same windows through a purely residual branch; Stage~1 assembles
window-local features into a chain-level conditioning representation; and a
rectified-flow decoder generates the final C$\alpha$ ensemble conditioned on
that representation. Each stage is trained with all upstream stages frozen and
their outputs pre-cached, so the four training records of Fig.~\ref{fig:curves}
are mutually independent optimisation problems rather than segments of a single
end-to-end run --- the property on which every attribution below depends. All
four stages share the cluster-level training/validation split of
Section~\ref{sec:meth_splits}, so the validation curves refer throughout to the
same held-out chains.
Throughout this section an epoch denotes one pass over the training subsample
consumed by that stage; because the stages consume different sampling units ---
sliding windows for Rounds~1 and~2, whole chains for Stage~1 --- epoch counts
are comparable within a stage but not between stages. Conformational latent
pretraining (Section~\ref{sec:meth_pretrain}) precedes these four stages and is
not one of them: it supplies a frozen conditioning input, its two probes are
training-only diagnostics that are discarded afterwards, and no convergence
record for it is reported here.

\subsubsection{Hierarchical local-geometry regression establishes window-level
	coordinate accuracy}

Round~1 processes each chain as a set of 15-residue sliding windows and
resolves, through seven scale-specific PairFormer
blocks\cite{jumper2021alphafold,abramson2024alphafold3}, the C$\alpha$--C$\alpha$ distances
at sequence separations $|i-j| = 1$ to 7 before decoding the windowed
C$\alpha$ trace --- an explicit reduction of three-dimensional
reconstruction to a set of one-dimensional distance regressions, in the spirit
of distogram- and orientation-based
predictors\cite{senior2020alphafold,yang2020trrosetta,ingraham2023chroma}. Over 80 epochs the
Kabsch-aligned\cite{kabsch1976solution} window RMSD fell from
4.26\,\AA{} to 2.97\,\AA{} on the training set and from 4.03\,\AA{} to 2.98\,\AA{}
on the validation set, reaching its best value of 2.96\,\AA{} at epoch 75 and
remaining flat to within 0.02\,\AA{} thereafter (Fig.~\ref{fig:curves}a). The learning curve separates into two regimes: a
plateau near 4.0\,\AA{} over the first ten epochs, followed by an approximately
linear descent that reaches a second plateau at epoch 70. This shape is
consistent with the design of the anchored distance heads, whose
tanh-bounded outputs concentrate the initial predictions near the empirical
anchor of each scale, so that the seven distance scales are calibrated first and
the coordinate decoder subsequently benefits from already-calibrated distance
features. Training and validation curves remain superimposed throughout,
differing by only 0.01\,\AA{} at the final epoch, indicating that the
43.3~M-parameter Round~1 does not overfit at the current data scale and that
model capacity is not the limiting factor at this stage --- a regime in which
structure-prediction trunks are known to generalise robustly even under
substantially reduced training-set size and diversity\cite{ahdritz2024openfold}.

\subsubsection{Gains in global alignment accuracy come at the cost of local
	geometric regularity}

Within the same run, the C$\alpha$--C$\alpha$ virtual bond-length
deviation follows the opposite trend to the global RMSD
(Fig.~\ref{fig:curves}b). On the validation set it fell rapidly from 0.42\,\AA{}
to 0.18\,\AA{} over the first ten epochs and then rose steadily---with
epoch-to-epoch fluctuation but no sustained recovery---to
0.30\,\AA{} by epoch 80, a 67\% increase over its minimum; the training set
reached its minimum of 0.21\,\AA{} at epoch 15 and rose to 0.32\,\AA{}, a 52\%
increase. Because both curves rise together, this is not overfitting but a
systematic preference of the objective: the seven distance scales are weighted
equally, so the nearest-neighbour term carries only one seventh of the distance
loss, and the gain the coordinate term obtains from better whole-window
superposition outweighs the degradation of individual bonds relative to the
ideal C$\alpha$ virtual-bond length of 3.80\,\AA{}\cite{levitt1976simplified}. As the
metric is computed from decoded coordinates rather than from the
nearest-neighbour distance head, its rise specifically reveals a growing
inconsistency between predicted distances and decoded coordinates, and shows
that window-level coordinate supervision alone does not preserve local chain
regularity. This observation directly motivated two downstream design choices:
the auxiliary geometry branch of Round~2, which predicts C$\alpha$
pseudo-bond and pseudo-torsion angles alongside per-scale residual distance
heads, and the always-on bond-length, bond-angle and steric-clash penalties of
the flow-matching objective --- the same class of polymer-statistics and
physical constraints that generative and coarse-grained models of protein
structure impose explicitly\cite{ingraham2023chroma,majewski2023cgpotentials}.

\subsubsection{Residual refinement nearly halves the window-level coordinate error}

Round~2 reads the complete Round~1 output as a side channel, re-injects it at
every layer of its cascade through feature-wise linear
modulation\cite{perez2018film}, and predicts only residual corrections to the
distances and coordinates. Because all modulation projections are
zero-initialised\cite{peebles2023dit} and the residual distance and coordinate
heads are initialised at $\sigma=10^{-3}$ with zero bias, the network reproduces
Round~1 to within numerical noise at initialisation, and the entire descent in
Fig.~\ref{fig:curves}c is attributable to the residual branch itself. The curve
is recorded from the end of the first epoch, by which point the residual branch
has already recovered a substantial part of its eventual gain, so its plotted
starting value lies below the Round~1 level rather than at it. Over 56
epochs the validation window RMSD fell from 1.85\,\AA{} to 1.59\,\AA{}, reaching
1.592\,\AA{} at epoch 54 and holding that value to the final epoch, a 46.3\%
reduction relative to the
2.96\,\AA{} of Round~1; the training set fell from 1.68\,\AA{} to 1.37\,\AA{}. In
contrast to Round~1, the two curves are separated throughout and the gap reaches
0.224\,\AA{} at the final epoch (validation 1.592\,\AA{}, training
1.368\,\AA{}), indicating that the 77.4~M parameters combined
with side-channel inputs exploit model capacity more fully while incurring mild
overfitting. The cosine learning-rate restart\cite{loshchilov2017sgdr} at epoch 31
raised the training error transiently by 0.053\,\AA{}, from 1.403\,\AA{} to
1.456\,\AA{}, after which it fell within 25 epochs to 1.368\,\AA{}, below the
pre-restart minimum. The validation error first rose to 1.690\,\AA{}, returned to
its pre-restart level only at epoch 47, and finally improved to 1.592\,\AA{},
exceeding the pre-restart best of 1.606\,\AA{} at epoch 23 by 0.014\,\AA{} --- a
difference that, with a single training run and no repeated seeds, we do not
interpret.

\subsubsection{Cross-window assembly lifts local geometry into a chain-level
	representation}

Stage~1 lifts every feature with window-local support to the chain index by
cross-window attention pooling and produces the conditioning representation
through a six-layer chain-level PairFormer trunk. Its parallel coordinate head
reached a chain-level Kabsch-aligned RMSD of 7.11\,\AA{} on the training set and
6.70\,\AA{} on the validation set at the end of the first epoch, for chains of at
most 200 residues (Fig.~\ref{fig:curves}d). The validation value lies below the
training value as a natural consequence of the evaluation protocol: the training
value is a running average over an epoch during which the model improves
continuously, whereas validation is a single evaluation at the epoch boundary.
That the chain-level error exceeds the window-level error by an order of
magnitude is a structural feature of this stage --- rigid-body superposition of
a 15-residue window requires only local conformation to be correct, whereas the
Kabsch RMSD of a chain of up to 200 residues is dominated by long-range assembly
error, the component that structure-prediction networks acquire last and least
readily\cite{ahdritz2024openfold,jumper2021alphafold}. We emphasise that the chain-level
coordinate head serves only as auxiliary supervision; the product of this stage
is the single representation that conditions the downstream decoder.

\subsubsection{The flow-matching decoder converges stably under chain-level
	conditioning}

\begin{figure*}[tp]
	\centering
	\includegraphics[width=\textwidth,height=\figmaxh,keepaspectratio]{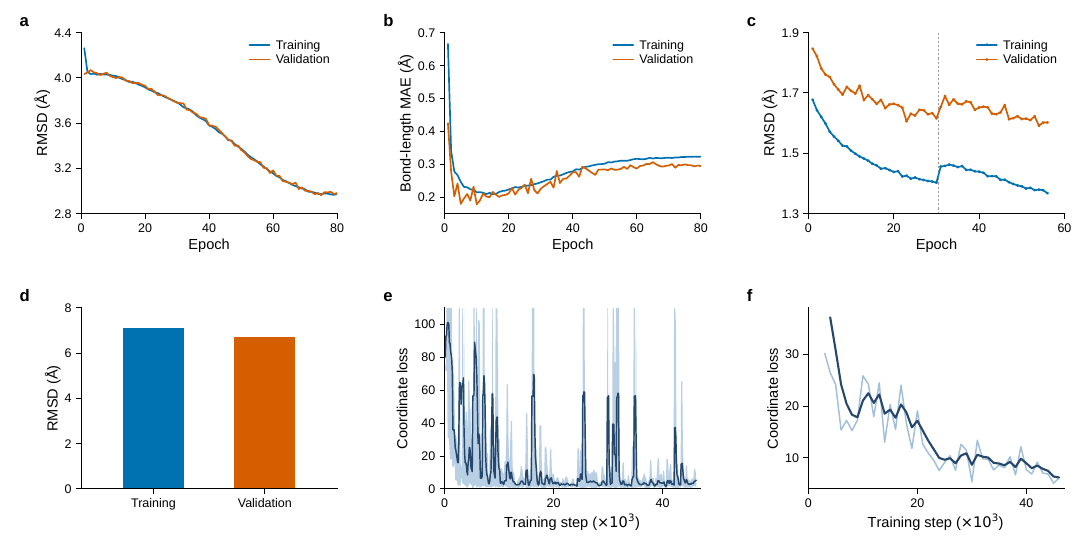}
	\caption{\textbf{Training convergence of the FoldScape cascade.}
		\textbf{a}, Window-level coordinate error of Round~1: root-mean-square deviation
		between predicted and reference C$\alpha$ traces within a 15-residue sliding
		window after optimal Kabsch superposition, for the training (blue) and
		validation (orange) sets over 80 epochs; the best validation value is 2.96\,\AA{}
		at epoch 75.
		\textbf{b}, Mean absolute deviation of C$\alpha$--C$\alpha$ virtual bond
		lengths, $\bigl|\,\lVert \mathrm{C\alpha}_i - \mathrm{C\alpha}_{i+1}\rVert -
		3.80\,\text{\AA}\,\bigr|$, computed from the decoded coordinates of the same run;
		the curves reach their minima at epoch 10 (validation, 0.18\,\AA{}) and epoch 15
		(training, 0.21\,\AA{}) and rise steadily thereafter to 0.30\,\AA{} and
		0.32\,\AA{} at the final epoch, the validation curve fluctuating from epoch to
		epoch about that rise.
		\textbf{c}, Window-level coordinate error of Round~2 over 56 epochs; the dotted
		vertical line marks a cosine learning-rate restart at epoch 31, and the best
		validation value is 1.59\,\AA{} at epoch 54.
		\textbf{d}, Chain-level Kabsch-aligned RMSD of the Stage~1 auxiliary coordinate
		head at the end of the first epoch (training 7.11\,\AA{}, validation 6.70\,\AA{};
		chains of at most 200 residues); the training value is a running average over
		the epoch and the validation value a single evaluation at its end.
		\textbf{e}, Training-set coordinate loss of the flow-matching decoder against
		optimisation step; the dark line is the exponential moving average and the pale
		band spans the range of the per-step raw loss, with spikes exceeding the axis
		limit truncated at 110.
		\textbf{f}, Validation-set coordinate loss of the same run, evaluated every
		1{,}000 steps (47 evaluations); the pale line shows the individual evaluations
		and the dark line their exponential moving average (smoothing factor 0.6).
		Evaluations before step 3{,}000 exceed the axis range and are not shown.
		Vertical axes in \textbf{a}--\textbf{d} are in \aa ngstr\"om; the coordinate
		loss in \textbf{e} and \textbf{f} is the weighted translation term of the
		flow-matching objective and is dimensionless. All curves come from a single
		training run per stage; no repeated seeds were performed, and no error bars are
		therefore shown.}
	\label{fig:curves}
\end{figure*}

The flow-matching decoder is trained on frozen upstream representations and
learns a rectified-flow conditional velocity field in coordinate
space\cite{liu2023rectified,lipman2023flow}. The training coordinate loss fell from above 110
to below 10 within the first 4{,}188 steps, and after 20{,}000 steps its
baseline stabilised at a median of 3.65, on top of which intermittent spikes
appear; 58 of the plotted intervals exceed the display range over 46{,}233 steps
(Fig.~\ref{fig:curves}e). The spike structure is consistent with the time
weighting $1/(1-t)^2$ of the rectified-flow translation term, which diverges as
the sampled flow time approaches 1 and is clamped numerically at $10^{4}$, so
that a single batch can exceed the baseline by one to two orders of magnitude;
the smoothed curve remains flat across the same intervals, indicating that the
spikes originate in time sampling rather than in optimisation
divergence\cite{karras2022edm}. The validation coordinate loss, evaluated every
1{,}000 steps, fell overall without any sustained upturn: it entered the display
range at approximately 30, fluctuated between 11.8 and 25.9 between 10{,}000 and
20{,}000 steps while the learning rate was near its peak, entered a smooth
descent after 25{,}000 steps, and reached 6.2 at 46{,}233 steps (5.6~h of
wall-clock training) (Fig.~\ref{fig:curves}f).

Taken together, the four stages define a clear chain of error propagation, in
which two different metrics must be kept apart.  Round~1 reduces the
window-level coordinate error to 2.96\,\AA{} and calibrates the seven distance
scales; Round~2 removes a further 46\% of that error, reaching 1.59\,\AA{}.  Both
are Kabsch RMSDs of 15-residue windows.  Stage~1 then lifts those features to
the chain index and exports the conditioning representation; the 6.70\,\AA{}
quoted above is the chain-level RMSD of its \emph{auxiliary} coordinate head
after a single epoch, and is therefore neither a converged value nor the output
of the pipeline.  The output of the pipeline is produced by the flow-matching
decoder, which converges on that representation without any sign of
overfitting; its accuracy is the quantity reported in
Section~\ref{sec:res_longchain} and probed against an external reference in
Section~\ref{sec:res_coverage}, and it is not bounded by the first-epoch value
of an auxiliary head. Nor is it bounded below by the window-level errors of
Rounds~1 and~2: those are the errors of intermediate modules supervised on
15-residue windows, whereas the decoder is supervised on chain coordinates and
reads the upstream stages only through the conditioning representation.
Fig.~\ref{fig:recon}a is the one place where the Stage~1 chain-level trace is
displayed directly, as an intermediate of the cascade rather than as the output
of the pipeline, and it is labelled as such.  Every stage converges on its own supervision target and every
inter-stage gain can be attributed segment by segment, which is precisely the
diagnostic advantage that hierarchical decomposition holds over end-to-end
training\cite{ahdritz2024openfold}.

\subsection{Constant-velocity SMD reproduces mechanical transition states
	at residue resolution}
\label{sec:res_phi}

Constant-velocity SMD is widely used to interpret force-spectroscopy
experiments, but residue-level conclusions are usually benchmarked only
against global observables such as rupture force or $\Delta x_{u}$, not
against site-specific data~\cite{rief1997reversible,carrionvazquez1999mechanical,lu2000key,isralewitz2001steered}.
We performed a direct site-by-site comparison, scoring each residue by
$\Phisim$, the fraction of its native contacts surviving in a
$\dpeak \pm 1.0$\,nm window around the first force peak, at each of the
four pulling speeds independently (Methods).

\begin{figure*}[tp]
	\centering
	\includegraphics[width=\textwidth,height=\figmaxh,keepaspectratio]{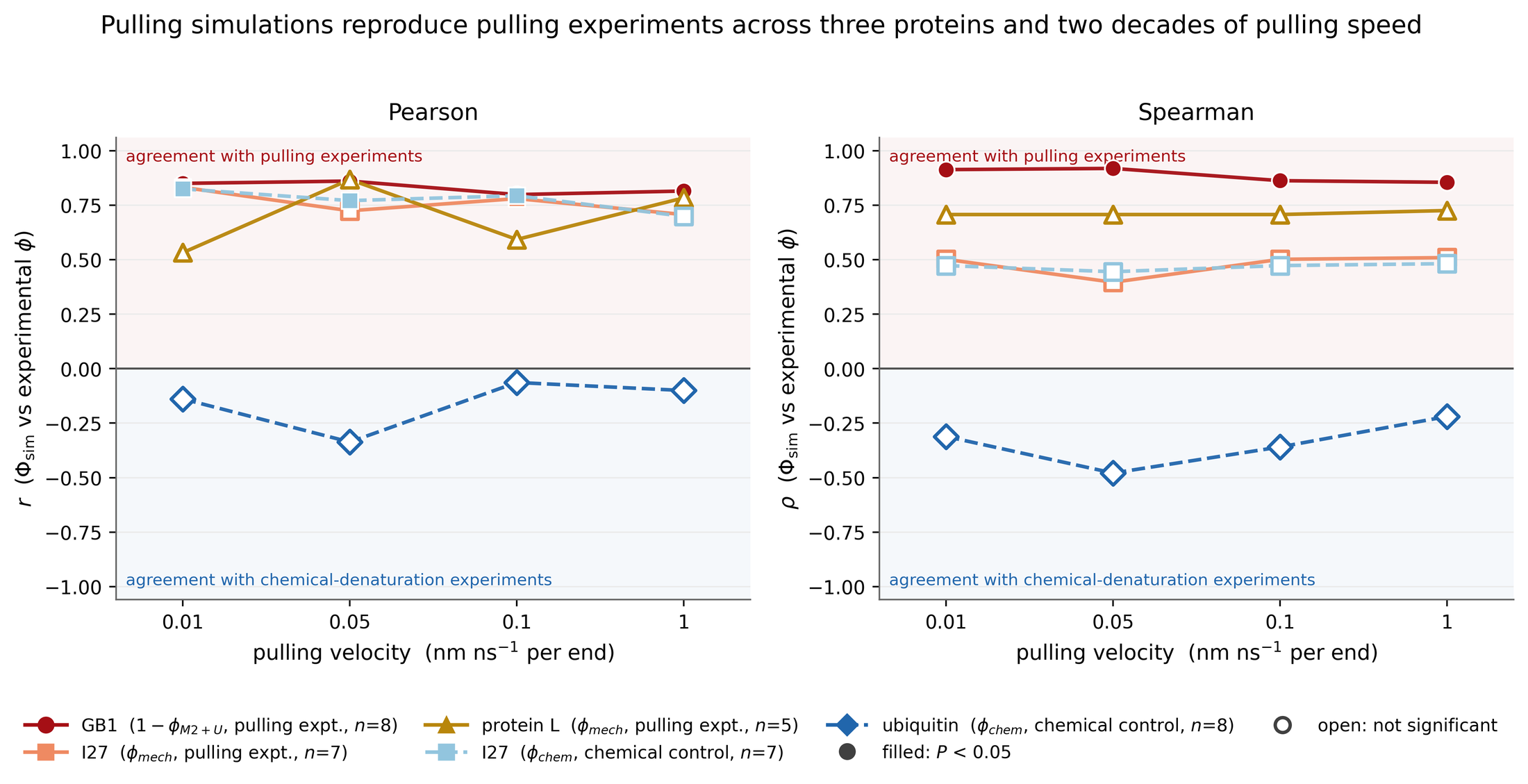}
	\caption{\textbf{Simulation--experiment agreement at residue resolution is
			robust across two decades of pulling speed.}
		Pearson $r$ (left) and Spearman $\rho$ (right) between $\Phisim$ at
		the force peak and experimental $\varphi$-values, plotted against pulling
		velocity (0.01, 0.05, 0.1, 1.0\,nm\,ns$^{-1}$ per end).
		Warm solid lines: mechanical $\varphi$-values (main evidence) ---
		GB1 ($1-\varphi_{\text{M2+U}}$, $n=8$), I27 ($\phimech$, $n=7$),
		protein~L ($\phimech$, $n=5$).
		Cold dashed lines: comparisons with chemical $\varphi$-values ---
		ubiquitin ($\phichem$, $n=8$), which is the specificity control, and
		I27 ($\phichem$, $n=7$), shown for completeness but not a control,
		because its $\phichem$ and $\phimech$ are collinear at the seven
		measured sites ($r = 0.989$).
		Filled symbols, $P<0.05$; open symbols, not significant.
		Pink shading: positive half-plane (agreement with pulling experiments);
		blue shading: negative half-plane (agreement with chemical
		denaturation).
		All 12 mechanical combinations fall above zero; all 4 ubiquitin
		combinations fall below zero.
		Bootstrap 95\% CIs (10,000 replicates) omitted for clarity.}
	\label{fig:phi_correlation}
\end{figure*}

\textbf{All 12 system$\,\times\,$speed combinations are concordant.}
Pearson $r$ and Spearman $\rho$ between $\Phisim$ and experimental
mechanical $\varphi$-values are positive for all three mechanically
characterised proteins across all four speeds
(Table~\ref{tab:phi_correlation}, Fig.~\ref{fig:phi_correlation}).
For GB1 ($n=8$ probes, $1-\varphi_{\text{M2+U}}$), $r$ ranges from
$+0.799$ to $+0.861$ and $\rho$ from $+0.855$ to $+0.919$; all are
significant at both metrics ($P_{r}<0.018$, $P_{\rho}<0.007$);
10,000-replicate bootstrap 95\% CIs exclude zero throughout.
For titin I27 ($\phimech$, $n=7$), mean $r = +0.761$
(+0.707 to +0.832); two of four speeds reach $P < 0.05$.
For protein~L ($\phimech$, $n=5$), mean $r = +0.694$
(+0.532 to +0.868), $\rho = +0.707$ to $+0.725$.

The key inferential weight rests on sign consistency rather than individual
$P$-values: three proteins with different fold topologies (I27,
$\beta$-sandwich immunoglobulin; protein~L and GB1, $\alpha+\beta$ grasp
fold), different pulling geometries, and different $\varphi$-measurement
strategies~\cite{best2002mechanical,best2003titin,sadler2009rheostat,shen2012bihistidine} all fall above zero
across all 12 combinations.

\textbf{Structural interpretation.}
For GB1, the three force-bearing probes (G4-51, G6-53, G8-55; the
$\beta_{1}$--$\beta_{4}$ shear interface) show the lowest $\Phisim$ at the
force peak, and the $\alpha$-helix probes (G24-28, G32-36) show the
highest --- exactly the experimental transition-state picture in which the
shear interface is largely disrupted while the helix remains
intact~\cite{shen2012bihistidine}.
For protein~L, I60V (interface of the two mechanical sub-domains) yields the
lowest $\Phisim$ at all four speeds, in agreement with
experiment: this is the only hydrophobic-core mutation that significantly
alters rupture force~\cite{sadler2009rheostat}.

\textbf{Specificity for mechanical unfolding.}
Ubiquitin is the specificity control. It has only chemical-denaturation
$\varphi$-values~\cite{went2005ubiquitin} and
gives $r = -0.34$ to $-0.06$ and $\rho = -0.48$ to $-0.22$ at all four
speeds --- uniformly negative and opposite to the three mechanical proteins
(Fig.~\ref{fig:phi_correlation}).
Negative correlation with chemical $\varphi$ is the required companion to the
main result, not an anomaly~\cite{carrionvazquez1999mechanical,fowler2002intermediate}.
I27 cannot serve as a specificity control because its $\phimech$ and
$\phichem$ are nearly collinear at the 7 measured sites
($r = 0.989$, $\rho = 0.982$, $P = 2.2\times10^{-5}$).

\textbf{Two alternative explanations are excluded.}
(i)~\textit{SMD follows the folding nucleus.}
Protein~L and GB1 share the same fold topology but have opposite nucleation
sites~\cite{kim2000specific,mccallister2000hairpin}; both are nonetheless positively
correlated with their respective mechanical $\varphi$-values.
(ii)~\textit{Trivial geometry.}
Sequence distance to the nearest pulled terminus gives $\rho = +0.916$ for
GB1 ($P=0.001$) --- numerically similar to $\Phisim$ --- but
partial-correlation analysis shows the information is not equivalent:
controlling for terminus distance, the partial correlation of $\Phisim$ with
experimental $\varphi$ remains $+0.676$ (GB1, $P=0.096$) and $+0.796$
(I27, $P=0.058$), whereas controlling for $\Phisim$ reduces the
terminal-distance partial correlation to $-0.114$ and $-0.395$,
respectively.

\textbf{Scope.}
Agreement is ordinal, not numerical: the span of $\Phisim$ across probe sites
is much smaller than the span of the experimental $\varphi$-values it is
compared with, and regression slopes are well below~1.
Within these limits, constant-velocity SMD trajectories encode mechanical
unfolding information at residue resolution, and this information is specific
to the force-spectroscopy route.

\begin{table*}[tp]
	\centering
	\caption{\textbf{Pearson $r$ and Spearman $\rho$ between $\Phisim$ at
			the force peak and experimental $\varphi$-values at four pulling speeds.}
		Asterisk: $P<0.05$.
		Speed units: nm\,ns$^{-1}$ per end.
		All 12 mechanical-protein combinations are positive; all 4 ubiquitin
		combinations are negative.}
	\label{tab:phi_correlation}
	\small
	\setlength{\tabcolsep}{4pt}
	\fitpage{%
		\renewcommand{\arraystretch}{1.3}
		\begin{tabular}{@{}lllllc@{}}
			\toprule
			System (quantity, $n$)
			& $v=0.01$ & $v=0.05$ & $v=0.1$ & $v=1.0$
			& $d_{\text{peak}}$ range \\
			\midrule
			\multicolumn{6}{l}{\textit{Main evidence: comparison with mechanical
					$\varphi$-values}} \\[2pt]
			Titin I27 ($\phimech$, 7)
			& $r{=}{+}0.83^{*}$, $\rho{=}{+}0.50$
			& $r{=}{+}0.72$,\quad $\rho{=}{+}0.40$
			& $r{=}{+}0.78^{*}$, $\rho{=}{+}0.50$
			& $r{=}{+}0.71$,\quad $\rho{=}{+}0.51$
			& 4.9--5.2\,nm \\
			Protein~L ($\phimech$, 5)
			& $r{=}{+}0.53$,\quad $\rho{=}{+}0.71$
			& $r{=}{+}0.87$,\quad $\rho{=}{+}0.71$
			& $r{=}{+}0.59$,\quad $\rho{=}{+}0.71$
			& $r{=}{+}0.78$,\quad $\rho{=}{+}0.73$
			& 3.7--4.5\,nm \\
			GB1 ($1{-}\varphi_{\text{M2+U}}$, 8)
			& $r{=}{+}0.85^{*}$, $\rho{=}{+}0.91^{*}$
			& $r{=}{+}0.86^{*}$, $\rho{=}{+}0.92^{*}$
			& $r{=}{+}0.80^{*}$, $\rho{=}{+}0.86^{*}$
			& $r{=}{+}0.82^{*}$, $\rho{=}{+}0.86^{*}$
			& 2.8--3.0\,nm \\[2pt]
			\multicolumn{6}{l}{\textit{Specificity control: comparison with chemical
					$\varphi$-values}} \\[2pt]
			Ubiquitin ($\phichem$, 8)
			& $r{=}{-}0.14$,\quad $\rho{=}{-}0.31$
			& $r{=}{-}0.34$,\quad $\rho{=}{-}0.48$
			& $r{=}{-}0.06$,\quad $\rho{=}{-}0.36$
			& $r{=}{-}0.10$,\quad $\rho{=}{-}0.22$
			& 4.6--4.9\,nm \\
			\bottomrule
	\end{tabular}}
\end{table*}

\subsection{Extension-conditioned ensembles are reproducible in their
	continuous observables across two decades of pulling velocity}
\label{sec:res_velocity}

SMD must be performed at velocities far above the experimental range because
all-atom simulations cannot access millisecond timescales~\cite{rico2013highspeed}.
If intermediate conformations changed systematically with speed, the
production-speed ensemble could not represent force-induced unfolding
intermediates.  We tested this directly.

\textbf{Experimental design.}
The four validation proteins --- ubiquitin (1UBQ, 76\,residues), titin I27
(1TIT, 89\,residues), protein~L (2PTL, 61\,residues) and protein~G B1
(1PGB, 56\,residues) --- were each simulated at 0.01, 0.05, 0.1 and
1.0\,nm\,ns$^{-1}$ per end with every other parameter held fixed, and all
cross-speed comparisons were made against end-to-end distance on a common
grid of 40 equal-width bins (Methods).
This design separates ``speed changed when a conformation appeared'' from
``speed changed the conformation itself,'' testing only the latter.

\begin{figure*}[tp]
	\centering
	\includegraphics[width=\textwidth,height=\figmaxh,keepaspectratio]{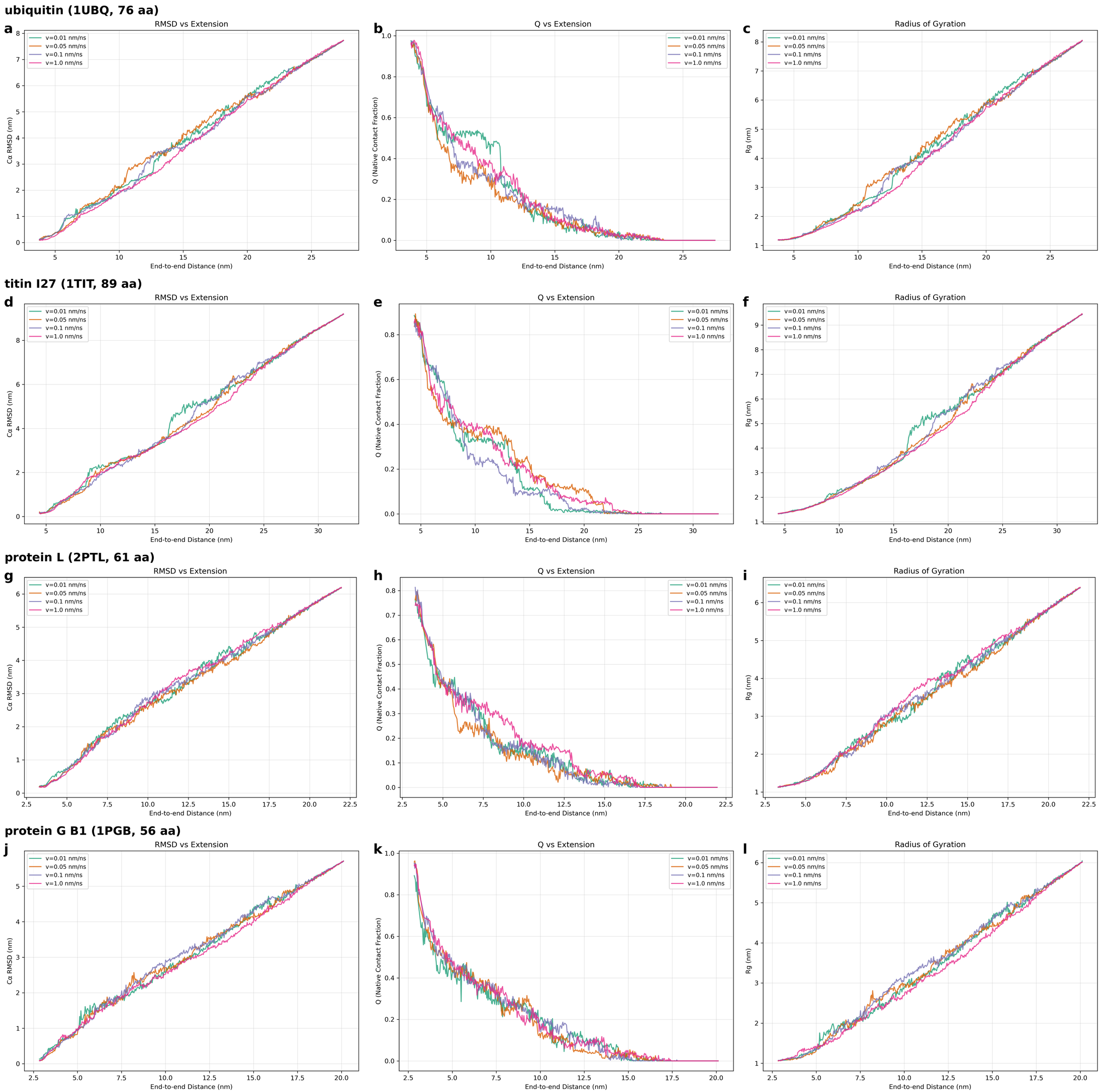}
	\caption{\textbf{Global conformational observables agree across two
			decades of pulling speed when plotted against extension.}
		Each row shows one protein: \textbf{a--c} ubiquitin (1UBQ);
		\textbf{d--f} titin I27 (1TIT); \textbf{g--i} protein~L (2PTL);
		\textbf{j--l} protein~G B1 (1PGB).
		Columns: C$\alpha$-RMSD relative to the native state after Kabsch
		superposition~\cite{kabsch1976solution} (\textbf{a,d,g,j}); native-contact
		fraction $Q$ (\textbf{b,e,h,k}); radius of gyration $\Rg$
		(\textbf{c,f,i,l}).
		Horizontal axis is end-to-end distance; four colours correspond to four
		pulling velocities.}
	\label{fig:global_observables}
\end{figure*}

\textbf{$Q$ and $\Rg$ agree to 2.9--7.0\% of their dynamic range.}
C$\alpha$-RMSD, native-contact fraction $Q$, and $\Rg$ traces overlap across
all four speeds (Fig.~\ref{fig:global_observables},
Table~\ref{tab:rms_diff}).
Summarising all 24 protein\,$\times$\,speed-pair combinations, the RMS
deviation in C$\alpha$-RMSD is 0.199\,nm (range 0.105--0.361); in $\Rg$,
0.196\,nm (0.122--0.351); in $Q$, 0.050 (0.026--0.085).
Normalised by the physical dynamic range, relative deviations are
2.9--3.5\% for $\Rg$ and 4.5--7.0\% for $Q$ --- consistent across four
proteins with different sizes and topologies.
Of 12 observable\,$\times$\,protein combinations, 8 yield positive and 4
negative Spearman correlations between $|\log_{10}(v_{2}/v_{1})|$ and RMS
deviation, with only 1 reaching $P<0.05$ (and in the opposite direction to
the majority).

\begin{figure*}[tp]
	\centering
	\includegraphics[width=\textwidth,height=\figmaxh,keepaspectratio]{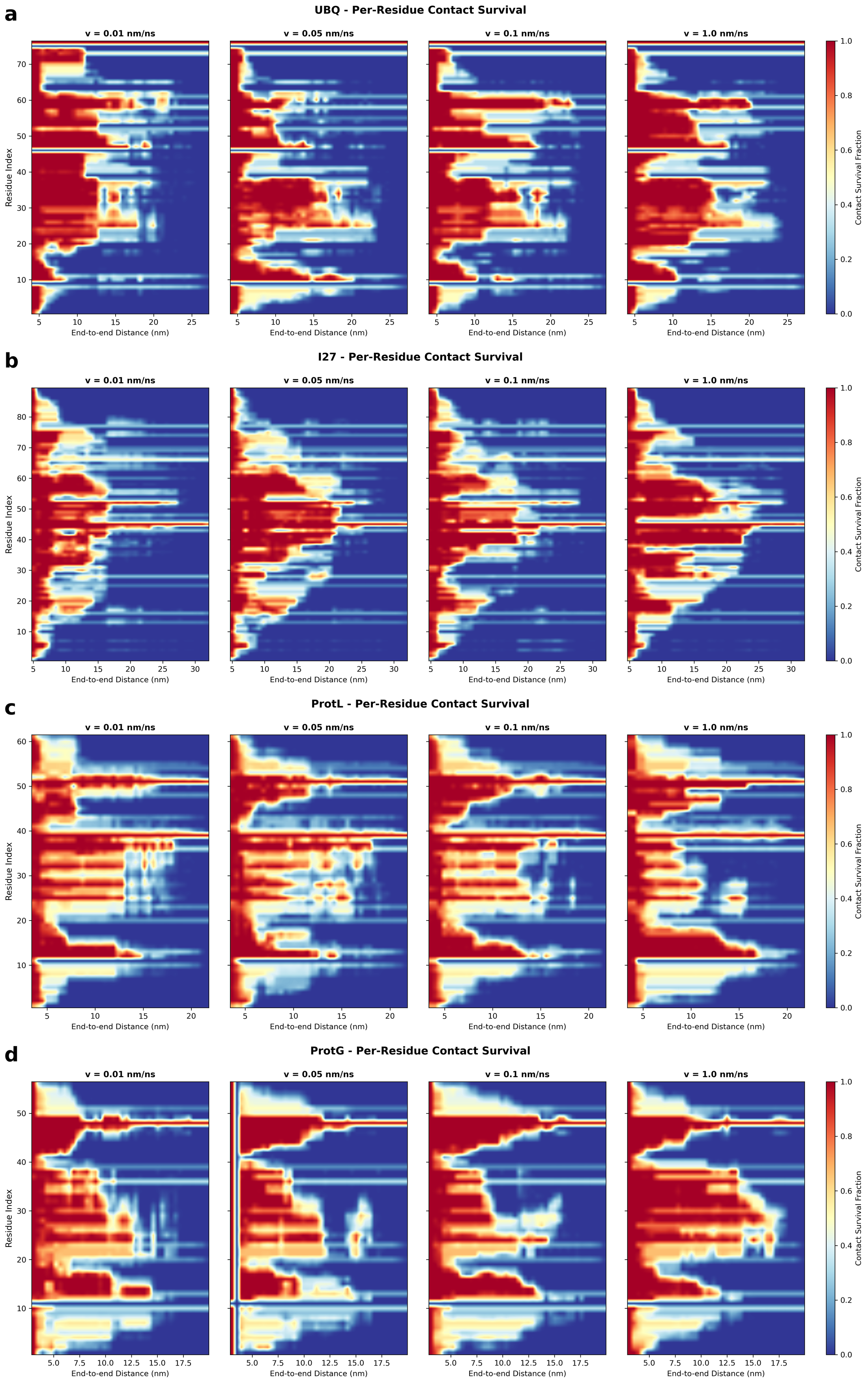}
	\caption{\textbf{Per-residue contact-survival
			maps at four pulling velocities.}
		\textbf{a} ubiquitin; \textbf{b} titin I27; \textbf{c} protein~L;
		\textbf{d} protein~G B1.
		Within each panel, columns correspond to $v = 0.01$, 0.05, 0.1,
		1.0\,nm\,ns$^{-1}$.
		Colour: fraction of native contacts retained
		(C$\alpha$--C$\alpha$ $<8.0$\,\AA, $|i-j|>3$, as in
		equation~(\ref{eq:Q})).
		Segments that disrupt early and those that persist occupy the same
		sequence positions at all speeds; disruption boundaries appear as sharp
		colour steps rather than gradual gradients.}
	\label{fig:contact_survival}
\end{figure*}

\textbf{Per-residue contact-survival patterns are preserved across speeds.}
Heatmaps of per-residue contact survival (Fig.~\ref{fig:contact_survival})
show the same block structure at all four speeds: segments that disrupt
early and segments that persist occupy the same sequence positions regardless
of speed.
Pearson $r$ averaged across the six speed pairs has a per-protein median of
0.809--0.923 (Table~\ref{tab:residue_consistency}); 93--98\% of residues
exceed $r>0.7$; Spearman $\rho$ gives nearly identical conclusions (mean
0.819--0.843).
Low-consistency residues cluster at mechanically critical sites: I27
residues 27, 28, 42, 44, 45, 52 flank the A--B and A$'$--G hydrogen-bond
networks that form the rate-limiting barrier~\cite{lu2000key,marszalek1999mechanical}.

\begin{figure*}[tp]
	\centering
	\includegraphics[width=\textwidth,height=\figmaxh,keepaspectratio]{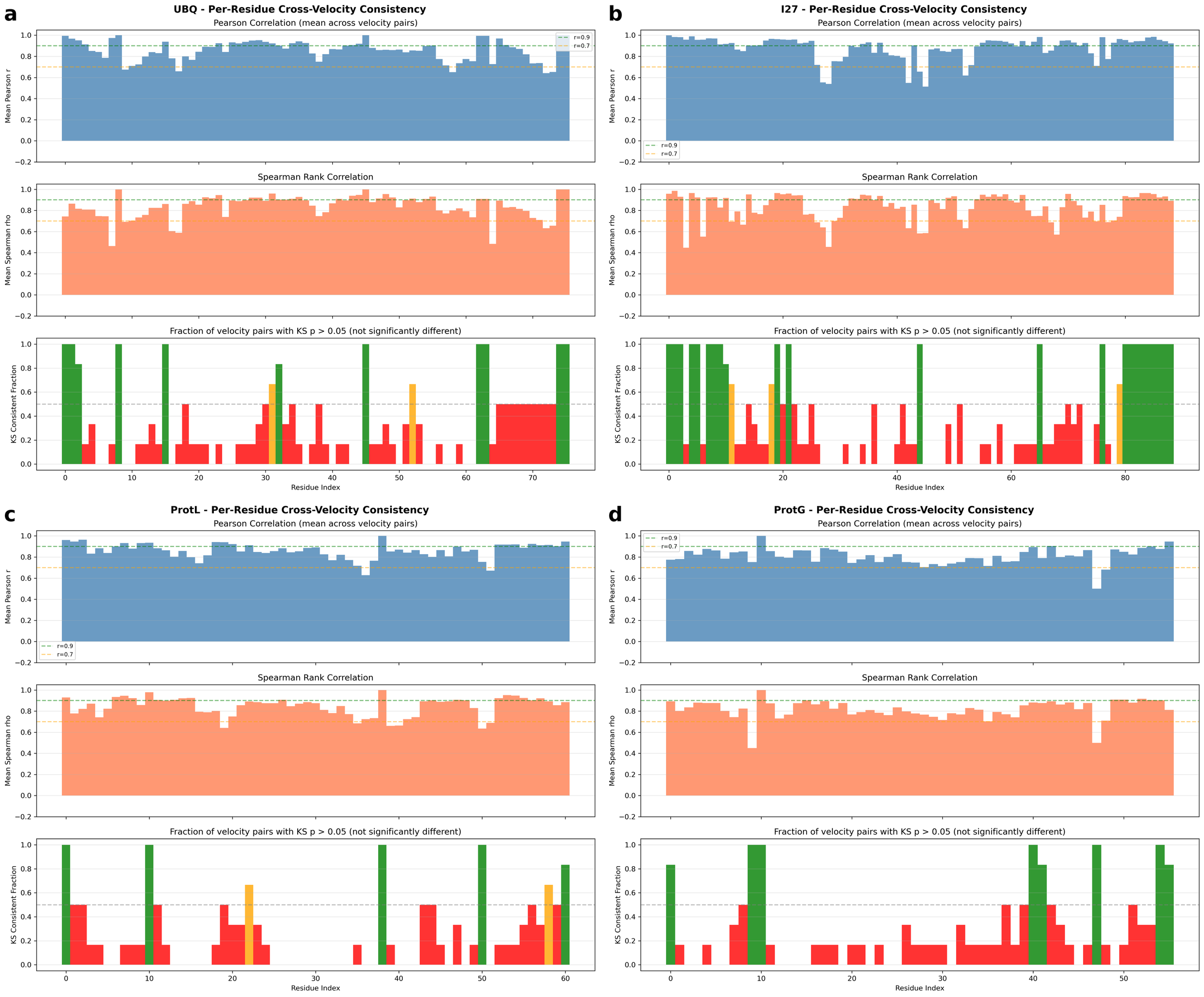}
	\caption{\textbf{Per-residue cross-speed
			consistency: shape is preserved, absolute distributions are not.}
		\textbf{a} ubiquitin; \textbf{b} titin I27; \textbf{c} protein~L;
		\textbf{d} protein~G B1.
		Top sub-panel: mean Pearson $r$ across six speed pairs.
		Middle: Spearman $\rho$.
		Bottom: fraction of speed pairs with KS $p>0.05$.
		Reference lines: $r=0.9$ (green dashed) and $r=0.7$ (orange dashed);
		KS panel 0.5 (grey dashed).
		High $r$/$\rho$ with low KS fraction confirms that contact-survival
		trends are speed-invariant but absolute values at a fixed extension are
		not.}
	\label{fig:residue_consistency}
\end{figure*}

Kolmogorov--Smirnov (KS) tests reveal that only 23--40\% of speed pairs are
indistinguishable in residue-level contact-survival \emph{distributions}
(Table~\ref{tab:residue_consistency}, Fig.~\ref{fig:residue_consistency}).
Speed preserves the \emph{shape} (monotonic decrease of contact survival
with extension) but not the absolute distribution at a given extension.

\begin{figure*}[tp]
	\centering
	\includegraphics[width=0.55\textwidth,height=\figmaxh,keepaspectratio]{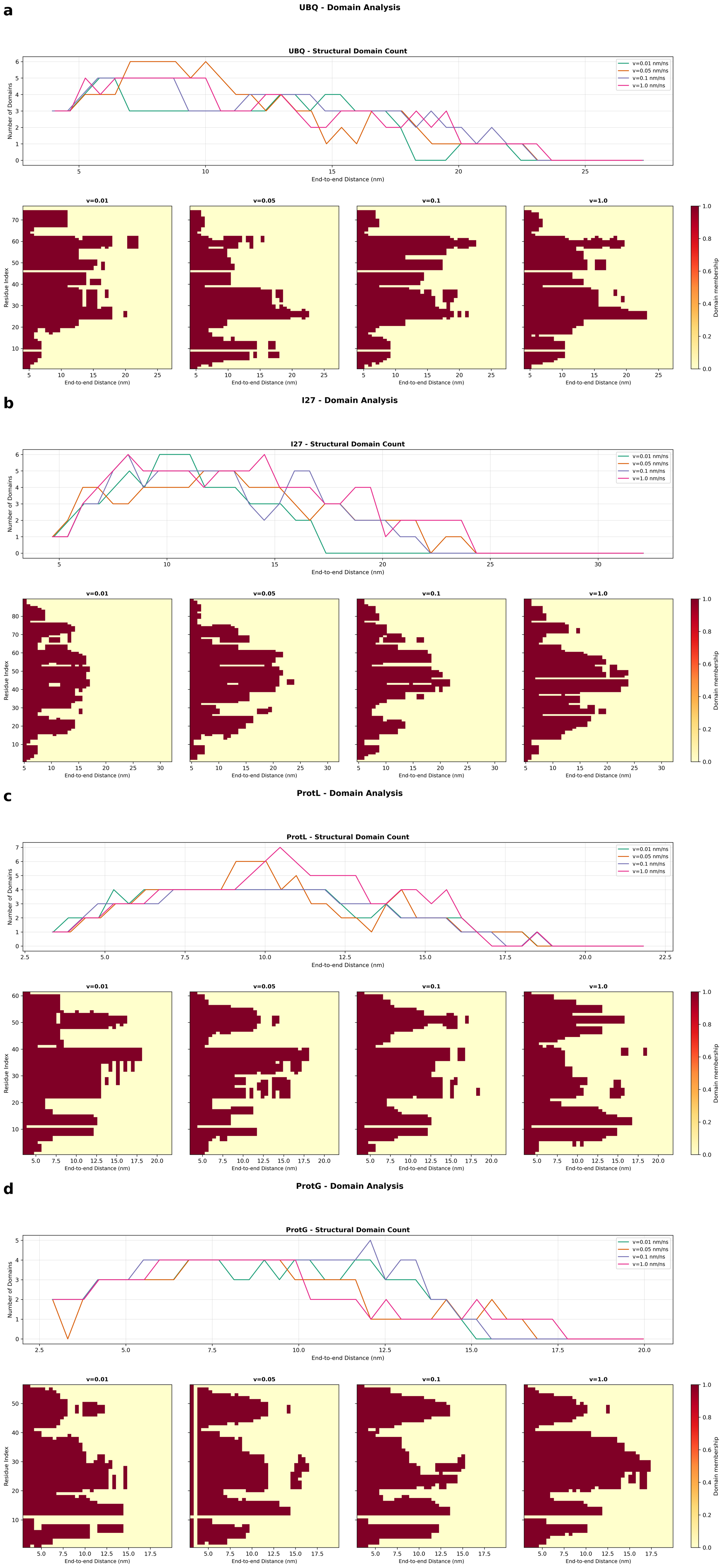}
	\caption{\textbf{Domain-organisation topology is
			speed-invariant; complete disruption is speed-delayed.}
		\textbf{a} ubiquitin; \textbf{b} titin I27; \textbf{c} protein~L;
		\textbf{d} protein~G B1.
		Upper sub-panel: structural domain count versus extension (four speeds
		overlaid).
		Lower sub-panel: per-residue domain-membership profile (dark: residue
		belongs to a sequentially contiguous, locally native domain).
		Domain profiles are significantly correlated across all 24 speed
		combinations ($r = 0.720$--$0.967$, mean 0.845, max
		$P = 3.0\times10^{-33}$).
		The extension at which the domain count first reaches zero shifts
		systematically to larger values at higher speed (most pronounced for
		I27, panel \textbf{b} upper).}
	\label{fig:domain_organization}
\end{figure*}

\textbf{Domain-organisation topology is conserved; disruption position shifts.}
Per-residue domain-assignment profiles are significantly correlated across
all 24 combinations ($r = 0.720$--$0.967$, mean 0.845, max
$P = 3.0\times10^{-33}$; Fig.~\ref{fig:domain_organization}).
Table~\ref{tab:disruption} reports a different quantity, the extension at
which the domain count first reaches zero.
A directed speed effect is nonetheless present: faster pulling displaces
the extension at which the chain loses all domain structure by a mean of
2.83\,nm (Spearman correlation with $\log v$: $+0.80$ for all four
proteins; with $n=4$ the exact two-sided permutation $P$ is 0.33 per protein,
but the direction is unanimous).
The effect is largest for titin I27 (16.67\,nm at 0.01 to 23.66\,nm at
1.0\,nm\,ns$^{-1}$; 50--71\% of $\Lc$) and negligible for protein~L
(18.03 to 18.50\,nm), consistent with their known mechanical
stabilities~\cite{lu2000key,best2003titin,brockwell2005pulling}.

\textbf{Implications for parameter choice.}
The quantities on which a conditional generative model is conditioned ---
continuous geometric and contact observables at a given extension --- are
reproducible across speeds here; the absolute residue-level contact-survival
distributions at a given extension are not, and are used neither as
conditioning variables nor as training targets.
The extension at which structures disrupt, which does shift with speed, is
not a model conditioning variable.
Trajectories at different speeds may therefore be pooled in a single training
set provided the training targets are restricted to continuous geometry and
contact observables at fixed reaction coordinate.

\begin{table}[!tb]
	\centering
	\caption{\textbf{RMS deviations between speed pairs (mean over six pairs;
			brackets: min--max).}}
	\label{tab:rms_diff}
	\small
	\setlength{\tabcolsep}{3pt}
	\renewcommand{\arraystretch}{1.2}
	\fitcolumn{\begin{tabular}{@{}lcccc@{}}
		\toprule
		Observable & Ubiquitin & I27 & Prot.~L & Prot.~G \\
		\midrule
		C$\alpha$-RMSD (nm)
		& 0.241 & 0.259 & 0.144 & 0.152 \\
		& [0.183--0.337] & [0.178--0.361] & [0.105--0.167] & [0.127--0.198] \\[2pt]
		$\Rg$ (nm)
		& 0.223 & 0.243 & 0.152 & 0.164 \\
		& [0.177--0.314] & [0.167--0.351] & [0.122--0.193] & [0.123--0.233] \\[2pt]
		$Q$
		& 0.060 & 0.058 & 0.044 & 0.038 \\
		& [0.042--0.085] & [0.039--0.082] & [0.026--0.059] & [0.031--0.044] \\[2pt]
		Norm.\ $\Delta\Rg$    & 3.2\% & 2.9\% & 2.9\% & 3.5\% \\
		Norm.\ $\Delta Q$     & 7.0\% & 6.8\% & 5.2\% & 4.5\% \\
		\bottomrule
	\end{tabular}}
\end{table}

\begin{table}[!tb]
	\centering
	\caption{\textbf{Per-residue cross-speed consistency
			(degenerate residues excluded).}
		$N$: residues retained.
		KS frac.: fraction of speed pairs with $p>0.05$.}
	\label{tab:residue_consistency}
	\small
	\setlength{\tabcolsep}{3pt}
	\renewcommand{\arraystretch}{1.2}
	\fitcolumn{\begin{tabular}{@{}lccccc@{}}
		\toprule
		Protein & $N$ & Med.\ $r$ & $r{>}0.7$ & Mean $\rho$ & KS frac. \\
		\midrule
		Ubiquitin  & 72 & 0.863 & 0.93 & 0.826 & 0.29 \\
		Titin I27  & 89 & 0.923 & 0.93 & 0.833 & 0.40 \\
		Protein~L  & 60 & 0.865 & 0.97 & 0.843 & 0.23 \\
		Protein~G  & 54 & 0.809 & 0.98 & 0.819 & 0.25 \\
		\bottomrule
	\end{tabular}}
\end{table}

\begin{table}[!tb]
	\centering
	\caption{\textbf{Extension at which all domain structure is irreversibly
			lost.}
		$\Lc = (N{-}1)\times0.38$\,nm: 28.50, 33.44, 22.80, 20.90\,nm.
		$\rho(\log v)$ computed from unrounded values. With $n=4$ the exact
		two-sided permutation $P$ is 0.33 per protein; it is reported for
		completeness, and the inference rests on the unanimous sign across the
		four proteins rather than on any single $P$ value.}
	\label{tab:disruption}
	\small
	\setlength{\tabcolsep}{3pt}
	\renewcommand{\arraystretch}{1.2}
	\fitcolumn{\begin{tabular}{@{}lcccccl@{}}
		\toprule
		Protein & 0.01 & 0.05 & 0.1 & 1.0 & $\rho$ & $\Lc$ frac. \\
		\midrule
		Ubiquitin & 21.87 & 22.53 & 22.51 & 23.08 & $+0.80$ & 0.77--0.81 \\
		I27       & 16.67 & 23.64 & 21.52 & 23.66 & $+0.80$ & 0.50--0.71 \\
		Protein~L & 18.03 & 18.03 & 18.48 & 18.50 & $+0.80$ & 0.79--0.81 \\
		Protein~G & 14.70 & 16.47 & 15.14 & 17.34 & $+0.80$ & 0.70--0.83 \\
		\bottomrule
	\end{tabular}}
\end{table}

\subsection{Force-induced unfolding proceeds in fragments: binary
	partition stability and a linear extension law}
\label{sec:res_segment}

Sections~\ref{sec:res_phi} and~\ref{sec:res_velocity} establish that the
trajectories are valid and the ensemble is speed-robust.
This section characterises the internal geometry of the intermediates to
determine whether a fragment-level modelling strategy is geometrically
coherent.

\textbf{Binary partition is stable across 500 proteins.}
Applied at fixed window width and threshold to all trajectory frames of
AF-Geo500 --- 500 proteins across 11 species --- the folded/unfolded
partition operator (Methods) ran without per-protein tuning and returned
557,550 contiguous unfolded segments of length $\geq 3$ with well-defined
boundaries.
The stability of the ``contiguous unfolded segment'' object under fixed
hyperparameters is itself a check on the binary abstraction: a broad
diffuse transition zone would prevent well-defined segments from forming.
This picture is consistent with the block structure visible in
Figs.~\ref{fig:contact_survival} and~\ref{fig:domain_organization}, in
which disruption boundaries appear as sharp colour steps rather than
gradual gradients.
The binary residue abstraction is a modelling assumption in WSME-class
statistical-mechanical
models~\cite{wako1978statistical1,wako1978statistical2,munoz1999simple,bruscolini2002exact,imparato2007ising}
and in the hydrogen-exchange foldon
picture~\cite{maity2005protein,englander2014nature}; the present measurement places it
on an all-atom footing.

\begin{figure*}[tp]
	\centering
	\includegraphics[width=\textwidth,height=\figmaxh,keepaspectratio]{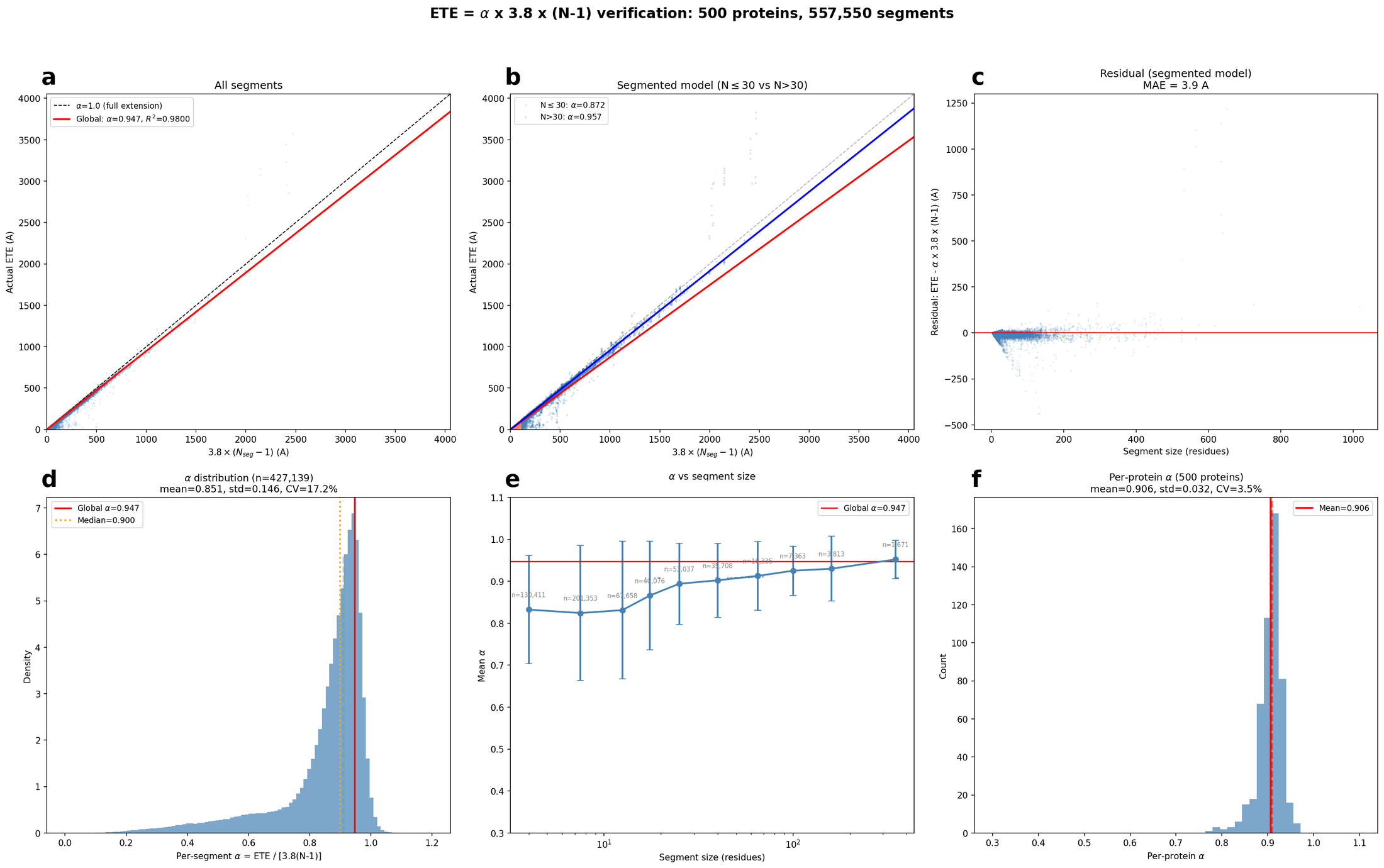}
	\caption{\textbf{The linear extension law
			$\ETE = \alpha \times 3.8\,(\Nseg{-}1)$ holds across 11 species,
			500 proteins, and 557,550 unfolded segments.}
		\textbf{a,} ETE versus $3.8\,(\Nseg{-}1)$ for all segments
		($10^{5}$ randomly drawn for display).
		Black dashed line: $\alpha = 1$ (full extension);
		red line: origin-constrained fit ($\alpha_{\text{global}}=0.947$,
		$R^{2}=0.980$).
		\textbf{b,} Two-piece fit: $\Nseg\leq30$ (coral,
		$\alpha=0.872$) and $\Nseg>30$ (steel-blue, $\alpha=0.957$).
		\textbf{c,} Residuals of the two-piece model; MAE\,$=3.9$\,\AA.
		\textbf{d,} Per-segment $\alpha_{k}$ distribution
		($n=427{,}139$; $\Nseg\geq5$, $\Lmax>1$\,\AA): mean 0.851,
		SD 0.146, CV 17.2\%, median 0.900 (orange dotted).
		\textbf{e,} $\alpha$ versus segment length (log-spaced bins);
		error bars: $\pm$1\,SD; grey numerals: per-bin counts.
		\textbf{f,} Per-protein $\alpha$ across 500 proteins:
		mean 0.906, SD 0.032, CV 3.5\%.}
	\label{fig:ete_scaling}
\end{figure*}

\textbf{End-to-end distance of unfolded segments obeys a linear extension
	law.}
Segment end-to-end distance was compared with the contour-length upper
bound $\Lmax = 3.8\,(\Nseg - 1)$\,\AA\ through the linear relation of
equation~(\ref{eq:ete}) (Section~\ref{sec:meth_segment}).
Origin-constrained least squares over all 557,550 segments gives
$\alpha_{\text{global}} = 0.947$ ($R^{2} = 0.980$;
Fig.~\ref{fig:ete_scaling}a).
The relationship holds from 3 to over 1,000 residues, spanning nearly three
decades in $\Lmax$.
A two-piece fit (Fig.~\ref{fig:ete_scaling}b) gives
$\alpha_{\text{short}} = 0.872$ ($\Nseg \leq 30$) and
$\alpha_{\text{long}} = 0.957$ ($\Nseg > 30$); residual MAE is 3.9\,\AA,
comparable to one virtual bond length
(Fig.~\ref{fig:ete_scaling}c).

\textbf{The scaling coefficient has a clear hierarchical structure.}
Three statistical levels must be distinguished
(Table~\ref{tab:alpha}, Fig.~\ref{fig:ete_scaling}d--f).
The global fit ($\alpha = 0.947$) is dominated by long segments (weight
$\propto \Lmax^{2}$).
The per-segment mean is 0.851 with CV 17.2\%, left-skewed (median 0.900).
The per-protein mean is 0.906 with CV only 3.5\%; the distribution is
unimodal and shows no systematic separation across 11 species spanning
prokaryotes (\textit{E.~coli}, \textit{B.~subtilis}), fungi
(\textit{C.~albicans}), plants (\textit{A.~thaliana}), invertebrates
(\textit{C.~elegans}, \textit{D.~melanogaster}, \textit{A.~gambiae},
\textit{A.~mellifera}), and vertebrates (\textit{D.~rerio},
\textit{B.~taurus}, \textit{C.~lupus}).

The 17.2\% per-segment CV is not noise: $\alpha$ rises systematically with
segment length, from $\approx 0.84$ across the shortest bins to $\approx 0.94$
for long segments, with bin-wise scatter narrowing as length increases
(Fig.~\ref{fig:ete_scaling}e).
Shorter segments retain more residual curvature immediately after leaving the
folded core; longer segments approach the full-extension limit.

\textbf{Universality is a protein-average, not a per-fragment, property.}
The two dispersion levels therefore license different uses of $\alpha$.
The 3.5\% coefficient of variation across 500 proteins and 11 species means
that a single value of $\alpha$ transfers between proteins and across the
sampled phylogenetic range.
The 17.2\% per-segment coefficient of variation, with its systematic
dependence on segment length, means that the same value does not transfer
between individual segments: treating $\alpha$ as a fixed scalar rather
than a length-dependent quantity introduces a $\sim$15\% systematic bias at
the level of a single segment.
Taken together with the residual MAE of 3.9\,\AA\ --- one virtual bond
length --- these measurements bound how far a fragment's end-to-end
geometry can be treated as determined by its own length alone; the
consequences for model design are taken up in the Discussion.

\begin{table}[!tb]
	\centering
	\caption{\textbf{Hierarchical statistics of the scaling coefficient
			$\alpha$ (500 proteins, 11 species).}}
	\label{tab:alpha}
	\small
	\setlength{\tabcolsep}{3pt}
	\renewcommand{\arraystretch}{1.2}
	\fitcolumn{\begin{tabular}{@{}lrcccc@{}}
		\toprule
		Level & $n$ & Mean & Median & SD & CV \\
		\midrule
		Per-segment & 427,139 & 0.851 & 0.900 & 0.146 & 17.2\% \\
		Per-protein & 500     & 0.906 & ---   & 0.032 &  3.5\% \\
		Global fit  & ---     & 0.947 & ---   & ---   &  ---   \\
		\bottomrule
	\end{tabular}}
\end{table}


\section{Discussion}

FoldScape is a sequence-conditioned generative model of protein conformation
whose defining claim is not accuracy at the native state but accuracy at every
point along the folding coordinate.  Three questions follow from that claim, and
we take them in order: whether the predictions are in fact uniformly accurate
across $Q$, and how such predictions should be scored; whether the contribution
of each stage of the cascade can be isolated, attributed and diagnosed; and
whether the training corpus that the model consumes, together with the
architecture built to consume it, rest on physically defensible foundations.

\subsection{Predictive accuracy is uniform along the folding coordinate}
The central point of Fig.~\ref{fig:recon} is that predictive accuracy is flat
along the folding coordinate, while the conventional global RMSD metric breaks
down of its own accord at the low-$Q$ end.  Across the evaluated partially folded and compact states, the RMSD of the
706-residue chain remains between 0.67 and 1.76\,\AA{} after excluding
the fully extended outlier, with no monotonic dependence on $Q$. These
reconstructions demonstrate accurate predictions at multiple folding stages,
consistent with FoldScape's sequence-conditioned generation throughout the
folding continuum.  This is a prerequisite for
extending a native-state predictor into a description of the folding process,
which requires a trustworthy ensemble across the coordinate,
not accuracy at $Q = 1$ alone\cite{bryngelson1995funnels,onuchic1997theory,dill2012folding}.  It
is this requirement that separates the present objective from the
single-structure task on which current predictors are
optimised\cite{jumper2021alphafold,lane2023frontier}, and that it shares with generative
approaches to equilibrium ensembles and Markov-state descriptions of folding
kinetics\cite{noe2019boltzmann,chodera2014msm,zheng2024distributional,lewis2025bioemu,waymentsteele2024conformations}.

The same trajectory also carries a transfer in chain length, since it is
reconstructed for a chain more than threefold longer than a single
chain-level forward pass accepts. That transfer matters beyond the raw
benchmark comparison\cite{lindorfflarsen2011fastfolding}: it shows that reconstructing
three-dimensional coordinates from a hierarchy of local one-dimensional
distances generalises well past the regime the model was built for, and that
long-range inter-domain orientation emerges from the accumulation of local
geometry along the chain rather than from direct prediction of long-range
contacts --- unlike predictors that make the contact map or distogram an
explicit object of inference\cite{senior2020alphafold,jumper2021alphafold}, and consonant with
fragment-based assembly, where global architecture is built from locally
determined geometry\cite{simons1997fragments}.  The independent collapse and later
docking of the two domains in b(iii)--b(iv) is a direct manifestation of this
mechanism\cite{han2007multidomain,borgia2011misfolding}.

\subsection{Global RMSD is ill-posed for extended conformations}
This mechanism's cost is exposed by b(i).  A fully extended chain is nearly
one-dimensional: rotation about its long axis is degenerate under optimal
superposition, so small per-window orientational errors accumulate linearly
along the contour and are amplified, over a chain of nearly 2{,}700\,\AA{},
into a global RMSD of 125.8\,\AA{}.  The predicted and reference traces
nevertheless follow the same local path, with the deviation appearing as a
slow bend of the whole chain rather than a breakdown of local geometry, and a
lateral offset of only a few per cent of the contour length.  The 125.8\,\AA{}
value therefore reflects an ill-posed metric for this conformational class,
not model failure at low $Q$ --- the extended-chain limit of a
long-recognised property of RMSD, which is dominated by the largest
displacements and scales with the size of the object compared, so that its
value is interpretable only within a restricted class of compact
structures\cite{maiorov1994significance,reva1998probability,carugo2001normalized}.

We therefore recommend scoring unfolded and early-collapse states with
superposition- or scale-free measures --- lDDT\cite{mariani2013lddt}, per-window
local RMSD, distance-matrix error, contact-area
scores\cite{olechnovic2013cadscore}, and the relative deviation of the radius of
gyration\cite{kohn2004random,riback2017scattering} --- and restricting global RMSD to
conformations that have already formed a compact core, where
superposition-based measures such as TM-score and GDT remain well
behaved\cite{zhang2004tmscore,zemla2003lga}.  This distinction matters for the central
claim of this work: a substantial part of the folding process takes place on the
expanded, unfolded side of the coordinate\cite{bryngelson1995funnels,onuchic1997theory},
where a single global RMSD criterion would yield misleading conclusions
precisely where confidence is most needed.

\subsection{Coverage is a second axis, independent of accuracy}
\label{sec:disc_coverage}
Flat accuracy along $Q$ is necessary for a generator of the folding process but
not sufficient: a method could be uniformly accurate on the conformations it emits
and still emit only conformations from a narrow region of the coordinate.
Section~\ref{sec:res_coverage} therefore measures the second axis directly.  At
matched sample size the median coverage of an AlphaFold~3 baseline ensemble is
one or two bins in twenty, whereas FoldScape occupies nearly all of the
radius-of-gyration axis and roughly three-quarters of the $Q$ axis, with the
same sign on every one of the twenty proteins and all three order parameters.
The two quantities are logically independent --- coverage says nothing about
whether a given frame is correct --- so the argument for FoldScape rests on
holding them jointly, not on either alone. They are, however, established on
different material: accuracy on the trajectory reconstructions of
Sections~\ref{sec:res_recon} and~\ref{sec:res_longchain}, coverage on the
twenty-protein panel of Section~\ref{sec:res_coverage}. No single target in
this work carries both measurements, because the coverage panel has no
reference folding trajectory against which per-conformation accuracy could be
scored; demonstrating the two properties simultaneously on one protein
requires a target for which both an explicit trajectory and an external
ensemble are available, and is the natural next experiment.

The comparison should be read for what it is.  The AF3 baseline varies only the
random seed, and its authors state that this does not approximate the solution
ensemble\cite{abramson2024alphafold3}; strategies that perturb the alignment
recover considerably more conformational
diversity\cite{waymentsteele2024conformations,delalamo2022sampling}, and
generative ensemble methods target that diversity
explicitly\cite{lewis2025bioemu,jing2024alphaflow,zheng2024distributional}.  The
narrow AF3 band is thus the expected behaviour of a native-state predictor, not
a defect.  The measurement demonstrates FoldScape's substantially broader conformational
coverage across the radius of gyration, native-contact fraction $Q$ and
secondary-structure fraction. At matched sample size, the FoldScape ensembles
span wider regions of all three order parameters across every evaluated protein,
whereas the seed-varied AF3 baseline concentrates near native structures.
This broader sampling opens access to conformations across diverse folding
stages, including unfolded, intermediate and compact structural states.

\subsection{Attributability, systematic trade-offs and the current bottleneck}
Decomposing three-dimensional coordinate reconstruction into a series of
one-dimensional distance regressions\cite{senior2020alphafold,yang2020trrosetta} rests on the
claim that each geometric degree of freedom can be solved, and diagnosed, on its
own.  The curves of Section~\ref{sec:res_training} provide direct evidence for
that claim: all four stages converge on their respective supervision targets,
and the gain contributed by each can be attributed segment by segment.  The
evidence from Round~2 is particularly clean.  Because its modulation projections
are zero-initialised\cite{perez2018film,peebles2023dit}, the network reproduces
Round~1 to within numerical noise at initialisation, so the 46\% reduction in error
comes entirely from the residual branch and is not confounded with the effect of
simply retraining a larger model.  Such attributability is difficult to obtain
within an end-to-end framework, where the learning process must instead be
reconstructed post hoc from intermediate structures\cite{ahdritz2024openfold}, and it is
the principal methodological return of the hierarchical design.

The most consequential systematic bias is the separation between global
alignment accuracy and local geometric regularity.  The rise in bond-length
deviation appears on the training and validation sides simultaneously and
therefore cannot be removed by early stopping or by conventional
regularisation; it reflects an implicit trade between whole-window
superposition and individual bond lengths that is built into the equally
weighted multi-scale loss.  The trade creates a real checkpoint dilemma: a
checkpoint at epoch 10--15 keeps the bond-length deviation at 0.18--0.21\,\AA{}
but leaves the window RMSD above 3.9\,\AA{}, whereas the final checkpoint buys
roughly 1\,\AA{} of RMSD at the cost of a 0.30\,\AA{} bond-length deviation.  The
auxiliary geometry branch of Round~2 and the physical penalties of the
flow-matching objective compensate for part of this cost
downstream\cite{ingraham2023chroma,majewski2023cgpotentials}, but the more fundamental remedy is
to resolve it inside Round~1 --- for example through a curriculum weight on the
nearest-neighbour scale that strengthens as training
proceeds\cite{bengio2009curriculum}, or through an explicit bond-length constraint, so
that both metrics can reach acceptable values at the same checkpoint.  We had
previously tested a static up-weighting of the short-range scales; it marginally
improved short-range fits while substantially degrading the
5-to-7-separation distances and the final trace, which suggests that any
reweighting must be dynamic rather than static.

Two observations from Round~2 point to the same conclusion: at the current
dataset scale, the marginal return on additional optimisation budget is already
low.  The first is the 0.224\,\AA{} training--validation gap at the final epoch;
the second is that the learning-rate restart bought only 0.014\,\AA{} of
validation improvement over 30 epochs while the training error continued to
fall.  This combination is the signature of a model beginning to absorb
dataset-specific detail, the regime in which added capacity or schedule length
yields far less than added data diversity\cite{ahdritz2024openfold}.  The next increment
in performance is therefore more likely to come from broadening trajectory
coverage --- for instance by moving the external molecular-dynamics resources
currently reserved for cross-validation into
training\cite{lindorfflarsen2011fastfolding,shaw2010atomic,hollingsworth2018md} --- than from
longer training or a larger model; recent ensemble emulators that absorb
hundreds of milliseconds of aggregate simulation make the same case at much
larger scale\cite{lewis2025bioemu}.

The gap between the converged window-level error of 1.59\,\AA{} and the
first-epoch chain-level error of 6.70\,\AA{} at the auxiliary Stage~1 head bounds
rather than measures the cost of assembly, but it points in a definite
direction: assembly, not local geometry prediction, is where the remaining
accuracy is to be found.  One structural reason is that cross-window pooling can
supply features only for residue pairs separated by less than the window length;
pairs at separations of at least the window length are initialised to zero in
the chain-level pair tensor, so long-range contacts must be recovered by the
chain-level trunk on its own from the sequence embeddings and the consensus
trace.  Enlarging the window, or supplying an independent prior for long-range
pairs --- the role played by triangle-based pair updates and by evolutionary or
language-model couplings in full-chain
predictors\cite{jumper2021alphafold,baek2021rosettafold,krishna2024rfaa} --- is the most direct route to
narrowing the window-to-chain gap.  It should be noted that the Stage~1
coordinate head is auxiliary supervision only, and its chain-level RMSD is not
equivalent to the structural accuracy of the complete pipeline, which is
determined by the sampling quality of the flow-matching decoder on the
conditioning representation.

The decoder curves expose a practical issue.  Even with the time weight clamped
at $10^{4}$, flow times sampled near the end of the trajectory still produce
one-to-two orders of magnitude of batch-loss fluctuation.  Such fluctuation does
not alter the downward trend of the smoothed curve, but it markedly increases
how often gradient clipping is triggered, and its net effect is an implicit
down-weighting of the high-flow-time regime --- precisely the regime in which
coordinate detail is finalised.  If fuller supervision is wanted there,
stratified time sampling or a bounded velocity parameterisation would be
preferable to numerical clamping alone; both the choice of loss weighting and
the choice of time-sampling density are known to dominate the training
behaviour of diffusion- and flow-based
generators\cite{karras2022edm,esser2024sd3}.

The cascade structure itself facilitates further improvement: each stage can be
replaced and evaluated independently while the upstream stages remain frozen, so
that changes targeting local geometry, long-range assembly and conditional
generation can be pursued independently of one another.  Once predictive
accuracy improves further, the same conditioning representation can serve
design-oriented
structure generation conditioned on the folding
fraction\cite{watson2023rfdiffusion,ingraham2023chroma}, and can be compared directly against
generative models of equilibrium
ensembles\cite{noe2019boltzmann,zheng2024distributional,jing2024alphaflow,lewis2025bioemu}.

\subsection{Prerequisites on the training corpus and the architecture}
The generative model described above was trained on steered molecular dynamics
(SMD) trajectories and is intended to predict residue-level conformations of
mechanical unfolding intermediates conditioned on the end-to-end elongation.
Realising that objective requires satisfying three independent methodological
conditions.  First, the SMD trajectories must sample the \emph{mechanical}
unfolding pathway rather than the time-reversal of the chemical folding pathway,
so that the learned representations carry force-relevant structural
information.  Second, the observables on which the ensemble conditioned on
elongation is trained must be reproducible across pulling speeds, so that
trajectories generated at a computationally tractable speed can stand in for
those that would be obtained at slower, more experiment-proximal rates; where
that reproducibility fails, the affected quantities must be excluded from the
training targets.  Third, the architecture
of the generative model must reflect the intrinsic physical organisation of
unfolding intermediates.  We address each condition in turn.

\subsection{SMD trajectories sample the mechanical unfolding pathway}
A central concern in using SMD data as a surrogate for experiment is whether
the computed residue-level unfolding order corresponds to the mechanical
transition state rather than the chemical folding nucleus.  Mechanical
unfolding proceeds along a specific shear coordinate defined by the
end-to-end pulling geometry, rupturing load-bearing hydrogen-bond networks
first; chemical unfolding is driven by bulk solvation and exposes the folding
nucleus last~\cite{matouschek1989mapping,fersht1995single,brockwell2003pulling}.  A simulation that correctly
captures the mechanical pathway should therefore correlate with mechanical
$\varphi$ values and be \emph{un}correlated with, or negatively correlated
with, chemical $\varphi$ values --- the two are not simply the reverse of each
other~\cite{carrionvazquez1999mechanical}.

We tested this prediction by comparing, for four pulling speeds, the
per-residue native-contact survival fraction at the first force peak
($\Phisim$) against independently measured mechanical and chemical
$\varphi$ values.  For GB1, $\Phisim$ correlated positively with
the mechanical $\varphi$ values $\varphi_{M2+U}$ measured via engineered
metal-chelation sites under AFM pulling conditions~\cite{shen2012bihistidine,cao2008chelation} across
all four speeds (Pearson $r = +0.80$--$+0.86$, $P < 0.02$; Spearman
$\rho = +0.86$--$+0.92$, $P < 0.007$).  The structural correspondence is
precise: the simulation identifies the $\beta_1$--$\beta_4$ shear interface as
the most disrupted element at the force peak and the $\alpha$-helix and first
$\beta$-hairpin as the most intact --- exactly the topology resolved by the
force-extension experiments.  By contrast, the simulation-to-chemical-$\varphi$
comparison for ubiquitin gave near-zero or negative correlations at all speeds
($r = -0.34$ to $-0.06$), serving as the requisite negative control: a method
that sampled chemical folding intermediates would produce the opposite
result~\cite{khorasanizadeh1993folding}.  That the direction of the correlation flips
across the two comparison sets, and does so consistently over all twelve
mechanical protein--speed combinations and all four chemical controls, is not a
statistical artefact but the expected signature of force-specific pathway
sampling~\cite{isralewitz2001steered}. We note that the four speeds of a given
protein share one force field, one starting structure and one probe set, so
they are not four independent replicates; the inferential weight rests on the
three mechanically characterised proteins, which have different folds and
different $\varphi$-measurement strategies.  A
partial-correlation analysis further confirms that $\Phisim$ carries
structural information beyond the geometric confound of endpoint
proximity~\cite{lu2000key}: after partialling out the sequence distance to the
nearest pulled terminus, the residual correlation between $\Phisim$
and the experimental $\varphi_{M2+U}$ for GB1 remains at $+0.68$ (df$\,=5$,
$P = 0.096$), whereas the endpoint-distance predictor loses nearly all
explanatory power once $\Phisim$ is controlled for.  The modest
significance level reflects the limited number of experimentally characterised
sites (7--8 per protein) rather than a weak effect.  These results establish
that the trajectories encode the mechanical unfolding pathway, validating their
use as training data for a force-conditioned generative model.

\subsection{Extension-conditioned ensembles are reproducible in their continuous observables across pulling speeds}
SMD trajectories are typically generated at pulling speeds several orders of
magnitude faster than AFM experiments, raising the question of whether the
sampled conformations are biased toward kinetically trapped, high-energy
structures.  Addressing this question requires distinguishing two effects of
speed: the elongation \emph{at} which a structural transition occurs, and the
conformational \emph{ensemble} that is sampled \emph{at} a given elongation.
Using end-to-end distance rather than simulation time as the reaction coordinate
decouples these effects~\cite{rief1997reversible,bustamante1994entropic}.

Across four pulling speeds spanning two orders of magnitude (0.01--1.0 nm/ns)
and four benchmark proteins with well-characterised mechanical properties
(ubiquitin, titin I27, protein L, and protein G B1 domain), the
native-contact fraction $Q$ and the radius of gyration $R_\mathrm{g}$ deviate
by only 2.9--7.0\% of their dynamic range when compared at the same
elongation, and the C$\alpha$-RMSD traces overlap over the shared extension
range (Table~\ref{tab:rms_diff}).  Residue-level contact-survival curves are highly
correlated across speed pairs (Pearson $r$ median 0.81--0.92; 93--98\% of
residues at $r > 0.7$), and the domain-assignment profiles along the sequence
are significantly correlated in all 24 protein--speed-pair combinations
($r = 0.72$--$0.97$, all $P < 3 \times 10^{-33}$).  The Spearman rank
correlations for contact-survival profiles are comparable (mean
$\rho = 0.82$--$0.84$), ruling out the possibility that the Pearson values
are driven by a small number of outlier residues.  There is no systematic
dependence of the deviations on the speed separation $|\log_{10}(v_2/v_1)|$:
of 12 (observable $\times$ protein) combinations, the Spearman correlation
between speed separation and RMS deviation is positive in 8 and negative in 4,
with only one reaching $P < 0.05$.

Pulling speed does produce one detectable systematic effect: faster pulling
delays complete structural dissolution to larger elongations by an average of
2.83 nm across the tested range.  This effect has a transparent physical
interpretation.  At constant velocity, the residence time in each elongation
window decreases with speed, so thermally activated barrier crossing occurs at
lower probability per unit elongation; residual structure is therefore dragged
to larger end-to-end distances before collapsing~\cite{evans1997dynamic,bell1978models}.
The effect is largest for titin I27, whose $\mathrm{A}'$--$\mathrm{G}$ strand
hydrogen-bond network constitutes a steep, localised barrier sensitive to
residence time, and smallest for protein L, whose shallower barrier is already
adequately sampled at all tested speeds~\cite{lu2000key,li2002reverse}.  This
speed dependence is an intrinsic feature of constant-velocity pulling, not a
defect of the force field or thermostat, and it does not compromise the
reproducibility, across speeds, of the continuous observables of the ensemble
\emph{conditioned on elongation}.  An important qualification follows from the Kolmogorov--Smirnov
analysis: although the shape of the contact-survival profile is conserved
across speeds, the absolute contact-survival \emph{values} at a given
elongation are not identically distributed (KS pass rate 23--40\%).  Speed
therefore shifts the absolute level of contact preservation without altering
the residue-level order of disruption.  This implies that trajectories from
different speeds may be combined as training data without introducing a
systematic conformational bias in the ordered disruption sequence, but that
absolute contact probabilities should not be read as quantitative $\varphi$
surrogates.

\subsection{Force field and simulation protocol}
All trajectories were generated with the AMBER ff14SB force field~\cite{maier2015ff14sb}
and OBC2 implicit solvent~\cite{onufriev2004exploring}, with a fixed spring constant,
Langevin thermostat, and time step applied uniformly across all pulling speeds.
The force field was chosen on the basis of its established accuracy for protein
backbone and side-chain geometry~\cite{maier2015ff14sb} and its widespread validation
in both equilibrium and non-equilibrium protein simulations~\cite{lindorfflarsen2012validation}.
Implicit solvent was adopted to allow systematic comparison across a wide range
of pulling speeds without the computational overhead of explicit solvent boxes
at the slower rates; the qualitative agreement between the simulated
force-peak structures and AFM-derived $\varphi$ values (discussed above) provides
an empirical upper bound on force-field-induced distortion at the level of
contact topology.  The internal speed-robustness analysis constitutes an
additional form of validation: a force field that introduced spurious
speed-dependent conformational artefacts would manifest as a systematic
increase in cross-speed deviation with speed separation, which is not observed.

\subsection{Rationale for the segmented generative architecture}
The deep generative model is organised around individual unfolded segments
(Rounds~1 and~2), rather than modelling the full-length partially unfolded
chain as a continuous object.  This architectural choice reflects a
quantitative property of the unfolding intermediates themselves.  A binary
partition of each trajectory frame into folded and unfolded regions, using a
local-RMSD criterion with fixed window width $w=7$ and threshold $\tau = 2.0$\,\AA{},
operates stably on 500 proteins spanning 11 species, yielding 557,550
segments of length $N_\mathrm{seg} \ge 3$ with well-defined boundaries.  The
stability of this binary operator across such a diverse sample is not trivial:
were the folded-to-unfolded transition spatially gradual rather than sharp,
the operator would produce diffuse boundaries, and the resulting segments would
not constitute geometrically coherent objects.  The sharp boundary indicates
that, on the length scale of a few residues, the transition between native
conformation and extended conformation is essentially discontinuous --- a
direct, data-level confirmation of the two-state assumption at the heart of
Wako--Sait\^{o}--Mu\~{n}oz--Eaton-class statistical mechanical models~\cite{wako1978statistical1,wako1978statistical2,munoz1999simple,bruscolini2002exact}.

The geometry of the unfolded segments further supports the segmented
architecture.  Section~\ref{sec:res_segment} shows that the end-to-end
distance of every segment is fixed by segment length through
equation~\eqref{eq:ete} to within one C$\alpha$--C$\alpha$ virtual bond,
and that the scaling coefficient is unimodal and narrow across proteins
while dispersing systematically with length across individual segments.
The interpretation of that result requires care, because a value of
$\alpha \approx 0.9$ invites two readings that it does not in fact
support.

Force spectroscopy conventionally fits chain extension with the worm-like
chain model~\cite{bustamante1994entropic}, converting contour-length increments into
unfolded residue counts through a per-residue contour length of
0.36--0.40\,nm~\cite{rief1997reversible,carrionvazquez1999mechanical} --- numerically close to the
3.8\,\AA{} virtual bond used here, but representing the asymptotic geometric
ceiling reached at infinite force.  At the opposite limit, the fully denatured
state is described as a self-avoiding walk whose radius of gyration scales as a
power law in chain length~\cite{kohn2004random}.  The ensemble measured here is
neither: $\alpha$ describes the extension actually realised at finite force
while a folded core still anchors one or both ends of the segment, and its
value falls quantitatively between a random coil and a fully straightened rod.
Equation~\eqref{eq:ete} therefore characterises a distinct ensemble rather
than restating either limiting case, which is what makes it usable as a
geometric prior over intermediates rather than over end states.

Three consequences follow for model design, and together they motivate
training Rounds~1 and~2 on short segments rather than on whole chains.
First, the learning unit is well posed: a contiguous unfolded segment is a
stable object under a fixed-hyperparameter operator across 500 proteins and 11
species, so a per-segment target is defined without protein-specific tuning,
whereas a diffuse crossover would have made segment boundaries threshold
artefacts and any per-segment supervision would have inherited that
arbitrariness.  Second, the target has low intrinsic variance and an
analytically known leading term: within a segment the dominant geometric
degree of freedom is fixed by segment length to 3.9\,\AA, so the network
learns the residual after that term is removed --- the component carrying
sequence and local-structure information --- rather than spending capacity
re-deriving a covalent geometric constraint from whole chains.  The
species-independence of the per-protein coefficient licenses $\alpha$ as a
shared prior, while its per-segment length dependence requires segment length
to enter as an explicit conditioning variable rather than as a fixed scalar.
Third, the harder long-range problem is deferred: equation~\eqref{eq:ete}
constrains geometry within a segment but says nothing about the relative
arrangement of two folded regions separated by an unfolded segment, nor about
contact probability as a function of separation.  These longer-range relations
are addressed by the later rounds, which operate on whole chains conditioned on
the reaction coordinate.  The use of small-fragment training in Rounds~1 and~2
is therefore not an approximation of convenience but a decomposition that
matches the physical modularity of the intermediates --- a modularity analogous
to the domain-by-domain unfolding picture established for multi-domain proteins
by single-molecule force spectroscopy~\cite{rief1997reversible,carrionvazquez1999mechanical}, and
consistent with the theoretical prediction that mechanical unfolding of
independently stable domains proceeds via sequential rather than concerted
transitions~\cite{imparato2007ising}.

\subsection{Limitations and future directions}
Several limitations should be noted.  First, each protein--speed combination
is represented by a single trajectory, so the cross-speed RMS deviations
reported in Section~\ref{sec:res_velocity} conflate genuine speed effects with
stochastic trajectory-to-trajectory variation.  To disentangle these, replicate
trajectories at the production speed with different random seeds would establish
a within-speed noise baseline; a cross-speed deviation that does not exceed
this baseline would elevate the current result from ``no detectable systematic
dependence'' to ``speed effect smaller than stochastic noise''.  Second, the
low KS pass rate (23--40\%) implies that absolute contact probabilities are
not quantitatively preserved across speeds.  Training targets should therefore
be restricted to continuous observables conditioned on elongation (geometry and
contact topology) rather than discrete labels derived from absolute thresholds,
such as ranked residue sets or contact-count categories.  Third, the binary
partition threshold $\tau = 2.0$\,\AA{} was chosen empirically and validated
globally; a systematic sensitivity analysis on $\tau$ and window width $w$
would consolidate the robustness claim.  Fourth, three additional quantitative
characterisations of the intermediates --- conservation of folded-region geometry
under stretching, near-independence of contacts within unfolded segments, and
power-law decay of inter-fragment contact probability with separation length
--- were formulated on a small set of proteins and are presented here without
full cross-species validation.  Confirming these properties at the 500-protein
scale would complete the quantitative framework and extend the theoretical
grounding of the segmented architecture.

A second group of limitations attaches to the training records themselves.  All
curves in Section~\ref{sec:res_training} were obtained from a single run per
stage without repeated random seeds, so small numerical differences --- in
particular the 0.014\,\AA{} improvement attributed to the learning-rate restart
--- should not be treated as statistically significant.  The RMSD reported for
Round~1 and Round~2 is defined on 15-residue windows and is therefore not
directly comparable with the chain-level or full-length RMSD quoted by other
methods\cite{jumper2021alphafold,abramson2024alphafold3}.  Stage~1 currently has only a
first-epoch record, and its full convergence behaviour awaits further training.
The decoder curve ends at 46{,}233 steps, short of the intended training budget,
and its validation loss was still falling at that point, so the reported value
should be read as an upper bound rather than a converged one. The
conformational latent pretraining stage that precedes the cascade is reported
only through its role in the pipeline: no reconstruction error for its Decoder
probe and no calibration for its Q-Predictor probe are shown, so the frozen
latent it exports is taken as given rather than independently benchmarked.

A third group attaches to the coverage benchmark.  Twenty proteins is a small
panel; the twenty sequences were not screened for homology against the training
corpus (Section~\ref{sec:meth_splits}), so the comparison measures
conformational spread rather than held-out generalisation; and the
AlphaFold~3 ensembles vary only the random seed; a baseline built
by subsampling or clustering the
alignment\cite{waymentsteele2024conformations,delalamo2022sampling}, or drawn
from a generative ensemble
method\cite{lewis2025bioemu,jing2024alphaflow}, would be a more demanding
comparison and is the natural next step.  The reference state for $Q$ is taken
from the AF3 ensemble itself rather than from an experimental structure, and no
reference ensemble from molecular dynamics was available for these twenty
sequences, so coverage here is measured against the joint range of the two
methods rather than against a known ground-truth range; whether the covered
region is populated with the correct Boltzmann weights is a separate question
that this statistic cannot address.

A fourth group concerns the evaluation of the reconstructions.  The six-protein
panel of Fig.~\ref{fig:recon}a is a qualitative topology comparison and is not
accompanied by per-panel chain lengths, $Q$ values and RMSDs; supplying them
would allow the panel to carry quantitative weight.  More fundamentally, the
125.8\,\AA{} outlier at $Q \to 0$ shows that global RMSD is the wrong yardstick
for extended conformations, so an accuracy claim spanning the whole of $Q$ will
require the superposition-free metrics recommended above to be adopted
systematically rather than reported alongside RMSD.  Finally, the scope of the
present claim should be stated plainly: FoldScape is a conformational generator,
not a thermodynamic estimator.  It produces conformations at a prescribed value
of the reaction coordinate, and this work makes no claim about the statistical
weight those conformations would carry in an equilibrium ensemble.  Because the
corpus is generated by mechanical rather than chemical denaturation, assigning
such weights would require an independent calibration against equilibrium
sampling, which we have not attempted.

\subsection{Conclusion}
This work joins four strands of evidence into a single argument: that
sequence-conditioned conformational ensembles can be generated across the
folding continuum --- from unfolded chains through folding intermediates
to the native folded state --- rather than predicting a single endpoint.

The first strand is the prediction itself. Across the evaluated stages of
a folding trajectory of a 706-residue chain --- more than threefold beyond
the chain length that a single chain-level forward pass accepts, and reached
by the stride-one crop-and-superpose scheme of
Section~\ref{sec:meth_longchain} --- the C$\alpha$ RMSD stays between
0.67 and 1.76\,\AA{} over sampled partially folded and compact
conformations, excluding the fully extended frame. Reconstruction accuracy
remains approximately stable along $Q$ within this evaluated range.  The single
apparent exception, the 125.8\,\AA{} value at the fully extended end, is a
property of global superposition on a nearly one-dimensional object rather than
a failure of the prediction, and it identifies a concrete methodological
requirement: unfolded and early-collapse states must be scored with
superposition-free measures if landscape-level claims are to be meaningful.

The second strand shows that the cascade producing those predictions can be
trained, and diagnosed, stage by stage.  Round~1 reduces the window-level
coordinate error to 2.96\,\AA{} while calibrating seven distance scales;
Round~2, whose zero-initialised modulation leaves it equal to Round~1 to within
numerical noise at initialisation, removes a further 46\% of that error to
reach 1.59\,\AA{};
Stage~1 completes the window-to-chain assembly and exports the conditioning
representation; and the flow-matching decoder converges on that representation
without overfitting.  Because every stage is trained with its upstream frozen,
each gain is attributable to a specific architectural choice rather than to
retraining at larger scale --- and the same decomposition exposes the two
systematic costs of the design, the trade between global alignment accuracy and
local bond regularity inside Round~1, and the window-to-chain gap that now
limits overall accuracy.

The third strand supplies the foundations.  The comparison of
$\Phisim$ with mechanical and chemical $\varphi$ values establishes
that constant-velocity SMD samples the force-specific unfolding pathway at
residue resolution, with all twelve mechanical protein--speed combinations
positive and all four chemical controls negative.  The speed-robustness
analysis demonstrates that the ensemble conditioned on elongation is
reproducible across two orders of magnitude of pulling speed in the
continuous observables the model is conditioned on, with deviations of
2.9--7.0\% of their dynamic range, allowing
trajectories at different speeds to be aggregated without introducing a
systematic conformational bias.  The large-scale characterisation of unfolding
intermediates across 500 proteins and 11 species reveals that partially
unfolded chains decompose into geometrically coherent segments whose extension
statistics are universal at the protein level and length-dependent at the
segment level --- the physical modularity that the segmented architecture was
built to match.

The fourth strand fixes the scale of the result against an external reference.
On twenty human proteins spanning 36--857 residues, and at matched sample size,
FoldScape ensembles cover a median of 0.99 of the radius-of-gyration axis and
0.72 of the $Q$ axis against 0.05 for AlphaFold~3 baseline ensembles on both,
with every protein pointing the same way on all three order parameters and the
conclusion unchanged across bin counts from 10 to 40.  Coverage and accuracy
are independent quantities, and it is holding them together --- a prediction
that is both correct at each folding stage and actually distributed across those
stages --- that a generator of the folding process requires.

Accurate reconstruction across folding stages is a defining requirement
for a generator of the folding process. FoldScape brings together controlled
sequence-conditioned conformational generation from unfolded states through
intermediates to native-state conformations, an architecture whose
contributions can be analysed individually, and training trajectories whose
physical validity has been independently examined. Together, these advances
extend sequence-based structure prediction to the folding continuum and
bring analyses of misfolding, cryptic intermediates and allosteric coupling
within reach at proteome scale.


\section*{Funding}
This project was supported by the Beijing Life Science Academy
(2025900CB0160).

\section*{Acknowledgements}
We thank the maintainers of the AlphaFold Protein Structure
Database and of the five public molecular dynamics repositories used for
cross-validation for making their data openly available.

\section*{Author contributions}
Contributions are described using the CRediT taxonomy.
\textbf{Conceptualization:} L.Z. and Z.W.X., who jointly formulated the
scientific question and designed the study.
\textbf{Methodology:} L.Z. and Z.W.X., who jointly developed the methodology
and the theoretical framework of FoldScape.
\textbf{Software:} L.Z., Q.L. and Z.N.X., who implemented the algorithmic
framework and carried out the deep learning training.
\textbf{Investigation:} L.Z., Y.Y. and Q.Y., who performed the molecular
dynamics simulations.
\textbf{Data curation:} L.Z., Y.Y. and Q.Y.
\textbf{Formal analysis:} L.Z., Y.Y. and Q.Y.
\textbf{Validation:} L.Z. and Z.W.X.
\textbf{Visualization:} L.Z.
\textbf{Writing --- original draft:} L.Z. and Y.Y.
\textbf{Writing --- review and editing:} Z.W.X., L.Z., P.L., Z.F., W.C. and
R.H.
\textbf{Resources:} Z.W.X., L.Z., P.L., Z.F., W.C. and R.H., who provided the
computational resources on which this work was carried out.
\textbf{Supervision and project administration:} Z.W.X. and L.Z.
\textbf{Funding acquisition:} Z.W.X., P.L., Z.F., W.C. and R.H.

L.Z., Q.L., Z.N.X., Y.Y. and Q.Y. contributed equally to this work. All
authors read and approved the final version of the manuscript.


\section*{Competing interests}
The authors declare no competing interests.

\section*{Data availability}
The sequence lists defining AF-Raw391K, AF-Core116K, AF-NR68K, AF-Sub10K,
AF-Sub1K and AF-Geo500, the coarse-grained and steered molecular dynamics
trajectories generated in this work, the AlphaFold~3 and FoldScape
conformational ensembles for the twenty benchmark proteins, the
per-conformation order-parameter table, the per-protein coverage table and the
complete statistical output in JSON will be deposited in a public repository
on publication, and are available from the corresponding author on reasonable
request until then. The input structures
were obtained from the AlphaFold Protein Structure Database
(\url{https://alphafold.ebi.ac.uk}); the external molecular dynamics
trajectories used for cross-validation are available from their original
repositories as cited in Section~\ref{sec:meth_externalmd}.

\section*{Code availability}
The FoldScape training and inference code, the simulation and analysis
pipelines, and the conformational coverage analysis script (random seed
20240521) will be released under an open-source licence in a public
repository on publication, and are available from the corresponding author on
reasonable request until then.

\section*{Author information}
\noindent\textbf{Corresponding author.} Correspondence to Zhengwei Xie
(\href{mailto:xiezhengwei@hsc.pku.edu.cn}{xiezhengwei@hsc.pku.edu.cn}).

\newpage


\newpage


\newpage

\section*{Supplementary Methods}
\addcontentsline{toc}{section}{Supplementary Methods}
\setcounter{table}{0}
\renewcommand{\thetable}{S\arabic{table}}

\subsection*{S1. Computational data generation}

\subsubsection*{S1.1 Structure retrieval and quality filtering}

Predicted protein structures in mmCIF format were downloaded from AlphaFold~DB\cite{varadi2024alphafold} via its programmatic API. Beyond the three primary filters (global pLDDT $\geq 70$, high-confidence residue fraction $\geq 50\%$, sequence length 50--2{,}000), a composite quality score $Q_\mathrm{qc}=0.4\,\mathrm{pLDDT}_\mathrm{global}+0.5\,S_\mathrm{domain}+0.1\,B_\mathrm{length}$ was used to rank borderline candidates. Each mmCIF file was parsed and converted to PDB format using BioPython 1.83\cite{cock2009biopython}, preserving atom coordinates, residue identifiers, chain assignments, occupancies and temperature factors.

\subsubsection*{S1.2 Iterative sequence clustering}

In the first MMseqs2\cite{steinegger2017mmseqs2} pass, AF-Raw391K was clustered at 65\% minimum identity with 80\% bidirectional coverage using the greedy incremental algorithm, yielding 116{,}405 representatives (AF-Core116K, 70.2\% reduction). In the second pass, 50 iterative rounds were applied with the identity threshold decreasing from 0.65 in steps of $\sim 0.05$ to a floor of 0.30. The dataset converged at 67{,}605 sequences after round~45, with rounds~46--50 confirming stability. Proportional random subsets (AF-Sub10K and AF-Sub1K) were drawn with stratified sampling to preserve the length-class distribution of AF-NR68K.

\subsubsection*{S1.3 Go
	Ca coarse-grained simulations}

Native-contact-based G\=o-like C$\alpha$ topologies were generated using GoCa\cite{walter2024goca}, which reduces each amino acid to a single C$\alpha$ bead and assigns non-bonded Lennard-Jones 12--10 interactions between native-contact pairs in the input structure. Bond, angle and dihedral potentials were derived from the native geometry. Each structure was placed in a cubic box with \SI{20}{\nano\metre} padding and energy-minimised by steepest descent (\num{50000}~steps; force tolerance \SI{100}{\kilo\joule\per\mol\per\nano\metre}). Production simulations used the leap-frog integrator with time step \SI{0.004}{\pico\second}, the Verlet neighbour-list scheme (update frequency 20 steps; buffer tolerance 0.005), non-bonded cutoffs of \SI{4.0}{\nano\metre} with the Potential-shift-Verlet modifier, and the v-rescale thermostat\cite{bussi2007canonical} ($\tau=\SI{0.05}{\pico\second}$) at $T^*=35$ in G\=o-model reduced units. Initial velocities were drawn from a Maxwell--Boltzmann distribution at $T^*=40$. Coordinates were written every 100~steps. The simulation length was determined from the residue count $N$ via piecewise linear interpolation between \SI{3}{\nano\second} at $N\leq100$ and \SI{35}{\nano\second} at $N\geq2{,}000$.

\subsubsection*{S1.4 Go-kit and OpenAWSEM topologies}

Go-kit\cite{neelamraju2019gokit} topologies use a heavy-atom contact cutoff of \SI{4.5}{\angstrom} with a contact-map scaling factor of 1.2 and a minimum sequence separation of $|i-j|\geq 3$; bonded force constants are $K_b=200$, $K_\theta=40$, $K_\phi=1$ in G\=o-model reduced units. The non-native hydrophobic option places attractive Lennard-Jones interactions on every hydrophobic-residue pair (Ala, Val, Leu, Ile, Pro, Phe, Met, Trp) with $|i-j|\geq 4$ and not already in the native-contact list, with $r_0=\SI{5.5}{\angstrom}$ and $\varepsilon_\mathrm{hp}=1.0$; glycine residues are excluded. AWSEM\cite{davtyan2012awsem,lu2021openawsem} uses three coarse-grained sites per residue (C$\alpha$, C$\beta$, O) and the default OpenAWSEM parameterisation, with backbone hydrogen bonding, direct-contact and water-mediated potentials, a burial term and a bioinformatically derived fragment-memory term\cite{papoian2004role}. The same constant-velocity SMD pulling protocol (spring constant, pulling rate, terminus definitions) used for all-atom OpenMM was applied; for Go-kit, Langevin dynamics was propagated at $T^*=35$ with friction \SI{1}{\per\pico\second} and time step \SI{0.004}{\pico\second}; for OpenAWSEM, at \SI{300}{\kelvin} with a \SI{2}{\femto\second} time step.

\subsubsection*{S1.5 External MD repositories}

Simulation engines, force fields, timescales and access interfaces of the five external repositories are those given in their respective primary publications\cite{mokhtari2026dynarepo,rodriguezespigares2020gpcrmd,roy2025mdrepo,beltran2024bioexcelcv19,pirtskhalava2021dbaasp}. Trajectories were resampled to a uniform 1-frame-per-50\,ps grid where necessary and re-aligned to a consistent coordinate convention; per-protein C$\alpha$ extraction matched the in-house pipeline.

\subsubsection*{S1.6 Validation simulation details}

Class-A equilibrium folding simulations used $5\!\times\!10^7$ steps at $T_f$, with $T_f$ determined by replica simulations at five temperatures bracketing the expected transition followed by heat-capacity analysis. Class-D rate-stability tests used four all-atom pulling velocities ($v=0.01$, $0.05$, $0.1$, $\SI{1.0}{\nano\metre\per\nano\second}$); class-D$'$ repeated the test in AWSEM at six velocities ($v=0.01$, $0.05$, $0.1$, $1.0$, $2.0$, $\SI{4.0}{\nano\metre\per\nano\second}$). In both classes the acceptance criterion is defined on the continuous observable---the root-mean-square difference between per-residue contact-survival curves at matched extension, required to fall within the band established by the all-atom analysis. Pairwise Spearman rank correlations of contact-rupture order between velocities were also computed, but are reported as a descriptive statistic only and carry no pass/fail threshold.

\subsection*{S2. Conformational latent pretraining}

\subsubsection*{S2.1 Deterministic Fourier embedding}

Each scalar $x$ is embedded as
\begin{equation}
	\fiteq{\mathrm{FE}_d(x)=\bigl[\sin(\omega_1 x),\ldots,\sin(\omega_{d/2}x),\;\cos(\omega_1 x),\ldots,\cos(\omega_{d/2}x)\bigr],}
\end{equation}
with $\omega_j=2\pi\exp(\mathrm{linspace}(\ln 1,\ln 1000,d/2))$. We use $d=128$ for $\tilde{R}_g$ and $f_{ss}$, $d=64$ for $Q$, producing a 320-dimensional concatenation as input to the conditioning MLP. Frequencies are precomputed as non-learnable buffers.

\subsubsection*{S2.2 ConformExpander parameter table}

\begin{table}[!tb]
\centering\small
\caption{ConformExpander layer dimensions and parameter counts. Entries are
exact counts including biases and sum to the stated total.}
\label{tab:conformexpander}
\fitcolumn{\begin{tabular}{@{}llr@{}}
\toprule
Layer & Dimensions & Parameters \\
\midrule
Fourier embedding   & $3\times\text{scalar}\to 320$ & 0 \\
Conditioning MLP    & $320\to 512\to 512$ (SiLU)    & 427{,}008 \\
FiLM $\bm{\gamma}$  & $512\to 1152$                 & 590{,}976 \\
FiLM $\bm{\beta}$   & $512\to 1152$                 & 590{,}976 \\
LayerNorm           & $1152$                        & 2{,}304 \\
Residual MLP        & $1152\xrightarrow{\mathrm{SiLU}}1152\to 1152$ & 2{,}656{,}512 \\
Gate $\alpha$       & $512\to 1$                    & 513 \\
\midrule
\textbf{Total} & & \textbf{4{,}268{,}289} \\
\bottomrule
\end{tabular}}
\end{table}

$W_\gamma$ and $W_\beta$ are initialised from $\mathcal{N}(0,0.01^2)$ with bias zero; with the $+\mathbf{1}$ offset applied after multiplication, the initial modulation is $\bm{\gamma}\approx\mathbf{1}$ and $\bm{\beta}\approx\mathbf{0}$, recovering an identity transformation that preserves the pretrained ESM-C embedding so that $\mathbf{v}\approx\bar{\mathbf{e}}$ before the conformational conditioning pathway has learned anything.

\subsubsection*{S2.3 Decoder architecture details}

The Decoder follows a broadcast-and-refine design. The global vector $\mathbf{v}\in\mathbb{R}^{1152}$ is projected to 256 dimensions and broadcast-added to per-residue embeddings $\mathbf{e}_i$ (also projected to 256 dimensions): $\mathbf{h}_i=W_\mathrm{cond}\mathbf{v}+W_\mathrm{emb}\mathbf{e}_i$. The result is processed by $B_D=4$ identical DecoderBlocks. Each block consists of: (i)~gated self-attention with 8 heads ($c_a=256$, head width 32) using PyTorch's \texttt{scaled\_dot\_product\_attention}\cite{dao2022flashattention} and a sigmoid gate $\Delta\mathbf{h}=W_o[\sigma(W_g\tilde{\mathbf{s}})\odot\mathrm{SDPA}(Q,K,V)]$; (ii)~SwiGLU transition\cite{shazeer2020glu} with expansion 2. Both sub-layers use pre-LayerNorm residual connections. The coordinate output head is a single linear projection $\mathbb{R}^{256}\to\mathbb{R}^3$; the structural-feature head applies masked mean-pooling followed by a two-layer MLP ($256\to 64\to 7$, SiLU intermediate, sigmoid output). The Decoder is deliberately under-parameterised (4 layers, 256-dim, no pair representation) so that $\mathcal{L}_\mathrm{recon}$ drives information compression into $\mathbf{v}$ rather than being trivially solved by the Decoder itself, analogous to linear probing in representation learning\cite{alain2016probes}.

\subsubsection*{S2.4 Structural-feature targets}

The seven targets $\mathbf{y}_\mathrm{true}\in[0,1]^7$ are computed by a parameter-free GlobalStatistics module from the C$\alpha$ distance matrix $D_{ij}=\|\mathbf{x}_i-\mathbf{x}_j\|$ and coordinates: $R_g$, mean off-diagonal pairwise distance $\bar{D}$, standard deviation $\sigma_D$, contact density $\rho_c$ (fraction of off-diagonal pairs with $D_{ij}<8$\,\AA{}), and end-to-end distance $d_{ee}$. Min-max normalisation gives $\mathbf{y}_\mathrm{true}=[Q,\tilde{R}_g^{(s)},f_{ss},\bar{D}_\mathrm{norm},\rho_c,\sigma_{D,\mathrm{norm}},d_{ee,\mathrm{norm}}]$ with $\tilde{R}_g^{(s)}=\mathrm{clamp}((R_g-2.0)/(80.0-2.0),0,1)$ (fixed bounds in \AA, deliberately distinct from the $N$-dependent $\tilde{R}_g$ used as the order-parameter input), $\bar{D}_\mathrm{norm}=\mathrm{clamp}((\bar{D}-3.0)/(100.0-3.0),0,1)$, $\sigma_{D,\mathrm{norm}}=\mathrm{clamp}((\sigma_D-0.5)/(40.0-0.5),0,1)$ and $d_{ee,\mathrm{norm}}=\mathrm{clamp}(d_{ee}/(3.8(N_\mathrm{eff}-1)),0,1)$. The gradient from $\mathcal{L}_\mathrm{struct}$ is detached from the target computation to prevent trivial solutions.

\subsubsection*{S2.5 Q-Predictor architecture}

The Q-Predictor mirrors the ConformExpander's FiLM mechanism at $c_h=256$. Fourier embeddings of $\tilde{R}_g$ and $f_{ss}$ (each 128-dim) are concatenated and processed by a two-layer MLP ($256\to 256\to 256$, SiLU) to give $\mathbf{c}_Q\in\mathbb{R}^{256}$. Modulation: $\bm{\gamma}_Q=W_\gamma^Q\mathbf{c}_Q$ (bias initialised to 1, equivalent to the ConformExpander's $+\mathbf{1}$ offset), $\bm{\beta}_Q=W_\beta^Q\mathbf{c}_Q$ (bias 0), applied as $\mathbf{h}_Q=\bm{\gamma}_Q\odot(W_e\bar{\mathbf{e}})+\bm{\beta}_Q$ with $W_e$ a bias-free linear $\mathbb{R}^{1152}\to\mathbb{R}^{256}$. Output: a linear head maps $\mathrm{SiLU}(\mathbf{h}_Q)$ to two scalars, $\hat{Q}=\sigma(\mathrm{out}_1)$ and $\hat{\sigma}=\exp(\mathrm{clamp}(\mathrm{out}_2,-4,4))$. Total: 558{,}594 parameters. By predicting $Q$ from $\bar{\mathbf{e}}$ conditioned on $(\tilde{R}_g,f_{ss})$ without access to coordinates, the Q-Predictor forces the ConformExpander to produce $\mathbf{v}$ vectors whose relationship to $(\tilde{R}_g,f_{ss})$ is informative about $Q$, complementing the coordinate-level signal from the Decoder.

\subsubsection*{S2.6 Kabsch alignment}

The Kabsch algorithm\cite{kabsch1976solution} computes the optimal rotation $\mathbf{R}^\star$ minimising $\sum_i\|\mathbf{R}\hat{\mathbf{x}}_i+\mathbf{t}-\mathbf{x}_i^\mathrm{true}\|^2$ via SVD of the cross-covariance matrix $\mathbf{H}=\hat{\mathbf{X}}_c^\top\mathbf{X}_c^\mathrm{true}$. To prevent reflections, the sign of the third singular vector is corrected based on $\det(\mathbf{V}^\top\mathbf{U}^\top)$. Numerical safeguards: (i)~all SVD operations are cast to float32 regardless of training precision (bfloat16) to prevent gradient degeneracy from near-singular covariance matrices; (ii)~if the coordinate variance falls below $10^{-6}$, alignment is skipped and only translation is applied; (iii)~SVD failures fall back to translation-only alignment.

\subsection*{S3. Round 1 (R1)}

\subsubsection*{S3.1 CondEncoder}

The CondEncoder combines the chain-level ESM-C global embedding with the three conformational descriptors used in conformational latent pretraining. Each scalar $x\in\{Q,\tilde{R}_g,f_{ss}\}$ is mapped to $\mathbb{R}^{64}$ by a deterministic Fourier embedding\cite{tancik2020fourier} with logarithmically spaced frequencies $\omega_j=2\pi\exp(\mathrm{linspace}(\ln 1,\ln 1000,32))$. The three 64-dimensional Fourier features are concatenated with the frozen conformational latent $\mathbf{v}\in\mathbb{R}^{1152}$ exported by the retained ConformExpander (Section~\ref{sec:meth_pretrain}; $\mathbf{v}\approx\bar{\mathbf{e}}$ before the conditioning pathway has been trained) into a 1{,}344-dimensional vector that is projected by a two-layer SiLU\cite{elfwing2018silu} MLP ($1344\to 384\to 384$) and layer-normalised\cite{ba2016layernorm} to yield $\mathbf{c}\in\mathbb{R}^{384}$. The same $\mathbf{c}$ is broadcast over all $W$ tokens in the single stream and projected (zero-bias) to $c_z=128$ for the pair stream.

\subsubsection*{S3.2 Pre-LN trunk Transformer}

The trunk is a 6-layer pre-LN Transformer encoder\cite{vaswani2017attention,xiong2020layernorm} with model width $c_s=384$, 16~attention heads (head width 24), feed-forward expansion 4 (hidden width 1{,}536), GELU activation\cite{hendrycks2016gelu} and uniform dropout 0.1; padding tokens are masked. Pre-LN was chosen to remove the need for Transformer-specific learning-rate warm-up and to stabilise mixed-precision training.

\subsubsection*{S3.3 PairFormer block}

Each of the $K=7$ scale-specific blocks follows the AlphaFold pair-track design\cite{jumper2021alphafold,abramson2024alphafold3}. Given $\mathbf{h}\in\mathbb{R}^{W\times c_s}$ and $\mathbf{z}\in\mathbb{R}^{W\times W\times c_z}$, the block updates $\mathbf{z}$ through (i)~an outer-product mean (OPM, projection width $c_m=32$), (ii)~outgoing and incoming triangle multiplicative updates, (iii)~triangle self-attention starting and ending at the corner node (4 heads of width 32), and (iv)~a Transition feed-forward block with expansion 4. The single stream is then updated by a gated single-token attention layer (16 heads of width 24) reading pair biases from $\mathbf{z}$, followed by a second Transition. Row-dropout is disabled (\texttt{drop\_row\_p}=0) because the windows are short enough that the regularisation gains are outweighed by sample-efficiency loss. Inputs to block~$k$ are augmented with the cumulative outputs of blocks $1,\ldots,k-1$ via MultiScaleAgg (S3.4) and with a zero-initialised conditioning re-injection $W_\mathrm{re}^{(k)}\mathbf{c}$.

\subsubsection*{S3.4 Multi-scale aggregation (MultiScaleAgg)}

A learned attention pooler aggregates the per-scale single-stream outputs $\{\mathbf{h}^{(k)}\}_{k=1}^{K}$. The list is stacked into a tensor of shape $(W,K,c_s)$, projected by a key matrix $W_\mathrm{key}\in\mathbb{R}^{c_s\times c_s}$ and combined with a learned query $\mathbf{q}\in\mathbb{R}^{c_s}$,
\begin{equation}
	\fiteq{\alpha_{i,k}=\mathrm{softmax}_k\bigl(\langle W_\mathrm{key}\mathbf{h}^{(k)}_i,\mathbf{q}\rangle\bigr),\qquad \bar{\mathbf{h}}_i=\sum_{k=1}^{K}\alpha_{i,k}\mathbf{h}^{(k)}_i,}
\end{equation}
followed by a final $W_\mathrm{out}$ projection and layer normalisation. Independent MultiScaleAgg modules with the same architecture are used inside every PairFormer block (with the partial list $\{\mathbf{h}^{(1)},\ldots,\mathbf{h}^{(k-1)}\}$) and once before coordinate readout (with the full list).

\subsubsection*{S3.5 Anchored distance heads}

For each scale $k$ the head reads the concatenation of single representations $\mathbf{h}^{(k)}_i,\mathbf{h}^{(k)}_{i+k}$ and the diagonal pair representation $\mathbf{z}^{(k)}_{i,i+k}$ (total dimension $2c_s+c_z=896$) and applies a 3-layer MLP ($896\to 256\to 64\to 1$, GELU\cite{hendrycks2016gelu}) followed by the bounded affine map $\hat{d}_{i,i+k}=a_k+s_k\tanh(\cdot)$. The anchor--scale pairs $(a_k,s_k)$ in \AA{} are
\[
	\fiteq{(3.80,0.50),(5.60,1.50),(6.50,2.50),(8.00,3.50),(9.50,5.00),(11.0,6.00),(12.5,7.00),}
\]
chosen so that $[a_k-s_k,a_k+s_k]$ brackets the 5th--95th percentile of empirical C$\alpha$ distances at separation $k$ in the training ensemble. The bounded $\tanh$ output guarantees that random or degenerate predictions never produce non-physical distances, which we found to be a major source of training instability in unbounded preliminary variants. The final linear layer of each head is initialised with $\sigma=10^{-3}$ and zero bias so that initial predictions concentrate at the anchor.

\subsubsection*{S3.6 Coordinate refinement and output}

The full list of seven block outputs is aggregated by the coordinate-stage MultiScaleAgg, broadcast-summed with the conditioning vector $\mathbf{c}$ and refined by an additional 2-layer pre-LN Transformer (same width and head configuration as the trunk). The refined token stream is layer-normalised and decoded by a single linear head ($\sigma=10^{-3}$, zero bias) to per-residue C$\alpha$ coordinates $\hat{\mathbf{X}}\in\mathbb{R}^{W\times 3}$, which are centred by subtracting their mask-weighted mean.

\subsubsection*{S3.7 Sliding-window deployment and consensus reassembly}

At training time R1 is supervised on a single window per forward pass; at inference and during R2 feature export the same module is unrolled over all $N-W+1$ windows of the chain in a windowed mini-batch (\texttt{win\_batch}=64) and the per-window outputs are stitched into chain-level tensors. Distance predictions $\hat{d}^{(k)}_{i,i+k}$ are indexed by their global residue pair via a precomputed gather index, and the windowed C$\alpha$ traces are stitched into a single chain by a parameter-free \emph{consensus} procedure: for each pair of overlapping windows $w$ and $w'$ the second is rigidly aligned to the first by Kabsch superposition\cite{kabsch1976solution} on the $W-1$ shared residues, contributions are accumulated by per-residue averaging, and the final consensus trace is centred. R1 trunk hidden states ($\mathbf{h}^{(7)}$, $\mathbf{h}_\mathrm{coord}$, $\mathbf{z}^{(7)}$) and per-scale head pre-activations are exported alongside the distances and coordinates and serve as direct inputs to R2.

\subsubsection*{S3.8 Loss}

\begin{equation}
	\fiteq{\mathcal{L}_{R1}=\sum_{k=1}^{K}\frac{1}{|\mathcal{P}_k|}\sum_{(i,j)\in\mathcal{P}_k}\mathrm{SL1}\bigl(\hat{d}_{ij}^{(k)},d_{ij}^\mathrm{true}\bigr)+\mathrm{SL1}\bigl(\hat{\mathbf{X}}-\mathrm{COM}(\hat{\mathbf{X}}),\mathbf{X}^\mathrm{true}-\mathrm{COM}(\mathbf{X}^\mathrm{true})\bigr),}
\end{equation}
with $\mathcal{P}_k=\{(i,j):j-i=k,\mathrm{mask}_i=\mathrm{mask}_j=1\}$, smooth-L1 transition point $\beta=1$. We tested weighting short-range scales more heavily---intuitively appealing because they are tighter and easier to predict---and found this marginally improved short-range fits at the cost of substantially worse 5--7-separation distances and a degraded final C$\alpha$ trace. Equal weighting was therefore used.

\subsubsection*{S3.9 Optimisation and training schedule}

R1 is trained from random initialisation on sliding windows distributed across GPUs via PyTorch DistributedDataParallel, AdamW\cite{loshchilov2019decoupled} ($\beta_1=0.9$, $\beta_2=0.999$, weight decay $0.01$) with peak learning rate $3\!\times\!10^{-4}$, 2{,}000-step linear warm-up followed by cosine annealing\cite{loshchilov2017sgdr} to the end of training, gradient-norm clipping at 1.0, bfloat16 mixed-precision autocast (fp32 SVD inside Kabsch alignment for numerical stability\cite{kabsch1976solution}). Consecutive non-finite loss/gradient-norm steps are detected and skipped without parameter updates; if the running count exceeds a fixed threshold the learning rate is multiplicatively reduced.

\begin{table}[!tb]
\centering\small
\caption{R1 parameter counts by component ($W=15$, $c_s=384$, $c_z=128$, $K=7$, ESM-C dimension 1{,}152). Component values are rounded to the nearest hundred; the total is the sum of the rounded components.}
\label{tab:r1params}
\fitcolumn{\begin{tabular}{@{}llr@{}}
\toprule
Component & Description & Parameters \\
\midrule
CondEncoder                  & $\mathbf{v}+\mathrm{FE}(Q,\tilde{R}_g,f_{ss})\to\mathbf{c}\in\mathbb{R}^{384}$ & 665{,}100 \\
$\mathbf{e}_i$ projection    & $\mathbb{R}^{1152}\to\mathbb{R}^{384}$               & 442{,}800 \\
Absolute positional embed.   & $\mathrm{Embedding}(512,384)$                       & 196{,}600 \\
Trunk Transformer            & 6 layers, pre-LN, 16 heads, exp.~4                  & 10{,}647{,}600 \\
Pair init.                   & rel.\ pos.\ + cond$\to\mathbf{z}$                   & 51{,}100 \\
PairFormer stack             & $7\times$ blocks (single + pair updates)            & 22{,}632{,}700 \\
MultiScaleAgg (intra-stack)  & $7\times$ attention poolers                         & 2{,}075{,}100 \\
Conditioning re-injection    & $7\times$ bias-free linears, zero-init              & 1{,}032{,}200 \\
Anchored distance heads      & $7\times$ MLP heads, $\tanh$ bounded                & 1{,}723{,}000 \\
Coord-stage MultiScaleAgg    & Final cross-scale pooling                           & 296{,}400 \\
Coordinate refinement        & 2-layer pre-LN Transformer                          & 3{,}548{,}900 \\
Coordinate head              & Linear $384\to 3$, $\sigma=10^{-3}$ init            & 1{,}200 \\
\midrule
\textbf{R1 total}            &                                                     & \textbf{$\approx 43{,}312{,}700$} \\
\bottomrule
\end{tabular}}
\end{table}

\subsection*{S4. Round 2 (R2)}

\subsubsection*{S4.1 R1-conditioned input fusion and trunk}

The R2 single-stream input is a sum of five learned projections plus a positional embedding,
\begin{equation*}
	\fiteq{\mathbf{h}_\mathrm{in}=W_e\mathbf{e}_i+W_{h^7}\mathbf{h}^{(7)}_\mathrm{R1}+W_{h_c}\mathbf{h}_\mathrm{coord,R1}+W_X\hat{\mathbf{X}}_\mathrm{R1}+\mathrm{pos\_emb}(W)+\mathbf{c}.}
\end{equation*}
Although $\hat{\mathbf{X}}_\mathrm{R1}\in\mathbb{R}^{W\times 3}$ is low-dimensional, projecting it into $c_s$-space and adding it to the token stream allows R2 to read R1's coordinate prediction as a feature alongside the higher-dimensional R1 hidden states, which we found stabilised the early training dynamics relative to using only R1 distances. The trunk is a 6-layer pre-LN Transformer (independent weights from R1, identical hyperparameters) followed by a final layer normalisation. The pair stream is initialised by re-projecting the R1 pair output, $\mathbf{z}^{(0)}=W_{z^7}\mathbf{z}^{(7)}_\mathrm{R1}+\mathrm{rel\_pos}(W)+W_{c\to z}\mathbf{c}$, masked element-wise by the outer product of the residue mask.

\subsubsection*{S4.2 Multi-source feature-wise linear modulation}

The architectural novelty of R2 is its dense use of FiLM\cite{perez2018film}. We use the standard FiLM transformation $\mathrm{FiLM}(\mathbf{h};\bm{\gamma},\bm{\beta})=\bm{\gamma}\odot\mathbf{h}+\bm{\beta}$ with $(\bm{\gamma},\bm{\beta})=\mathrm{Linear}(\mathbf{u})$ from a per-residue side-channel $\mathbf{u}$. Every projection that produces $(\bm{\gamma},\bm{\beta})$ is initialised with zero weights, so at the start of training every FiLM channel acts as the identity; together with the near-zero initialisation of the residual heads (S4.3, S4.5) this leaves the network output equal to R1's to within numerical noise. Modulation strength is then learned data-adaptively. The FiLM input at layer $\ell\in\{1,\ldots,8\}$ is the sum of four streams,
\begin{equation}
	\fiteq{\delta\mathbf{h}^{(\ell)}=\sum_{k=1}^{7}\mathrm{FiLM}\bigl(\mathbf{g}^{(k)}_\mathrm{R1}\!\to\!W\bigr)+\sum_{k=1}^{\ell-1}\mathrm{FiLM}\bigl(\mathbf{g}^{(k)}_\mathrm{R2}\!\to\!W\bigr)+\mathbb{1}_{\ell\geq 3}\mathrm{FiLM}(\mathbf{g}^{\theta\tau}_\mathrm{R2})+\mathbb{1}_{\ell\geq 5}\mathrm{FiLM}(\mathbf{g}^{\mu\sigma}_\mathrm{R2}).}
\end{equation}
The first two terms inject R1 and partial R2 distance-head pre-activations from all $K=7$ scales, padded from their natural shape $(W-k,c_g)$ back to $(W,c_g)$ by a parameter-free pad-and-symmetrise rule that places each scale-$k$ feature on both endpoints of every $(i,i+k)$ pair; an affine-free layer-norm\cite{ba2016layernorm} is shared across the four streams to keep their gradients on the same scale.

\subsubsection*{S4.3 Cascade and residual distance heads}

Inside the cascade, the block input is
\begin{equation*}
	\fiteq{\mathbf{h}^{(k)}_\mathrm{in}=\mathbf{h}^{(k-1)}+\mathrm{MSAgg}_k\bigl(\{\mathbf{h}^{(j)}\}_{j<k}\bigr)+\mathrm{CondFiLM}_k(\mathbf{c})+\delta\mathbf{h}^{(k)},}
\end{equation*}
followed by the multi-source FiLM update. After block~4, four additional pair-FiLM modules conditioned on the pair-branch hidden state $\mathbf{g}^D_\mathrm{R2}$ (one per remaining block plus the coordinate layer) modulate $\mathbf{z}$. The residual distance head at scale $k$ reads the same triple of features used by R1 and additionally concatenates $\mathbf{c}$,
\begin{equation*}
	\fiteq{\delta\hat{d}^{(k)}_{i,i+k}=\mathrm{MLP}\bigl(\mathbf{h}^{(k)}_i,\mathbf{h}^{(k)}_{i+k},\mathbf{z}^{(k)}_{i,i+k},\mathbf{c}\bigr),\quad \hat{d}^{(k)}_\mathrm{R2}=\hat{d}^{(k)}_\mathrm{R1}+\delta\hat{d}^{(k)};}
\end{equation*}
the MLP shape is $(2c_s+c_z+c_s)\to c_g\to 64\to 1$, with the final layer initialised at $\sigma=10^{-3}$ and zero bias.

\subsubsection*{S4.4 Auxiliary supervision branches}

Four auxiliary branches are attached at fixed points in the cascade. Each is a depth-$d$ feed-forward MLP with width $c_a=128$ ($c_\mathrm{dij}=32$ for the pair branch), GELU activation, with the penultimate hidden state used as FiLM source.
\noindent\emph{Geometry (after block 2; depth 3, output 3):} predicts per-residue C$\alpha$ pseudo-bond angle $\theta_i$ (sigmoid-bounded on $[0,\pi]$) and pseudo-torsion components $\sin\tau_i,\cos\tau_i$\cite{yang2020trrosetta}.
\noindent\emph{Profile (after block 4; depth 2, output 2):} predicts per-residue mean $\mu_i$ and standard deviation $\sigma_i$ of empirical C$\alpha$--C$\alpha$ distance distributions in the window centred at residue~$i$.
\noindent\emph{Pair (after block 4; depth 2, output 1):} predicts the full $W\!\times\!W$ pairwise distance map via a softplus output\cite{senior2020alphafold}; this provides a long-range distance signal that complements the seven short-range scale-specific predictions.
\noindent\emph{Global (after block 7; depth 3, output 4):} predicts radius of gyration, end-to-end distance, mean contact order and fraction of compactness as ablation diagnostics.

\subsubsection*{S4.5 Coordinate refinement}

The coordinate stage takes $\bar{\mathbf{h}}=\mathrm{MSAgg}_\mathrm{coord}(\{\mathbf{h}^{(k)}\}_{k=1}^{7})$, applies a final round of multi-source FiLM modulation (the $\ell{=}8$ position) and a CondFiLM re-injection of $\mathbf{c}$, and refines through $L=4$ CoordRefineLayers; each combines a CondFiLM block, a single-token attention layer biased by the FiLM-modulated pair stream and a feed-forward Transition with expansion 4. The final layer-norm output is decoded by a linear head ($\sigma=10^{-3}$, zero bias) to a coordinate residual $\delta\mathbf{X}$ that yields $\hat{\mathbf{X}}_\mathrm{R2}=\hat{\mathbf{X}}_\mathrm{R1}+\delta\mathbf{X}$ after centroid removal.

\subsubsection*{S4.6 Loss}

The total loss aggregates seven distance terms, one coordinate term and four auxiliary terms,
\begin{equation*}
	\fiteq{\mathcal{L}_{R2}=\sum_{k=1}^{7}\lambda_k^d\mathrm{SL1}(\hat{d}^{(k)}_\mathrm{R2},d^{(k)}_\star)+\lambda^x\mathrm{SL1}(\hat{\mathbf{X}}_\mathrm{R2},\mathbf{X}_\star^\mathrm{aligned})+\sum_{a\in\{\theta\tau,\mu\sigma,D,\mathrm{glob}\}}\lambda^a\mathcal{L}_a,}
\end{equation*}
with smooth-L1 cutoffs $\beta=1$\,\AA{} for distances and $\beta=0.5$ for coordinates. Per-scale distance weights are uniform; the four auxiliary terms together carry weight $0.25$ ($\lambda^{\theta\tau}=0.10$, $\lambda^{\mu\sigma}=0.05$, $\lambda^D=0.05$, $\lambda^\mathrm{glob}=0.05$).

\subsubsection*{S4.7 Optimisation and feature caching}

R1 is frozen and its outputs (distances, coordinates and the four hidden-state tensors) are pre-computed once per training shard with bfloat16 precision and stored as memory-mapped arrays on shared scratch storage; this removes any forward-pass cost of R1 during R2 optimisation. R2 is optimised with AdamW\cite{loshchilov2019decoupled} (peak learning rate $2\!\times\!10^{-4}$, 2{,}000-step warm-up, cosine annealing\cite{loshchilov2017sgdr} to $5\!\times\!10^{-6}$, gradient-norm clipping at 1.0, bfloat16 mixed precision with full-precision optimiser state). Training runs on 8 NVIDIA H100 GPUs (data-parallel, no model sharding) and reaches its plateau in approximately 18 hours. We early-stop on the held-out smooth-L1 distance loss averaged over all seven scales.

\begin{table}[!tb]
\centering\small
\caption{R2 parameter breakdown ($W=15$, $c_s=384$, $c_z=128$, $c_g=256$, $c_a=128$, 7 cascade blocks). Sub-module values are rounded independently to $0.01$\,M and therefore need not sum exactly to the total.}
\label{tab:r2params}
\fitcolumn{\begin{tabular}{@{}lr@{}}
\toprule
Sub-module & Parameters \\
\midrule
Trunk + input projections (CondEncoder, $W_e,W_{h^7},W_{h_c},W_X$, pos\_emb, 6-layer trunk) & 12.25\,M \\
Pair init.\ ($W_{z^7}$, rel\_pos, cond\_to\_z) & 0.07\,M \\
Cascade: 7 PairFormer blocks + 7 MultiScaleAgg + 7 CondFiLM & 26.77\,M \\
Multi-source FiLM (R1+R2+geometry+profile, 8 cascade positions) & 23.59\,M \\
Pair-stream FiLM (4 modules, blocks 5--7 + coord layer) & 0.03\,M \\
Residual distance heads (7 heads, $\delta$-MLP including $\mathbf{c}$ injection) & 2.41\,M \\
Auxiliary branches (geometry, profile, pair, global) & 0.40\,M \\
Coordinate refinement (final MSAgg + CondFiLM + 4 CoordRefineLayer + linear head) & 11.83\,M \\
\midrule
\textbf{Total R2} & \textbf{77.36\,M} \\
\bottomrule
\end{tabular}}
\end{table}

\subsection*{S5. Stage 1}

\subsubsection*{S5.1 CrossWindowPool}

The chain-assembly module relies on a single primitive, the CrossWindowPool, which converts a variable-length list of per-window feature vectors at a given chain index into a single $c_a=256$-dimensional summary. Given $\mathbf{x}\in\mathbb{R}^{n\times m\times c_\mathrm{in}}$ ($n$ chain indices, $m$ candidate windows per index, $c_\mathrm{in}$ raw features) and a Boolean validity mask $\mathbf{v}\in\{0,1\}^{n\times m}$, we compute
\begin{equation}
	\fiteq{\mathbf{h}=\mathrm{Linear}_\mathrm{in}(\mathbf{x}),\quad \alpha_{nm}=\frac{\exp(\langle W_\mathrm{key}\mathbf{h}_{nm},\mathbf{q}\rangle)\mathbf{v}_{nm}}{\sum_{m'}\exp(\langle W_\mathrm{key}\mathbf{h}_{nm'},\mathbf{q}\rangle)\mathbf{v}_{nm'}},\quad \mathbf{y}_n=\mathrm{LN}\bigl(W_\mathrm{out}\textstyle\sum_m\alpha_{nm}\mathbf{h}_{nm}\bigr),}
\end{equation}
with one learned query $\mathbf{q}\in\mathbb{R}^{c_a}$, key projection $W_\mathrm{key}$ and output projection $W_\mathrm{out}$ followed by layer normalisation\cite{ba2016layernorm}. This is the AlphaFold-style attention pooler\cite{bahdanau2015attention,vaswani2017attention,jumper2021alphafold} restricted to a single learnable query: every chain index learns a window-selection policy that down-weights edge windows and up-weights centred ones. Empirically the learned attention concentrates $>80\%$ of its mass on the central window for centre-eligible pairs and degrades smoothly for edge pairs.

\subsubsection*{S5.2 Near-range and long-range pair features}

\noindent\emph{Near-range ($k\leq 7$, $c_\mathrm{in}=804$).} For each window $w$ that contains pair $(i,j=i+k)$ with local indices $(\ell_i,\ell_j)$, the feature vector concatenates: R1 anchored distance $\hat{d}^{(k)}_{\mathrm{R1},w,\ell_i}$, R2 refined distance $\hat{d}^{(k)}_{\mathrm{R2},w,\ell_i}$, the two coordinate-derived distances, the R1 distance-head pre-activation $\mathbf{g}^{(k)}_{\mathrm{R1},w,\ell_i}\in\mathbb{R}^{c_g}$, the R2 pre-activation, the two pair features $\mathbf{z}^{(7)}_{\cdot,w,\ell_i,\ell_j}\in\mathbb{R}^{c_z}$ and a learned 32-dim embedding of the in-window starting position $p_w$. Both anchored and coordinate-derived distances are kept because the anchored heads regress directly to a target distance whereas coordinate-derived distances inherit Kabsch-stitching artefacts; the network learns to weight them adaptively.

\noindent\emph{Long-range ($k\geq 8$, $c_\mathrm{in}=290$).} The feature vector drops the anchored predictions and the distance-head pre-activations (which do not exist for $k>K$) and concatenates only the R1/R2 coordinate-derived distances, the R1/R2 pair features at $(\ell_i,\ell_j)$ and a window-position embedding.

After pooling, near and far per-pair vectors are written into the chain-level pair tensor $\mathbf{z}_\mathrm{pool}\in\mathbb{R}^{N\times N\times c_a}$ at $(i,j)$ and its mirror $(j,i)$; entries with $|i-j|\geq W$ remain zero, reflecting the absence of any window simultaneously containing both residues.

\subsubsection*{S5.3 Single-stream cross-window pooling}

A third CrossWindowPool ($c_\mathrm{in}=4c_s+32=1568$) produces per-residue summaries by gathering, for every chain residue $i$, the four trunk hidden states from every window containing it---R1 last-block single $\mathbf{h}^{(7)}_\mathrm{R1}$, R1 coordinate-stage $\mathbf{h}_\mathrm{coord,R1}$ and their R2 counterparts---together with a 32-dim embedding of the residue's local position within each window, producing $\mathbf{s}_h\in\mathbb{R}^{N\times c_a}$.

\subsubsection*{S5.4 Pair-stream initialisation}

\begin{equation*}
	\fiteq{\mathbf{z}^{(0)}=W_\mathrm{p2z}\mathbf{z}_\mathrm{pool}+\mathrm{rel\_pos}_{\max_N}+W_\mathrm{c2z}\mathbf{c}+W_\mathrm{rbf}\,\mathrm{RBF}(\|\hat{\mathbf{X}}_{\mathrm{cons},i}-\hat{\mathbf{X}}_{\mathrm{cons},j}\|),}
\end{equation*}
with codebook size $\max_N=200$ and $n_\mathrm{rbf}=24$ Gaussian bases linearly spaced over $[0,30]$\,\AA. Element-wise masking by the residue-mask outer product zeros padded positions.

\subsubsection*{S5.5 Five-source gated fusion}

The chain-level single representation must integrate signals of very different statistical character: high-dimensional ESM-C embeddings, pooled R1/R2 hidden states, an explicit distance signal, the global conditioning vector and the consensus coordinates. We merge them with a five-source gated fusion that learns, per residue, how much to trust each source\cite{srivastava2015highway,vaswani2017attention}:
\begin{align*}
\mathbf{u}^{(1)}&=W_e\mathbf{e}_i+\mathrm{pos\_emb}_{\max_N}\quad\text{(ESM-C+pos)},\\
\mathbf{u}^{(2)}&=\mathrm{bilateral\_proj}(\mathbf{s}_d)\quad\text{(distance signal)},\\
\mathbf{u}^{(3)}&=W_h\mathbf{s}_h\quad\text{(pooled R1/R2 hidden)},\\
\mathbf{u}^{(4)}&=\mathbf{c}\quad\text{(broadcast conditioning)},\\
\mathbf{u}^{(5)}&=\mathrm{MLP}(\hat{\mathbf{X}}_\mathrm{cons})\quad\text{(consensus coordinate projection)},
\end{align*}
where $\mathbf{s}_d\in\mathbb{R}^{N\times c_a}$ is the bilateral broadcast of the pair-pooled features back onto both endpoints of every $(i,j)$. The fusion applies a learned sigmoid gate to each source and projects the concatenation back to $c_s$,
\begin{equation*}
	\fiteq{\mathbf{s}^{(0)}=W_\mathrm{fuse}\,\mathrm{cat}\bigl(\sigma(W_g^{(1)}\mathbf{u}^{(1)})\odot\mathbf{u}^{(1)},\ldots,\sigma(W_g^{(5)}\mathbf{u}^{(5)})\odot\mathbf{u}^{(5)}\bigr).}
\end{equation*}

\subsubsection*{S5.6 Full-chain PairFormer trunk}

$L=6$ identical PairFormer blocks act on $(\mathbf{s},\mathbf{z})$. Block $\ell$ updates the single stream as $\mathbf{s}\leftarrow\mathrm{CondFiLM}_\ell^c(\mathbf{s},\mathbf{c})$, $\mathbf{s}\leftarrow\mathrm{CondFiLM}_\ell^d(\mathbf{s},\mathbf{s}_d)$, $\mathbf{s}\leftarrow\mathbf{s}+W_\ell^\mathrm{esm}\mathbf{u}^{(1)}$ ($W_\ell^\mathrm{esm}$ initialised to zero), then applies the standard PairFormer update of $(\mathbf{s},\mathbf{z})$. Every other block ($\ell\in\{2,4,6\}$) re-injects the conditioner's pair output into $\mathbf{z}$ via a zero-initialised linear projection multiplied by the pair mask. We chose $L=6$ because the trunk acts on full chains and is not parameterised per-scale.

\subsubsection*{S5.7 Coordinate head and single-representation export}

The single-stream output of block $L$ is layer-normalised and decoded by a linear head ($c_s\to 3$, $\sigma=10^{-2}$, zero bias) to a residual $\delta\mathbf{X}$; the same output (without the coordinate head) is exported as the single representation $\mathbf{s}\in\mathbb{R}^{N\times c_s}$ that conditions the downstream FM decoder. Both pathways are masked element-wise and centroid-corrected, $\hat{\mathbf{X}}_{S1}=\hat{\mathbf{X}}_\mathrm{cons}+\delta\mathbf{X}-\mathrm{mean}_\mathrm{mask}(\hat{\mathbf{X}}_\mathrm{cons}+\delta\mathbf{X})$. Although flow matching ultimately decodes coordinates from $\mathbf{s}$ alone, providing the auxiliary coordinate head during Stage~1 training measurably improves the structural quality of $\mathbf{s}$.

\subsubsection*{S5.8 Loss, feature caching and parameter table}

Loss: $\mathcal{L}_{S1}=\mathrm{SL1}(\hat{\mathbf{X}}_{S1},\mathbf{X}_\star^\mathrm{aligned})$ with smooth-L1 cutoff $\beta=0.5$\,\AA, ground truth Kabsch-aligned\cite{kabsch1976solution} to the consensus trace before the loss is taken. Optimisation: AdamW\cite{loshchilov2019decoupled} (peak learning rate $1\!\times\!10^{-4}$), 1{,}000-step warm-up, cosine annealing\cite{loshchilov2017sgdr} to $5\!\times\!10^{-6}$, gradient-norm clipping at 1.0, bfloat16 mixed precision; effective batch size 32 chains on 8 H100 GPUs (length-bucketed). The training record reported in Section~\ref{sec:res_training} covers the first epoch of this run.

\begin{table}[!tb]
\centering\small
\caption{Stage~1 parameter breakdown ($N\leq 200$, $c_s=384$, $c_z=128$, $c_a=256$, $W=15$, $K=7$, $L=6$).}
\label{tab:s1params}
\fitcolumn{\begin{tabular}{@{}lr@{}}
\toprule
Component & Parameters \\
\midrule
Chain-assembly module: 3 $\times$ CrossWindowPool + projections + RBF features & 1.31\,M \\
Single-stream initialisation (ESM proj, pos, CondEncoder, $\hat{\mathbf{X}}_\mathrm{cons}$ MLP, GatedFusion) & 2.74\,M \\
Pair-stream initialisation (rel\_pos, cond\_to\_z, RBF projection) & 0.08\,M \\
6-layer chain-level PairFormer trunk & 19.40\,M \\
Per-layer FiLM (CondFiLM$^c$, CondFiLM$^d$) and ESM/pair re-injections & 4.47\,M \\
Coord head (LN $\to c_s\to 3$ linear) & $<0.01$\,M \\
\midrule
\textbf{Total Stage 1} & \textbf{28.00\,M} \\
\bottomrule
\end{tabular}}
\end{table}

\subsection*{S6. Flow matching}

\subsubsection*{S6.1 Coordinate normalisation and zero-COM constraint}

Training chains are mean-centred to zero centre-of-mass before the flow-matching interpolation. The noise prior is sampled as $\mathbf{X}_0\sim\mathcal{N}(\mathbf{0},\mathbf{I}_3)^N$ and projected to the zero-COM subspace, so both endpoints of $\mathbf{X}_t=(1-t)\mathbf{X}_0+t\mathbf{X}_1$ lie on the same affine subspace and the interpolant remains zero-COM throughout training\cite{lipman2023flow,yim2023se3}. We do not impose rotational gauge fixing because the decoder is invariant only to translations; the training set is augmented with random global rotations applied jointly to $\mathbf{X}_0$ and $\mathbf{X}_1$.

\subsubsection*{S6.2 Time embedding and AdaLN conditioning}

Flow time $t\in[0,1]$ is encoded by a deterministic Fourier embedding\cite{tancik2020fourier} with 32 logarithmically spaced frequencies; the 64-dim sin/cos vector is projected by a 2-layer MLP to a $c_t=384$-dim time embedding, summed with a per-residue broadcast of $\mathbf{s}$ to give the layer-wise conditioning vector $\mathbf{c}_\ell\in\mathbb{R}^{N\times 384}$. Each Transformer block applies AdaLN-Zero\cite{peebles2023dit} using $(\bm{\gamma},\bm{\beta},\bm{\alpha})=\mathrm{Linear}(\mathbf{c}_\ell)$ to scale, shift and gate the attention and transition sub-layers, with $\bm{\alpha}$ initialised to zero so the network outputs the identity in the residual branches at the start of training. Two cond-side Transition layers (expansion 2) refine the conditioning signal non-linearly before modulation.

\subsubsection*{S6.3 Coordinate decoder}

\noindent\emph{Token stream.} The token-stream input is the sum of a linear projection of $\mathbf{X}_t$ ($3\to c_s$, no bias) and a feature-factory output that concatenates a chain-break indicator and---when self-conditioning is enabled---the self-conditioning coordinates $\mathbf{X}_\mathrm{sc}$ (also linearly projected). At each layer, the token stream is updated by a residual MHA-with-pair-bias block followed by a residual Transition, both AdaLN-modulated by $\mathbf{c}_\ell$; query/key layer normalisation\cite{henry2020qkln} and rotary positional encoding\cite{su2024rope} are applied within attention.

\noindent\emph{Pair representation.} A pair representation $\mathbf{P}\in\mathbb{R}^{N\times N\times 128}$ is built once from a feature factory that concatenates: relative sequence separation (127-bin one-hot), Gaussian-RBF features of $\|\mathbf{X}_{t,i}-\mathbf{X}_{t,j}\|$ (64 bins, range 0.1--3.0 in scaled units) and Gaussian-RBF features of $\|\mathbf{X}_{\mathrm{sc},i}-\mathbf{X}_{\mathrm{sc},j}\|$ (128 bins, same range) when self-conditioning is enabled. The pair representation is constant across layers; inside each MHA, a learned linear projection $\mathbb{R}^{128}\to\mathbb{R}^{n_\mathrm{heads}}$ converts $\mathbf{P}_{ij}$ into a per-head additive bias added to the attention logits. We do not propagate triangle multiplication or triangle attention updates inside the FM trunk.

\noindent\emph{Registers.} Following recent work on register tokens for vision and protein Transformers\cite{darcet2024registers}, 32 learned $c_s$-dimensional register tokens are prepended to the token stream before the trunk and removed before the coordinate readout, absorbing high-norm outlier tokens that otherwise emerge in the early layers.

\subsubsection*{S6.4 Coordinate readout and clean-coordinate parametrisation}

The single-stream output of the last block is decoded by a linear head $W_\mathrm{coord}\in\mathbb{R}^{c_s\times 3}$ to a clean-coordinate prediction $\hat{\mathbf{X}}_1$. We adopt the clean-coordinate parametrisation rather than directly predicting the velocity, because in our regime the variance of the coordinate target is bounded by data statistics and is more numerically stable than the velocity target which diverges as $t\to 1$\cite{lipman2023flow,liu2023rectified}; the velocity field is recovered exactly as $\hat{\mathbf{v}}=(\hat{\mathbf{X}}_1-\mathbf{X}_t)/(1-t)$. The output is masked element-wise and centroid-corrected before the loss is taken.

\subsubsection*{S6.5 Self-conditioning}

We include the self-conditioning trick from\cite{chen2023bitdiff,watson2023rfdiffusion}: with probability $0.5$ a training step first runs the decoder once with $\mathbf{X}_\mathrm{sc}=\mathbf{0}$ to obtain a stop-gradient prediction $\hat{\mathbf{X}}_1^{\,\mathrm{sc}}$, then runs the decoder a second time with $\mathbf{X}_\mathrm{sc}=\hat{\mathbf{X}}_1^{\,\mathrm{sc}}$ to obtain the loss-bearing prediction; with probability $0.5$ the network is trained with $\mathbf{X}_\mathrm{sc}=\mathbf{0}$. At sampling time the previous integration step's clean-coordinate prediction is fed in as $\mathbf{X}_\mathrm{sc}$ at every step.

\subsubsection*{S6.6 Composite training loss}

The full loss combines the rectified-flow translation loss with three auxiliary loss groups, each weighted by a global coefficient and optionally by the cosine time schedule.

\noindent\emph{Translation loss.}
\begin{equation*}
	\fiteq{\mathcal{L}_\mathrm{trans}=w_\mathrm{FM}(t)\cdot\frac{1}{3|\mathcal{M}|}\sum_{i\in\mathcal{M}}\|\hat{\mathbf{X}}_{1,i}-\mathbf{X}_{1,i}\|^2,\quad w_\mathrm{FM}(t)=\min\!\Bigl(\frac{1}{(1-t)^2+10^{-5}},10^4\Bigr).}
\end{equation*}

\noindent\emph{Global structure losses (early denoising; multiplied by $w_g(t)$).} Per-sample C$\alpha$-RMSD after Kabsch alignment\cite{kabsch1976solution} ($\lambda_\mathrm{recon}=1$), differentiable lDDT score on C$\alpha$ atoms\cite{mariani2013lddt} ($\lambda_\mathrm{lDDT}=2$), contact-map BCE at the 8\,\AA{} cutoff ($\lambda_\mathrm{contact}=0.5$), differentiable $R_g$ mismatch ($\lambda_{R_g}=1$), differentiable $f_{ss}$ mismatch ($\lambda_{f_{ss}}=1$), end-to-end distance mismatch ($\lambda_\mathrm{e2e}=0.5$).

\noindent\emph{Local distance loss (late denoising; multiplied by $w_l(t)$).} Smooth-L1 distance consistency on all $K=7$ short sequence separations,
\begin{equation*}
	\fiteq{\mathcal{L}_\mathrm{local}=\frac{1}{K}\sum_{k=1}^{K}\frac{1}{|\mathcal{M}_k|}\sum_{(i,i+k)\in\mathcal{M}_k}\mathrm{SL1}_{\beta=1}\bigl(\|\hat{\mathbf{X}}_{1,i}-\hat{\mathbf{X}}_{1,i+k}\|,\|\mathbf{X}_{1,i}-\mathbf{X}_{1,i+k}\|\bigr),\;\lambda_\mathrm{dist}=1.}
\end{equation*}

\noindent\emph{Physical constraints (always on).} Bond-length constraint $\sum_i\mathrm{SL1}(\|\hat{\mathbf{X}}_{1,i+1}-\hat{\mathbf{X}}_{1,i}\|,3.80\,\text{\AA})$ ($\lambda_\mathrm{bond}=1$), bond-angle constraint at the C$\alpha_{i-1}$--C$\alpha_i$--C$\alpha_{i+1}$ pseudo-angle ($\lambda_\mathrm{angle}=0.3$), soft clash penalty $\sum_{|i-j|\geq 4}\max(0,c-\|\hat{\mathbf{X}}_{1,i}-\hat{\mathbf{X}}_{1,j}\|)^2$ at $c=4.0$\,\AA{} ($\lambda_\mathrm{clash}=0.5$).

\noindent\emph{Outer schedule.}
\begin{equation*}
	\fiteq{\mathcal{L}_\mathrm{FM}=\lambda_\mathrm{trans}\mathcal{L}_\mathrm{trans}+\lambda_\mathrm{global}w_g(t)\mathcal{L}_\mathrm{global}+\lambda_\mathrm{local}w_l(t)\mathcal{L}_\mathrm{local}+\lambda_\mathrm{neutral}\mathcal{L}_\mathrm{phys},}
\end{equation*}
with $\lambda_\mathrm{trans}=1.0$, $\lambda_\mathrm{global}=0.5$, $\lambda_\mathrm{local}=1.0$, $\lambda_\mathrm{neutral}=1.0$.

\subsubsection*{S6.7 Optimisation}

R1, R2 and Stage~1 are frozen; Stage~1 single representations are pre-cached as bfloat16 memory-mapped tensors keyed by chain ID. The decoder is trained with AdamW\cite{loshchilov2019decoupled} ($\beta_1=0.9$, $\beta_2=0.95$, $\epsilon=10^{-8}$, weight decay $0.01$), peak learning rate $3\!\times\!10^{-4}$, 5{,}000-step warm-up, cosine annealing\cite{loshchilov2017sgdr} to $1\!\times\!10^{-5}$, gradient-norm clipping at 1.0, bfloat16 mixed precision. Training: effective batch size 64 chains on 8 NVIDIA H100 GPUs (data-parallel, length-bucketed); the run reported in Section~\ref{sec:res_training} was stopped at 46{,}233 steps, short of the configured budget. Time samples are drawn uniformly $t\sim\mathcal{U}(0,1)$; we tested logit-normal\cite{esser2024sd3} and stratified time schedules and observed no improvement at our chain-length range.

\subsubsection*{S6.8 Sampling}

At inference, conformations are generated by integrating $\mathrm{d}\mathbf{X}_t/\mathrm{d}t=\hat{\mathbf{v}}(\mathbf{X}_t,t,\mathbf{s})$ from $t=0$ to $t=1$ using forward Euler with 100 uniform steps ($\Delta t=0.01$), starting from $\mathbf{X}_0\sim\mathcal{N}(\mathbf{0},\mathbf{I}_3)$ projected to the zero-COM subspace. Self-conditioning is enabled at sampling time. Conformational diversity is achieved by drawing many independent $\mathbf{X}_0$ for the same $\mathbf{s}$, which corresponds to drawing many independent samples from $p_\mathrm{FM}(\mathbf{X}\mid\mathbf{s},Q,\tilde{R}_g,f_{ss})$. The score-based sampling mode of the underlying R$^3$N flow matcher\cite{karras2022edm,zheng2023guided} is supported but produces no measurable gain over independent prior samples in our regime.

\begin{table}[!tb]
\centering\small
\caption{FoldScape parameter budget. Values are rounded to the nearest
0.1~M and are computed from the layer dimensions of each module; they are
therefore estimates of the deployed parameter count rather than a runtime
census. The two pretraining probes exist only during conformational latent
pretraining and are discarded afterwards, so they do not enter the inference
forward pass. The flow-matching decoder is the smallest module in that pass
($\sim$8\% of the total); the bulk of the structural reasoning is performed by
R1, R2 and Stage~1 in producing $\mathbf{s}$.}
\label{tab:pipelineparams}
\fitcolumn{\begin{tabular}{@{}lr@{}}
\toprule
Pipeline component & Parameters \\
\midrule
\multicolumn{2}{@{}l}{\emph{Training-time only (discarded after pretraining)}} \\
\quad Coordinate-reconstruction Decoder probe & 3.5\,M \\
\quad Q-Predictor probe & 0.6\,M \\
\midrule
\multicolumn{2}{@{}l}{\emph{Inference forward pass}} \\
\quad ConformExpander (frozen) & 4.3\,M \\
\quad R1 (frozen) & 43.3\,M \\
\quad R2 (frozen) & 77.4\,M \\
\quad Stage 1 (frozen) & 28.0\,M \\
\quad Flow-matching decoder & $\approx 13.0$\,M \\
\midrule
\textbf{Total FoldScape (forward pass at inference)} & $\approx 166.0$\,M \\
\textbf{Total instantiated during training} & $\approx 170.1$\,M \\
\bottomrule
\end{tabular}}
\end{table}

\subsection*{S7. All-atom reconstruction}

\subsubsection*{S7.1 Overview and rationale}

C$\alpha$-only ensembles are interpretable for global topology and large-scale conformational dynamics, but quantitative analyses requiring side-chain or backbone-amide degrees of freedom---hydrogen-bond networks, packing, contact geometry, docking against small-molecule ligands---require all-atom coordinates\cite{wassenaar2014backward,badaczewska2020reconstruction}. No single back-mapping scheme is systematically optimal across all secondary-structure classes, loop conformations and intrinsically disordered segments: geometry-based methods perform well on regular secondary structure but can mis-assign side-chain rotamers in crowded environments, whereas learned methods inherit the bias of their training distribution but generalise better to unusual local geometries. We therefore adopted a dual-pipeline strategy and used the agreement between the two reconstructions as an internal consistency check.

\subsubsection*{S7.2 cg2all (deep-learning reconstruction)}

cg2all\cite{heo2024oneparticle} is an SE(3)-equivariant graph neural network that maps a C$\alpha$-only trace to a full heavy-atom protein in a single deterministic forward pass. Each residue is represented as a node whose scalar and vector features are updated through equivariant message-passing layers, so that rotations and translations of the input are transmitted exactly through the network and emerge as the same rigid motion of the output coordinates\cite{kmiecik2016coarsegrained,ingolfsson2014power}. We used the pretrained CalphaBasedModel checkpoint distributed with cg2all, trained on high-resolution Protein Data Bank structures and designed specifically for one-C$\alpha$-per-residue inputs. Reconstruction was performed frame-by-frame on every snapshot of the predicted CG ensembles without post-hoc energy minimisation, in order to preserve the statistics of the upstream CG model. Because the equivariant network is deterministic, repeated inference reproduces exactly the same heavy-atom coordinates, allowing the reconstruction step to be treated as a fixed transformation of the CG ensemble.

\subsubsection*{S7.3 PULCHRA (geometry-based reconstruction)}

PULCHRA\cite{rotkiewicz2008pulchra} reconstructs an all-atom protein in three stages: (i)~backbone N, C and O atoms are rebuilt by searching a curated tetrapeptide-fragment library and selecting, for each residue, the fragment whose C$\alpha$ geometry best matches the local four-residue window; (ii)~side-chain atoms are placed using a backbone-dependent rotamer library\cite{shapovalov2011rotamer}, with the rebuilt $\varphi,\psi$ dihedrals selecting the most probable rotamer for each residue; (iii)~a short local relaxation removes the worst steric clashes while leaving the overall backbone trace essentially unchanged. PULCHRA's input-preservation option was enabled throughout, so the C$\alpha$ coordinates of the output match those of the CG input to numerical precision, ensuring that PULCHRA and cg2all share the same C$\alpha$ frame for cross-method comparison.

\subsubsection*{S7.4 Input preparation, frame selection and stereochemical validation}

Each CG structure was supplied to both reconstructors in PDB format with one C$\alpha$ atom per residue and chain identifiers retained from the upstream model. Residues flagged as unresolved in the CG predictions were omitted rather than extrapolated, so the reconstructed ensembles contain no atoms for which the upstream model did not commit to a C$\alpha$ position. For multi-chain assemblies, reconstruction was performed on the full complex in a single pass; in all cases the inter-chain C$\alpha$ distances of the input were preserved.

All reconstructed models were validated with MolProbity\cite{williams2018molprobity}: all-atom clashscore (number of serious steric overlaps per 1{,}000 atoms), Ramachandran-favoured and -outlier percentages, side-chain rotamer-outlier percentage, percentage of C$\beta$ deviations $>0.25$\,\AA{} and the composite MolProbity score. Frames with MolProbity score $>3.0$ are flagged. Distributions of per-frame scores accompany the deposited analysis outputs.

\subsubsection*{S7.5 Cross-method consistency analysis}

For every CG snapshot the cg2all and PULCHRA models were superposed on their C$\alpha$ atoms---a trivial operation, since both pipelines preserve the input C$\alpha$ positions---and compared at the heavy-atom level. We quantified agreement through three complementary metrics: the all-heavy-atom RMSD; the paired distributions of backbone $\varphi,\psi$ dihedrals, compared by Kullback--Leibler divergence on a $10^\circ$ Ramachandran grid; and per-residue $\chi_1$ rotamer assignments drawn from the same rotamer library\cite{shapovalov2011rotamer}. A residue was flagged as rotamer-ambiguous if its $\chi_1$ rotamer state differed between the two reconstructions in more than 30\% of the ensemble frames; such residues were excluded from any subsequent analysis that depends on side-chain orientation, including hydrogen-bond networks and ligand-pocket identification.

\subsubsection*{S7.6 Loops, disordered segments and limitations}

For loop and disordered segments where neither method has a clear ``ground truth'', cg2all and PULCHRA may select different rotamer or backbone-conformation modes; we retain the cg2all output as the canonical reconstruction (because of its deterministic equivariant properties) but flag those regions for downstream interpretation. Both schemes preserve the C$\alpha$ coordinates and cannot recover atomic detail that conflicts with the upstream C$\alpha$-only prediction; cross-method agreement does not certify physical correctness, only consistency. The workflow reconstructs heavy atoms only; polar hydrogens were added with standard chemistry-aware protonation tools at neutral pH in a separate step downstream of the pipeline. Related back-mapping strategies at other resolutions\cite{souza2021martini3} and the all-atom structure module of AlphaFold\cite{jumper2021alphafold} were not benchmarked against the present workflow.


\begin{thebibliography}{999}
\small
\setlength{\itemsep}{1pt}

\bibitem{anfinsen1973principles}
Anfinsen, C. B. Principles that govern the folding of protein
chains. Science 181, 223--230 (1973).

\bibitem{bryngelson1995funnels}
Bryngelson, J. D., Onuchic, J. N., Socci, N. D. \& Wolynes, P. G.
Funnels, pathways, and the energy landscape of protein folding: a
synthesis. Proteins 21, 167--195 (1995).

\bibitem{onuchic1997theory}
Onuchic, J. N., Luthey-Schulten, Z. \& Wolynes, P. G. Theory of
protein folding: the energy landscape perspective. Annu. Rev. Phys.
Chem. 48, 545--600 (1997).

\bibitem{wolynes1997folding}
Wolynes, P. G. Folding funnels and energy landscapes of larger
proteins within the capillarity approximation. Proc. Natl Acad. Sci. USA
94, 6170--6175 (1997).

\bibitem{abramson2024alphafold3}
Abramson, J. et al. Accurate structure prediction of biomolecular
interactions with AlphaFold 3. Nature 630, 493--500 (2024).

\bibitem{baek2021rosettafold}
Baek, M. et al. Accurate prediction of protein structures and
interactions using a three-track neural network. Science 373, 871--876
(2021).

\bibitem{lin2023esmfold}
Lin, Z. et al. Evolutionary-scale prediction of atomic-level protein
structure with a language model. Science 379, 1123--1130 (2023).

\bibitem{dill2012folding}
Dill, K. A. \& MacCallum, J. L. The protein-folding problem, 50 years
on. Science 338, 1042--1046 (2012).

\bibitem{outeiral2022physics}
Outeiral, C., Nissley, D. A. \& Deane, C. M. Current structure
predictors are not learning the physics of protein folding.
Bioinformatics 38, 1881--1887 (2022).

\bibitem{henzlerwildman2007dynamics}
Henzler-Wildman, K. \& Kern, D. Dynamic personalities of proteins.
Nature 450, 964--972 (2007).

\bibitem{shaw2010atomic}
Shaw, D. E. et al. Atomic-level characterization of the structural
dynamics of proteins. Science 330, 341--346 (2010).

\bibitem{lindorfflarsen2011fastfolding}
Lindorff-Larsen, K., Piana, S., Dror, R. O. \& Shaw, D. E. How
fast-folding proteins fold. Science 334, 517--520 (2011).

\bibitem{clementi2000topological}
Clementi, C., Nymeyer, H. \& Onuchic, J. N. Topological and
energetic factors: what determines the structural details of the
transition state ensemble and ``en-route'' intermediates for protein
folding? An investigation for small globular proteins. J. Mol. Biol.
298, 937--953 (2000).

\bibitem{davtyan2012awsem}
Davtyan, A. et al. AWSEM-MD: protein structure prediction using
coarse-grained physical potentials and bioinformatically based local
structure biasing. J. Phys. Chem. B 116, 8494--8503 (2012).

\bibitem{torrie1977umbrella}
Torrie, G. M. \& Valleau, J. P. Nonphysical sampling distributions
in Monte Carlo free-energy estimation: umbrella sampling. J. Comput.
Phys. 23, 187--199 (1977).

\bibitem{kumar1992wham}
Kumar, S., Rosenberg, J. M., Bouzida, D., Swendsen, R. H. \&
Kollman, P. A. The weighted histogram analysis method for free-energy
calculations on biomolecules. I. The method. J. Comput. Chem. 13,
1011--1021 (1992).

\bibitem{noe2019boltzmann}
Noé, F., Olsson, S., Köhler, J. \& Wu, H. Boltzmann generators:
sampling equilibrium states of many-body systems with deep learning.
Science 365, eaaw1147 (2019).

\bibitem{lewis2025bioemu}
Lewis, S. et al. Scalable emulation of protein equilibrium
ensembles with generative deep learning. Science 389, eadv9817 (2025).

\bibitem{jing2024alphaflow}
Jing, B., Berger, B. \& Jaakkola, T. AlphaFold meets flow matching
for generating protein ensembles. In Proc. 41st International Conference
on Machine Learning vol. 235, 22277--22303 (PMLR, 2024).

\bibitem{zheng2024distributional}
Zheng, S. et al. Predicting equilibrium distributions for molecular
systems with deep learning. Nat. Mach. Intell. 6, 558--567 (2024).

\bibitem{lee2026stability}
Lee, H. et al. Protein folding stability estimation with explicit
consideration of unfolded states. Nat. Commun. 17, 1883 (2026).

\bibitem{chang2026pathways}
Chang, L. \& Perez, A. Rapid estimation of protein folding pathways
from sequence alone using AlphaFold2. Nat. Commun. 17, 170 (2026).

\bibitem{das2006landscapes}
Das, P., Moll, M., Stamati, H., Kavraki, L. E. \& Clementi, C.
Low-dimensional, free-energy landscapes of protein-folding reactions by
nonlinear dimensionality reduction. Proc. Natl Acad. Sci. USA 103,
9885--9890 (2006).

\bibitem{perez2018film}
Perez, E., Strub, F., de Vries, H., Dumoulin, V. \& Courville, A.
FiLM: visual reasoning with a general conditioning layer.
In \textit{Proc.\ AAAI Conf.\ Artif.\ Intell.} (AAAI 2018), 3942--3951 (2018).

\bibitem{jumper2021alphafold}
Jumper, J. \emph{et al.}
Highly accurate protein structure prediction with AlphaFold.
\textit{Nature} \textbf{596}, 583--589 (2021).

\bibitem{geffner2025proteina}
Geffner, T. \emph{et al.}
Proteina: scaling flow-based protein structure generative models.
In \textit{Proc.\ Int.\ Conf.\ Learn.\ Represent.} (ICLR 2025) (2025).

\bibitem{lipman2023flow}
Lipman, Y., Chen, R.\ T.\ Q., Ben-Hamu, H., Nickel, M. \& Le, M.
Flow matching for generative modeling.
In \textit{Proc.\ Int.\ Conf.\ Learn.\ Represent.} (ICLR 2023) (2023).

\bibitem{liu2023rectified}
Liu, X., Gong, C. \& Liu, Q.
Flow straight and fast: learning to generate and transfer data with rectified flow.
In \textit{Proc.\ Int.\ Conf.\ Learn.\ Represent.} (ICLR 2023) (2023).

\bibitem{albergo2023stochastic}
Albergo, M.\ S. \& Vanden-Eijnden, E.
Building normalizing flows with stochastic interpolants.
In \textit{Proc.\ Int.\ Conf.\ Learn.\ Represent.} (ICLR 2023) (2023).

\bibitem{varadi2024alphafold}
Varadi, M. \emph{et al.}
AlphaFold Protein Structure Database in 2024: providing structure coverage for over 214 million protein sequences.
\textit{Nucleic Acids Res.} \textbf{52}, D368--D375 (2024).

\bibitem{steinegger2017mmseqs2}
Steinegger, M. \& S\"oding, J.
MMseqs2 enables sensitive protein sequence searching for the analysis of massive data sets.
\textit{Nat.\ Biotechnol.} \textbf{35}, 1026--1028 (2017).

\bibitem{cock2009biopython}
Cock, P.\ J.\ A. \emph{et al.}
Biopython: freely available Python tools for computational molecular biology and bioinformatics.
\textit{Bioinformatics} \textbf{25}, 1422--1423 (2009).

\bibitem{walter2024goca}
Walter, L.\ J., Quoika, P.\ K. \& Zacharias, M.
Structure-based protein assembly simulations including various binding sites and conformations.
\textit{J.\ Chem.\ Inf.\ Model.} \textbf{64}, 3465--3476 (2024).

\bibitem{abraham2015gromacs}
Abraham, M.\ J. \emph{et al.}
GROMACS: high performance molecular simulations through multi-level parallelism from laptops to supercomputers.
\textit{SoftwareX} \textbf{1--2}, 19--25 (2015).

\bibitem{bussi2007canonical}
Bussi, G., Donadio, D. \& Parrinello, M.
Canonical sampling through velocity rescaling.
\textit{J.\ Chem.\ Phys.} \textbf{126}, 014101 (2007).

\bibitem{eastman2017openmm}
Eastman, P. \emph{et al.}
OpenMM 7: rapid development of high performance algorithms for molecular dynamics.
\textit{PLoS Comput.\ Biol.} \textbf{13}, e1005659 (2017).

\bibitem{maier2015ff14sb}
Maier, J.\ A. \emph{et al.}
ff14SB: improving the accuracy of protein side chain and backbone parameters from ff99SB.
\textit{J.\ Chem.\ Theory Comput.} \textbf{11}, 3696--3713 (2015).

\bibitem{onufriev2004exploring}
Onufriev, A., Bashford, D. \& Case, D.\ A.
Exploring protein native states and large-scale conformational changes with a modified generalized Born model.
\textit{Proteins} \textbf{55}, 383--394 (2004).

\bibitem{sotomayor2007single}
Sotomayor, M. \& Schulten, K.
Single-molecule experiments in vitro and in silico.
\textit{Science} \textbf{316}, 1144--1148 (2007).

\bibitem{lu1998unfolding}
Lu, H., Isralewitz, B., Krammer, A., Vogel, V. \& Schulten, K.
Unfolding of titin immunoglobulin domains by steered molecular dynamics simulation.
\textit{Biophys.\ J.} \textbf{75}, 662--671 (1998).

\bibitem{neelamraju2019gokit}
Neelamraju, S., Wales, D.\ J. \& Gosavi, S.
Go-kit: a tool to enable energy landscape exploration of proteins.
\textit{J.\ Chem.\ Inf.\ Model.} \textbf{59}, 1703--1708 (2019).

\bibitem{lu2021openawsem}
Lu, W. \emph{et al.}
OpenAWSEM with Open3SPN2: a fast, flexible, and accessible framework for large-scale coarse-grained biomolecular simulations.
\textit{PLoS Comput.\ Biol.} \textbf{17}, e1008308 (2021).

\bibitem{papoian2004role}
Papoian, G.\ A., Ulander, J., Eastwood, M.\ P., Luthey-Schulten, Z. \& Wolynes, P.\ G.
Water in protein structure prediction.
\textit{Proc.\ Natl Acad.\ Sci.\ USA} \textbf{101}, 3352--3357 (2004).

\bibitem{mokhtari2026dynarepo}
Mokhtari, O., Bignon, E., Khakzad, H. \& Karami, Y.
DynaRepo: the repository of macromolecular conformational dynamics.
\textit{Nucleic Acids Res.} \textbf{54}, D393--D401 (2026).

\bibitem{rodriguezespigares2020gpcrmd}
Rodr{\'\i}guez-Espigares, I. \emph{et al.}
GPCRmd uncovers the dynamics of the 3D-GPCRome.
\textit{Nat.\ Methods} \textbf{17}, 777--787 (2020).

\bibitem{roy2025mdrepo}
Roy, A. \emph{et al.}
MDRepo---an open data warehouse for community-contributed molecular dynamics simulations of proteins.
\textit{Nucleic Acids Res.} \textbf{53}, D477--D486 (2025).

\bibitem{beltran2024bioexcelcv19}
Beltr{\'a}n, D., Hospital, A., Gelp{\'\i}, J.\ L. \& Orozco, M.
A new paradigm for molecular dynamics databases: the COVID-19 database, the legacy of a titanic community effort.
\textit{Nucleic Acids Res.} \textbf{52}, D393--D403 (2024).

\bibitem{pirtskhalava2021dbaasp}
Pirtskhalava, M. \emph{et al.}
DBAASP v3: database of antimicrobial/cytotoxic activity and structure of peptides as a resource for development of new therapeutics.
\textit{Nucleic Acids Res.} \textbf{49}, D288--D297 (2021).

\bibitem{hayes2025esm3}
Hayes, T. \emph{et al.}
Simulating 500 million years of evolution with a language model.
\textit{Science} \textbf{387}, 850--858 (2025).

\bibitem{best2013native}
Best, R.\ B., Hummer, G. \& Eaton, W.\ A.
Native contacts determine protein folding mechanisms in atomistic simulations.
\textit{Proc.\ Natl Acad.\ Sci.\ USA} \textbf{110}, 17874--17879 (2013).

\bibitem{kohn2004random}
Kohn, J.\ E. \emph{et al.}
Random-coil behavior and the dimensions of chemically unfolded proteins.
\textit{Proc.\ Natl Acad.\ Sci.\ USA} \textbf{101}, 12491--12496 (2004).

\bibitem{kneller2015ss}
Kneller, G.\ R. \& Hinsen, K.
Protein secondary-structure description with a coarse-grained model.
\textit{Acta Crystallogr.\ D Biol.\ Crystallogr.} \textbf{71}, 1411--1422 (2015).

\bibitem{tancik2020fourier}
Tancik, M. \emph{et al.}
Fourier features let networks learn high frequency functions in low dimensional domains.
In \textit{Adv.\ Neural Inf.\ Process.\ Syst.} (NeurIPS 2020), 7537--7547 (2020).

\bibitem{vaswani2017attention}
Vaswani, A. \emph{et al.}
Attention is all you need.
In \textit{Adv.\ Neural Inf.\ Process.\ Syst.} (NeurIPS 2017), 5998--6008 (2017).

\bibitem{elfwing2018silu}
Elfwing, S., Uchibe, E. \& Doya, K.
Sigmoid-weighted linear units for neural network function approximation in reinforcement learning.
\textit{Neural Netw.} \textbf{107}, 3--11 (2018).

\bibitem{dhariwal2021diffusion}
Dhariwal, P. \& Nichol, A.
Diffusion models beat GANs on image synthesis.
In \textit{Adv.\ Neural Inf.\ Process.\ Syst.} (NeurIPS 2021), 8780--8794 (2021).

\bibitem{shazeer2020glu}
Shazeer, N.
GLU variants improve Transformer.
Preprint at \url{https://arxiv.org/abs/2002.05202} (2020).

\bibitem{kabsch1976solution}
Kabsch, W.
A solution for the best rotation to relate two sets of vectors.
\textit{Acta Crystallogr.\ A} \textbf{32}, 922--923 (1976).

\bibitem{nix1994estimating}
Nix, D.\ A. \& Weigend, A.\ S.
Estimating the mean and variance of the target probability distribution.
In \textit{Proc.\ IEEE Int.\ Conf.\ Neural Networks} vol.~1, 55--60 (IEEE, 1994).

\bibitem{loshchilov2019decoupled}
Loshchilov, I. \& Hutter, F.
Decoupled weight decay regularization.
In \textit{Proc.\ Int.\ Conf.\ Learn.\ Represent.} (ICLR 2019) (2019).

\bibitem{loshchilov2017sgdr}
Loshchilov, I. \& Hutter, F.
SGDR: stochastic gradient descent with warm restarts.
In \textit{Proc.\ Int.\ Conf.\ Learn.\ Represent.} (ICLR 2017) (2017).

\bibitem{xiong2020layernorm}
Xiong, R. \emph{et al.}
On layer normalization in the Transformer architecture.
In \textit{Proc.\ Int.\ Conf.\ Mach.\ Learn.} (ICML 2020), 10524--10533 (2020).

\bibitem{girshick2015fastrcnn}
Girshick, R.
Fast R-CNN.
In \textit{Proc.\ IEEE Int.\ Conf.\ Comput.\ Vis.} (ICCV 2015), 1440--1448 (2015).

\bibitem{huber1964robust}
Huber, P.\ J.
Robust estimation of a location parameter.
\textit{Ann.\ Math.\ Stat.} \textbf{35}, 73--101 (1964).

\bibitem{caruana1997multitask}
Caruana, R.
Multitask learning.
\textit{Mach.\ Learn.} \textbf{28}, 41--75 (1997).

\bibitem{he2016resnet}
He, K., Zhang, X., Ren, S. \& Sun, J.
Deep residual learning for image recognition.
In \textit{Proc.\ IEEE Conf.\ Comput.\ Vis.\ Pattern Recognit.} (CVPR 2016), 770--778 (2016).

\bibitem{yang2020trrosetta}
Yang, J. \emph{et al.}
Improved protein structure prediction using predicted interresidue orientations.
\textit{Proc.\ Natl Acad.\ Sci.\ USA} \textbf{117}, 1496--1503 (2020).

\bibitem{senior2020alphafold}
Senior, A.\ W. \emph{et al.}
Improved protein structure prediction using potentials from deep learning.
\textit{Nature} \textbf{577}, 706--710 (2020).

\bibitem{bahdanau2015attention}
Bahdanau, D., Cho, K. \& Bengio, Y.
Neural machine translation by jointly learning to align and translate.
In \textit{Proc.\ Int.\ Conf.\ Learn.\ Represent.} (ICLR 2015) (2015).

\bibitem{peebles2023dit}
Peebles, W. \& Xie, S.
Scalable diffusion models with transformers.
In \textit{Proc.\ IEEE/CVF Int.\ Conf.\ Comput.\ Vis.} (ICCV 2023), 4195--4205 (2023).

\bibitem{chen2023bitdiff}
Chen, T., Zhang, R. \& Hinton, G.
Analog bits: generating discrete data using diffusion models with self-conditioning.
In \textit{Proc.\ Int.\ Conf.\ Learn.\ Represent.} (ICLR 2023) (2023).

\bibitem{darcet2024registers}
Darcet, T., Oquab, M., Mairal, J. \& Bojanowski, P.
Vision Transformers need registers.
In \textit{Proc.\ Int.\ Conf.\ Learn.\ Represent.} (ICLR 2024) (2024).

\bibitem{mariani2013lddt}
Mariani, V., Biasini, M., Barbato, A. \& Schwede, T.
lDDT: a local superposition-free score for comparing protein structures and models using distance difference tests.
\textit{Bioinformatics} \textbf{29}, 2722--2728 (2013).

\bibitem{wassenaar2014backward}
Wassenaar, T.\ A., Pluhackova, K., B{\"o}ckmann, R.\ A., Marrink, S.\ J. \& Tieleman, D.\ P.
Going backward: a flexible geometric approach to reverse transformation from coarse-grained to atomistic models.
\textit{J.\ Chem.\ Theory Comput.} \textbf{10}, 676--690 (2014).

\bibitem{heo2024oneparticle}
Heo, L. \& Feig, M.
One particle per residue can describe all-atom protein structures.
\textit{Structure} \textbf{32}, 97--111.e6 (2024).

\bibitem{rotkiewicz2008pulchra}
Rotkiewicz, P. \& Skolnick, J.
Fast procedure for reconstruction of full-atom protein models from reduced representations.
\textit{J.\ Comput.\ Chem.} \textbf{29}, 1460--1465 (2008).

\bibitem{shapovalov2011rotamer}
Shapovalov, M.\ V. \& Dunbrack, R.\ L. Jr.
A smoothed backbone-dependent rotamer library for proteins derived from adaptive kernel density estimates and regressions.
\textit{Structure} \textbf{19}, 844--858 (2011).

\bibitem{williams2018molprobity}
Williams, C.\ J. \emph{et al.}
MolProbity: more and better reference data for improved all-atom structure validation.
\textit{Protein Sci.} \textbf{27}, 293--315 (2018).

\bibitem{fowler2001mapping}
Fowler, S.\ B. \& Clarke, J.
Mapping the folding pathway of an immunoglobulin domain: structural detail from phi value analysis and movement of the transition state.
\textit{Structure} \textbf{9}, 355--366 (2001).

\bibitem{kim2000specific}
Kim, D.\ E., Fisher, C. \& Baker, D.
A breakdown of symmetry in the folding transition state of protein L.
\textit{J.\ Mol.\ Biol.} \textbf{298}, 971--984 (2000).

\bibitem{best2002mechanical}
Best, R.\ B., Fowler, S.\ B., Toca-Herrera, J.\ L. \& Clarke, J.
A simple method for probing the mechanical unfolding pathway of proteins in detail.
\textit{Proc.\ Natl Acad.\ Sci.\ USA} \textbf{99}, 12143--12148 (2002).

\bibitem{nauli2001crystal}
Nauli, S., Kuhlman, B. \& Baker, D.
Computer-based redesign of a protein folding pathway.
\textit{Nat.\ Struct.\ Biol.} \textbf{8}, 602--605 (2001).

\bibitem{went2005ubiquitin}
Went, H.~M. \& Jackson, S.~E.
Ubiquitin folds through a highly polarized transition state.
\textit{Protein Eng. Des. Sel.} \textbf{18}, 229--237 (2005).

\bibitem{carrionvazquez1999mechanical}
Carri{\'o}n-V{\'a}zquez, M. \emph{et al.}
Mechanical and chemical unfolding of a single protein: a comparison.
\textit{Proc.\ Natl Acad.\ Sci.\ USA} \textbf{96}, 3694--3699 (1999).

\bibitem{schlierf2004force}
Schlierf, M., Li, H. \& Fernandez, J.\ M.
The unfolding kinetics of ubiquitin captured with single-molecule force-clamp techniques.
\textit{Proc.\ Natl Acad.\ Sci.\ USA} \textbf{101}, 7299--7304 (2004).

\bibitem{brockwell2005pulling}
Brockwell, D.\ J. \emph{et al.}
Mechanically unfolding the small, topologically simple protein L.
\textit{Biophys.\ J.} \textbf{89}, 506--519 (2005).

\bibitem{cao2007polyprotein}
Cao, Y. \& Li, H.
Polyprotein of GB1 is an ideal artificial elastomeric protein.
\textit{Nat.\ Mater.} \textbf{6}, 109--114 (2007).

\bibitem{shen2012bihistidine}
Shen, T., Cao, Y., Zhuang, S. \& Li, H.
Engineered bi-histidine metal chelation sites map the structure of the
mechanical unfolding transition state of an elastomeric protein domain
GB1.
\textit{Biophys. J.} \textbf{103}, 807--816 (2012).

\bibitem{karanicolas2002origins}
Karanicolas, J. \& Brooks, C.\ L. III.
The origins of asymmetry in the folding transition states of protein L and protein G.
\textit{Protein Sci.} \textbf{11}, 2351--2361 (2002).

\bibitem{lu2000key}
Lu, H. \& Schulten, K.
The key event in force-induced unfolding of titin's immunoglobulin
domains.
\textit{Biophys. J.} \textbf{79}, 51--65 (2000).

\bibitem{isralewitz2001steered}
Isralewitz, B., Gao, M. \& Schulten, K.
Steered molecular dynamics and mechanical functions of proteins.
\textit{Curr. Opin. Struct. Biol.} \textbf{11}, 224--230 (2001).

\bibitem{matouschek1989mapping}
Matouschek, A., Kellis, J.~T. Jr, Serrano, L. \& Fersht, A.~R.
Mapping the transition state and pathway of protein folding by protein
engineering.
\textit{Nature} \textbf{340}, 122--126 (1989).

\bibitem{duarte2010optimal}
Duarte, J.~M., Sathyapriya, R., Stehr, H., Filippis, I. \& Lappe, M.
Optimal contact definition for reconstruction of contact maps.
\textit{BMC Bioinformatics} \textbf{11}, 283 (2010).

\bibitem{sadler2009rheostat}
Sadler, D.~P. \emph{et al.}
Identification of a mechanical rheostat in the hydrophobic core of protein~L.
\textit{J. Mol. Biol.} \textbf{393}, 237--248 (2009).

\bibitem{cao2008chelation}
Cao, Y., Yoo, T. \& Li, H.
Single molecule force spectroscopy reveals engineered metal chelation is
a general approach to enhance mechanical stability of proteins.
\textit{Proc. Natl Acad. Sci. USA} \textbf{105}, 11152--11157 (2008).

\bibitem{rico2013highspeed}
Rico, F., Gonzalez, L., Casuso, I., Puig-Vidal, M. \& Scheuring, S.
High-speed force spectroscopy unfolds titin at the velocity of molecular
dynamics simulations.
\textit{Science} \textbf{342}, 741--743 (2013).

\bibitem{wako1978statistical1}
Wako, H. \& Sait\^o, N.
Statistical mechanical theory of the protein conformation. I. General
considerations and the application to homopolymers.
\textit{J. Phys. Soc. Jpn.} \textbf{44}, 1931--1938 (1978).

\bibitem{wako1978statistical2}
Wako, H. \& Sait\^o, N.
Statistical mechanical theory of the protein conformation. II. Folding
pathway for protein.
\textit{J. Phys. Soc. Jpn.} \textbf{44}, 1939--1945 (1978).

\bibitem{munoz1999simple}
Mu\~noz, V. \& Eaton, W.~A.
A simple model for calculating the kinetics of protein folding from
three-dimensional structures.
\textit{Proc. Natl Acad. Sci. USA} \textbf{96}, 11311--11316 (1999).

\bibitem{bruscolini2002exact}
Bruscolini, P. \& Pelizzola, A.
Exact solution of the Mu\~noz--Eaton model for protein folding.
\textit{Phys. Rev. Lett.} \textbf{88}, 258101 (2002).

\bibitem{imparato2007ising}
Imparato, A., Pelizzola, A. \& Zamparo, M.
Ising-like model for protein mechanical unfolding.
\textit{Phys. Rev. Lett.} \textbf{98}, 148102 (2007).

\bibitem{maity2005protein}
Maity, H., Maity, M., Krishna, M.~M.~G., Mayne, L. \& Englander, S.~W.
Protein folding: the stepwise assembly of foldon units.
\textit{Proc. Natl Acad. Sci. USA} \textbf{102}, 4741--4746 (2005).

\bibitem{englander2014nature}
Englander, S.~W. \& Mayne, L.
The nature of protein folding pathways.
\textit{Proc. Natl Acad. Sci. USA} \textbf{111}, 15873--15880 (2014).

\bibitem{kabsch1978discussion}
Kabsch, W.
A discussion of the solution for the best rotation to relate two sets of
vectors.
\textit{Acta Crystallogr. A} \textbf{34}, 827--828 (1978).

\bibitem{levitt1976simplified}
Levitt, M.
A simplified representation of protein conformations for rapid simulation
of protein folding.
\textit{J. Mol. Biol.} \textbf{104}, 59--107 (1976).

\bibitem{oldfield1994analysis}
Oldfield, T.~J. \& Hubbard, R.~E.
Analysis of C$\alpha$ geometry in protein structures.
\textit{Proteins} \textbf{18}, 324--337 (1994).

\bibitem{kryshtafovych2023casp15}
Kryshtafovych, A., Schwede, T., Topf, M., Fidelis, K. \& Moult, J.
Critical assessment of methods of protein structure prediction (CASP)---Round~XV.
\textit{Proteins} \textbf{91}, 1539--1549 (2023).

\bibitem{lane2023frontier}
Lane, T.\ J.
Protein structure prediction has reached the single-structure frontier.
\textit{Nat.\ Methods} \textbf{20}, 170--173 (2023).

\bibitem{han2007multidomain}
Han, J.-H., Batey, S., Nickson, A.\ A., Teichmann, S.\ A. \& Clarke, J.
The folding and evolution of multidomain proteins.
\textit{Nat.\ Rev.\ Mol.\ Cell Biol.} \textbf{8}, 319--330 (2007).

\bibitem{borgia2011misfolding}
Borgia, A. \emph{et al.}
Single-molecule fluorescence reveals sequence-specific misfolding in multidomain proteins.
\textit{Nature} \textbf{474}, 662--665 (2011).

\bibitem{ingraham2023chroma}
Ingraham, J.\ B. \emph{et al.}
Illuminating protein space with a programmable generative model.
\textit{Nature} \textbf{623}, 1070--1078 (2023).

\bibitem{ahdritz2024openfold}
Ahdritz, G. \emph{et al.}
OpenFold: retraining AlphaFold2 yields new insights into its learning mechanisms and capacity for generalization.
\textit{Nat.\ Methods} \textbf{21}, 1514--1524 (2024).

\bibitem{majewski2023cgpotentials}
Majewski, M. \emph{et al.}
Machine learning coarse-grained potentials of protein thermodynamics.
\textit{Nat.\ Commun.} \textbf{14}, 5739 (2023).

\bibitem{karras2022edm}
Karras, T., Aittala, M., Aila, T. \& Laine, S.
Elucidating the design space of diffusion-based generative models.
In \textit{Adv.\ Neural Inf.\ Process.\ Syst.} (NeurIPS 2022) (2022).

\bibitem{rief1997reversible}
Rief, M., Gautel, M., Oesterhelt, F., Fernandez, J.~M. \& Gaub, H.~E.
Reversible unfolding of individual titin immunoglobulin domains by AFM.
\textit{Science} \textbf{276}, 1109--1112 (1997).

\bibitem{best2003titin}
Best, R.~B., Fowler, S.~B., Toca-Herrera, J.~L., Steward, A., Paci, E. \&
Clarke, J.
Mechanical unfolding of a titin Ig domain: structure of the transition
state revealed by combining atomic force microscopy, protein engineering
and molecular dynamics simulations.
\textit{J. Mol. Biol.} \textbf{330}, 867--877 (2003).

\bibitem{fowler2002intermediate}
Fowler, S.~B. \emph{et al.}
Mechanical unfolding of a titin Ig domain: structure of the unfolding
intermediate revealed by combining AFM, molecular dynamics simulations, NMR
and protein engineering.
\textit{J. Mol. Biol.} \textbf{322}, 841--849 (2002).

\bibitem{mccallister2000hairpin}
McCallister, E.~L., Alm, E. \& Baker, D.
Critical role of $\beta$-hairpin formation in protein~G folding.
\textit{Nat. Struct. Biol.} \textbf{7}, 669--673 (2000).

\bibitem{marszalek1999mechanical}
Marszalek, P.~E. \emph{et al.}
Mechanical unfolding intermediates in titin modules.
\textit{Nature} \textbf{402}, 100--103 (1999).

\bibitem{chodera2014msm}
Chodera, J.\ D. \& No\'e, F.
Markov state models of biomolecular conformational dynamics.
\textit{Curr.\ Opin.\ Struct.\ Biol.} \textbf{25}, 135--144 (2014).

\bibitem{waymentsteele2024conformations}
Wayment-Steele, H.\ K. \emph{et al.}
Predicting multiple conformations via sequence clustering and AlphaFold2.
\textit{Nature} \textbf{625}, 832--839 (2024).

\bibitem{simons1997fragments}
Simons, K.\ T., Kooperberg, C., Huang, E. \& Baker, D.
Assembly of protein tertiary structures from fragments with similar local sequences using simulated annealing and Bayesian scoring functions.
\textit{J.\ Mol.\ Biol.} \textbf{268}, 209--225 (1997).

\bibitem{maiorov1994significance}
Maiorov, V.\ N. \& Crippen, G.\ M.
Significance of root-mean-square deviation in comparing three-dimensional structures of globular proteins.
\textit{J.\ Mol.\ Biol.} \textbf{235}, 625--634 (1994).

\bibitem{reva1998probability}
Reva, B.\ A., Finkelstein, A.\ V. \& Skolnick, J.
What is the probability of a chance prediction of a protein structure with an rmsd of 3\,\AA?
\textit{Fold.\ Des.} \textbf{3}, 141--147 (1998).

\bibitem{carugo2001normalized}
Carugo, O. \& Pongor, S.
A normalized root-mean-square distance for comparing protein three-dimensional structures.
\textit{Protein Sci.} \textbf{10}, 1470--1473 (2001).

\bibitem{olechnovic2013cadscore}
Olechnovi\v{c}, K., Kulberkyt\.{e}, E. \& Venclovas, \v{C}.
CAD-score: a new contact area difference-based function for evaluation of protein structural models.
\textit{Proteins} \textbf{81}, 149--162 (2013).

\bibitem{riback2017scattering}
Riback, J.\ A. \emph{et al.}
Innovative scattering analysis shows that hydrophobic disordered proteins are expanded in water.
\textit{Science} \textbf{358}, 238--241 (2017).

\bibitem{zhang2004tmscore}
Zhang, Y. \& Skolnick, J.
Scoring function for automated assessment of protein structure template quality.
\textit{Proteins} \textbf{57}, 702--710 (2004).

\bibitem{zemla2003lga}
Zemla, A.
LGA: a method for finding 3D similarities in protein structures.
\textit{Nucleic Acids Res.} \textbf{31}, 3370--3374 (2003).

\bibitem{bengio2009curriculum}
Bengio, Y., Louradour, J., Collobert, R. \& Weston, J.
Curriculum learning.
In \textit{Proc.\ Int.\ Conf.\ Mach.\ Learn.} (ICML 2009), 41--48 (2009).

\bibitem{hollingsworth2018md}
Hollingsworth, S.\ A. \& Dror, R.\ O.
Molecular dynamics simulation for all.
\textit{Neuron} \textbf{99}, 1129--1143 (2018).

\bibitem{krishna2024rfaa}
Krishna, R. \emph{et al.}
Generalized biomolecular modeling and design with RoseTTAFold All-Atom.
\textit{Science} \textbf{384}, eadl2528 (2024).

\bibitem{esser2024sd3}
Esser, P. \emph{et al.}
Scaling rectified flow Transformers for high-resolution image synthesis.
In \textit{Proc.\ Int.\ Conf.\ Mach.\ Learn.} (ICML 2024) (2024).

\bibitem{watson2023rfdiffusion}
Watson, J.\ L. \emph{et al.}
De novo design of protein structure and function with RFdiffusion.
\textit{Nature} \textbf{620}, 1089--1100 (2023).

\bibitem{fersht1995single}
Fersht, A.~R., Itzhaki, L.~S., ElMasry, N.~F., Matthews, J.~M. \& Otzen, D.~E.
Single versus parallel pathways of protein folding and fractional formation of
structure in the transition state.
\textit{Proc. Natl Acad. Sci. USA} \textbf{92}, 10426--10429 (1995).

\bibitem{brockwell2003pulling}
Brockwell, D.~J. \emph{et al.}
Pulling geometry defines the mechanical resistance of a $\beta$-sheet protein.
\textit{Nat. Struct. Biol.} \textbf{10}, 731--737 (2003).

\bibitem{khorasanizadeh1993folding}
Khorasanizadeh, S., Peters, I.~D., Butt, T.~R. \& Roder, H.
Folding and stability of a tryptophan-containing mutant of ubiquitin.
\textit{Biochemistry} \textbf{32}, 7054--7063 (1993).

\bibitem{bustamante1994entropic}
Bustamante, C., Marko, J.~F., Siggia, E.~D. \& Smith, S.
Entropic elasticity of $\lambda$-phage DNA.
\textit{Science} \textbf{265}, 1599--1600 (1994).

\bibitem{evans1997dynamic}
Evans, E. \& Ritchie, K.
Dynamic strength of molecular adhesion bonds.
\textit{Biophys. J.} \textbf{72}, 1541--1555 (1997).

\bibitem{bell1978models}
Bell, G.~I.
Models for the specific adhesion of cells to cells.
\textit{Science} \textbf{200}, 618--627 (1978).

\bibitem{li2002reverse}
Li, H. \emph{et al.}
Reverse engineering of the giant muscle protein titin.
\textit{Nature} \textbf{418}, 998--1002 (2002).

\bibitem{lindorfflarsen2012validation}
Lindorff-Larsen, K., Maragakis, P., Piana, S., Eastwood, M.\ P., Dror, R.\ O. \& Shaw, D.\ E.
Systematic validation of protein force fields against experimental data.
\textit{PLoS ONE} \textbf{7}, e32131 (2012).

\bibitem{dao2022flashattention}
Dao, T., Fu, D.\ Y., Ermon, S., Rudra, A. \& R{\'e}, C.
FlashAttention: fast and memory-efficient exact attention with IO-awareness.
In \textit{Adv.\ Neural Inf.\ Process.\ Syst.} (NeurIPS 2022), 16344--16359 (2022).

\bibitem{alain2016probes}
Alain, G. \& Bengio, Y.
Understanding intermediate layers using linear classifier probes.
In \textit{Proc.\ Int.\ Conf.\ Learn.\ Represent.\ (Workshop)} (2017).

\bibitem{ba2016layernorm}
Ba, J.\ L., Kiros, J.\ R. \& Hinton, G.\ E.
Layer normalization.
Preprint at \url{https://arxiv.org/abs/1607.06450} (2016).

\bibitem{hendrycks2016gelu}
Hendrycks, D. \& Gimpel, K.
Gaussian error linear units (GELUs).
Preprint at \url{https://arxiv.org/abs/1606.08415} (2016).

\bibitem{srivastava2015highway}
Srivastava, R.\ K., Greff, K. \& Schmidhuber, J.
Training very deep networks.
In \textit{Adv.\ Neural Inf.\ Process.\ Syst.} (NeurIPS 2015), 2377--2385 (2015).

\bibitem{yim2023se3}
Yim, J. \emph{et al.}
SE(3) diffusion model with application to protein backbone generation.
In \textit{Proc.\ Int.\ Conf.\ Mach.\ Learn.} (ICML 2023), 40001--40039 (2023).

\bibitem{henry2020qkln}
Henry, A., Dachapally, P.\ R., Pawar, S.\ S. \& Chen, Y.
Query-key normalization for Transformers.
In \textit{Findings of EMNLP 2020}, 4246--4253 (2020).

\bibitem{su2024rope}
Su, J., Lu, Y., Pan, S., Murtadha, A., Wen, B. \& Liu, Y.
RoFormer: enhanced Transformer with rotary position embedding.
\textit{Neurocomputing} \textbf{568}, 127063 (2024).

\bibitem{zheng2023guided}
Zheng, K., Lu, C., Chen, J. \& Zhu, J.
Improved techniques for maximum likelihood estimation for diffusion ODEs.
In \textit{Proc.\ Int.\ Conf.\ Mach.\ Learn.} (ICML 2023), 42363--42389 (2023).

\bibitem{badaczewska2020reconstruction}
Badaczewska-Dawid, A.\ E., Kolinski, A. \& Kmiecik, S.
Computational reconstruction of atomistic protein structures from coarse-grained models.
\textit{Comput.\ Struct.\ Biotechnol.\ J.} \textbf{18}, 162--176 (2020).

\bibitem{kmiecik2016coarsegrained}
Kmiecik, S. \emph{et al.}
Coarse-grained protein models and their applications.
\textit{Chem.\ Rev.} \textbf{116}, 7898--7936 (2016).

\bibitem{ingolfsson2014power}
Ing{\'o}lfsson, H.\ I. \emph{et al.}
The power of coarse graining in biomolecular simulations.
\textit{WIREs Comput.\ Mol.\ Sci.} \textbf{4}, 225--248 (2014).

\bibitem{souza2021martini3}
Souza, P.\ C.\ T. \emph{et al.}
Martini 3: a general purpose force field for coarse-grained molecular dynamics.
\textit{Nat.\ Methods} \textbf{18}, 382--388 (2021).

\bibitem{delalamo2022sampling}
del Alamo, D., Sala, D., Mchaourab, H.\ S. \& Meiler, J.
Sampling alternative conformational states of transporters and receptors with AlphaFold2.
\textit{eLife} \textbf{11}, e75751 (2022).

\bibitem{kabsch1983dssp}
Kabsch, W. \& Sander, C.
Dictionary of protein secondary structure: pattern recognition of
hydrogen-bonded and geometrical features.
\textit{Biopolymers} \textbf{22}, 2577--2637 (1983).

\bibitem{bentley1975kdtree}
Bentley, J.\ L.
Multidimensional binary search trees used for associative searching.
\textit{Commun.\ ACM} \textbf{18}, 509--517 (1975).

\bibitem{hurlbert1971rarefaction}
Hurlbert, S.\ H.
The nonconcept of species diversity: a critique and alternative parameters.
\textit{Ecology} \textbf{52}, 577--586 (1971).

\bibitem{gotelli2001richness}
Gotelli, N.\ J. \& Colwell, R.\ K.
Quantifying biodiversity: procedures and pitfalls in the measurement and
comparison of species richness.
\textit{Ecol.\ Lett.} \textbf{4}, 379--391 (2001).

\bibitem{efron1979bootstrap}
Efron, B.
Bootstrap methods: another look at the jackknife.
\textit{Ann.\ Stat.} \textbf{7}, 1--26 (1979).

\bibitem{ernst2004permutation}
Ernst, M.\ D.
Permutation methods: a basis for exact inference.
\textit{Stat.\ Sci.} \textbf{19}, 676--685 (2004).

\bibitem{mann1947utest}
Mann, H.\ B. \& Whitney, D.\ R.
On a test of whether one of two random variables is stochastically larger
than the other.
\textit{Ann.\ Math.\ Stat.} \textbf{18}, 50--60 (1947).

\bibitem{wilcoxon1945signedrank}
Wilcoxon, F.
Individual comparisons by ranking methods.
\textit{Biom.\ Bull.} \textbf{1}, 80--83 (1945).

\bibitem{holm1979bonferroni}
Holm, S.
A simple sequentially rejective multiple test procedure.
\textit{Scand.\ J.\ Stat.} \textbf{6}, 65--70 (1979).

\bibitem{harris2020numpy}
Harris, C.\ R. \emph{et al.}
Array programming with NumPy.
\textit{Nature} \textbf{585}, 357--362 (2020).

\bibitem{virtanen2020scipy}
Virtanen, P. \emph{et al.}
SciPy 1.0: fundamental algorithms for scientific computing in Python.
\textit{Nat.\ Methods} \textbf{17}, 261--272 (2020).

\bibitem{hunter2007matplotlib}
Hunter, J.\ D.
Matplotlib: a 2D graphics environment.
\textit{Comput.\ Sci.\ Eng.} \textbf{9}, 90--95 (2007).

\end{thebibliography}
\end{document}